\documentclass[10pt]{article}
\usepackage{amssymb,amsmath,fullpage,times,epsf}
\usepackage{CJK}
\usepackage{mathrsfs}
\usepackage{stfloats}
\usepackage{lipsum}
\usepackage{amssymb}
\usepackage{subfigure}
\usepackage{float}
\usepackage{graphics}
\usepackage{graphicx}
\usepackage{color}
\usepackage{bm}
\usepackage{amsfonts}
\usepackage{color}
\usepackage{comment}
\usepackage{extarrows}
\usepackage{epsfig}

\DeclareSymbolFont{largesymbol}{OMX}{yhex}{m}{n}
\DeclareMathAccent{\Widehat}{\mathord}{largesymbol}{"62}

\numberwithin{equation}{section}

\allowdisplaybreaks

\def\beginproof{\par\noindent\textbf{Proof.}~~}
\def\endproof{\ \vbox{\hrule\hbox{\vrule height1.0ex\hskip1.0ex\vrule}\hrule }\par\medskip}

\def\tr{\operatorname{tr}}

\newtheorem{theorem}{Theorem}[section]

\newtheorem{lemma}[theorem]{Lemma}

\newtheorem{proposition}[theorem]{Proposition}
\newtheorem{corollary}[theorem]{Corollary}

\newtheorem{remark}[theorem]{Remark}

\newcommand{\mi}{\mathrm{i}}

\def\diag{\mathrm{diag}}

\def\mb{\mathbf}

\def\Re{\mathrm{Re}}
\def\sech{\mathrm{sech}}

\def\log{\mathrm{log}}
\def\exp{\mathrm{exp}}

\def\Res{\mathrm{Res}}
\def\det{\mathrm{det}}
\def\max{\mathrm{max}}
\def\PT{$\mathcal{P}\mathcal{T}$}

\def\eref#1{(\ref{#1})}

\usepackage[square,numbers,sort&compress]{natbib}
\def\Im{\mathrm{Im}}

\begin{document}
\begin{CJK}{GBK}{song}
\date{}

\title{Inverse scattering transform for the defocusing nonlinear Schr\"{o}dinger equation with local and nonlocal nonlinearities under non-zero boundary conditions}

\author{Chuanxin Xu$^{1}$, Tao Xu$^{1,}$\thanks{Corresponding author, e-mail: xutao@cup.edu.cn.}\,, Min Li$^{2}$\\
	{\em 1. College of Science, China University of Petroleum, Beijing
		102249, China} \\
{\em 2. Department of Mathematics and Physics,}\\
{\em North China Electric Power University, Beijing 102206,
	China}}
\maketitle

\date{}
\vspace{-5mm}

\begin{abstract}


Within the framework of the Riemann-Hilbert problem, the theory of inverse scattering transform is established for the defocusing nonlinear Schr\"{o}dinger equation with local and nonlocal nonlinearities (which originates from the parity-symmetric reduction of the Manakov
system) under  non-zero boundary conditions. First, the adjoint Lax pair and auxiliary eigenfunctions are introduced for the direct scattering,
and the analyticity, symmetries of eigenfunctions and scattering matrix are studied in detail.
Then, the distribution of discrete eigenvalues is examined, and the asymptotic behaviors of the eigenfunctions and scattering coefficients are analyzed
rigorously. Compared with the Manakov system, the reverse-space nonlocality introduces an additional symmetry, leading to stricter constraints on
eigenfunctions, scattering coefficients and norming constants. Further, the Riemann-Hilbert problem is formulated for the the inverse problem with the scattering coefficients admitting an arbitrary number of simple zeros. For the reflectionless case, the  N-soliton solutions are presented in the determinant form. With N=1, the dark and beating one-soliton solutions are obtained, which are respectively associated with a pair of discrete eigenvalues lying on and off the circle on the spectrum plane. Via the asymptotic  analysis,  the two-soliton solutions are found to admit the interactions between two dark solitons or two beating solitons, as well as the superpositions of two beating solitons or one beating soliton and one dark soliton.

\vspace{5mm}


\end{abstract}

\newpage

\section{Introduction}
Over the past ten years, the nonlocal  nonlinear evolution equations (NEEs) have attracted much attention in the area of integrable systems.  In 2013, Ablowitz and Musslimani first proposed the following reverse-space nonlocal nonlinear Schr\"{o}dinger (NLS) equation~\cite{Ablowitz1}:
\begin{equation}
	\label{NNLSS1}
\mi q_t(x,t) = q_{xx}(x,t) + 2 \sigma q^2(x,t) q^*(-x,t),
\end{equation}
where the complex-valued function $q(x,t)$ depends on real variables $x$ and $t$, the asterisk denotes complex conjugate, $\sigma=  1$ and $\sigma= - 1$ represent the focusing $(+)$ and defocusing $(-)$ nonlinearities, respectively. In contrast to the celebrated NLS equation, the nonlocality of Eq.~\eref{NNLSS1} lies in that the nonlinear term depends on the values of the solutions both at the positions $x$ and $-x$. Also, this equation is  said to be parity-time (PT) symmetric  since the self-induced potential $ 2 \sigma q(x,t) q^*(-x,t)$ in Eq.~\eref{NNLSS1}
is generally complex and PT-symmetric~\cite{Ablowitz1,Ablowitz119, lnverse,Ablowitz11}.
Remarkably, Eq.~\eref{NNLSS1} can arise from  the Ablowitz-Kaup-Newell-Segur spectral problem with
the PT symmetric reduction (space reversal and complex conjugation), and thus it admits the Lax pair and an infinite number of
conservation laws. Recently, much attention was paid to its  mathematical properties~\cite{Ablowitz1,lnverse,Santini,response4,Genoud,feg6,Shepelsky2023,RY1,RY2,RY3,VSA} and various localized-wave solutions~\cite{Ablowitz1,lnverse,YJK3,Ablowitz119,Ablowitz11,FBF,LML, LiXu, Michor, Wen2, YZY, LiXu1,XCX,TSF6, Gurses,WPZ,ZDJ}. At the same time,
the  reverse-space, reverse-time and reverse-space-time nonlocal reductions were examined in the  traditional soliton equation hierarchies.
As a consequence, researchers discovered a number of nonlocal integrable NEEs, spanning from 1+1 and 1+2 dimensions to semi-discrete settings~\cite{Ablowitz21,Ablowitz119,Ablowitz6,Ablowitz7,Ablowitz8,Fokas,vector,Sinha,DNLS,Rao,mKdV,NNWave,Lou2,Cen,JKY,NKdV,W}.

It has been an important concern to establish the physical relevance of nonlocal integrable NEEs~\cite{Lou2,TAG,LiXu4,MJA2019}. In 2018, with a clear physical motivation, Yang proposed the following integrable NLS equation with local and nonlocal nonlinearities~\cite{JKY}:
\begin{align}
	\label{NNLS} \mi q_t(x,t) + q_{xx}(x,t) + 2 \sigma q(x,t) \big( |q(x,t)|^2  + |q(-x,t)|^2 \big) =0 \quad (\sigma= \pm 1),
\end{align}
where the self-induced potential $2 \sigma \big( |q(x,t)|^2  + |q(-x,t)|^2 \big)$ is real and even-symmetric in $x$.
In fact, Eq.~\eref{NNLS} can be reduced from the Manakov system~\cite{Manakov1}
\begin{subequations}
	\label{Manakov}
	\begin{align}
		&\mi q_t(x,t) + q_{xx}(x,t) + 2 \sigma q(x,t) \big( |q(x,t)|^2  + |r(x,t)|^2 \big) =0, \\
		&\mi r_t(x,t) + r_{xx}(x,t) + 2 \sigma r(x,t) \big( |q(x,t)|^2  + |r(x,t)|^2 \big) =0,
	\end{align}
\end{subequations}
with the symmetric constraint
\begin{align}
r(x,t) = q(-x,t). \label{PS}
\end{align}
As we know, the Manakov system is an integrable coupled model which  has significant applications in various physical systems with two components/fields interact nonlinearly~\cite{GPA,Manakov5,Manakov1,Kivshar,Manakov2,WZY,Manakov7,Manakov8}, e.g., the propagation of two orthogonally polarized optical pulses in birefringent fibers~\cite{GPA,Manakov5}, the interaction of two incoherent light beams in crystals~\cite{Kivshar,Manakov2}, and the evolution of two atomic clouds with different species in Bose-Einstein condensates~\cite{Manakov7,Manakov8}. Particularly, when the initial conditions satisfy a parity symmetry constraint between two components/fields, the resultant nonlinear wave dynamics can be governed by the  NLS equation with local and nonlocal nonlinearities~\eref{NNLS}.  Therefore, the physical meanings of Eq.~\eref{NNLS} become evident through its connection with the Manakov system~\eref{Manakov}.




As a physically important nonlocal model, the integrable properties and explicit solutions of Eq.~\eref{NNLS} were explored in recent years.
On the basis of the solutions derived from the Manakov system,
Yang studied the symmetry relations of discrete scattering data and obtained the bright one- and two-soliton solutions~\cite{JKY}. However, it is hard to derive the N-soliton solutions because the symmetry relations of eigenvectors depend on the number and locations of eigenvalues in a very intricate way. Subsequently, Wu developed the theory of inverse scattering transform (IST) for the initial-value problem of focusing Eq.~\eref{NNLS} under zero boundary condition (ZBC)~\cite{Wu}. In the framework of Riemann-Hilbert problem (RHP), Ref.~\cite{Wu} performed the spectral analysis from the $t$-part of the Lax pair instead of the $x$-part, and derived the N-soliton solutions in the reflectionless cases. Different from the soliton behaviors in Eq.~\eref{NNLSS1}, the bright-soliton solutions of Eq.~\eref{NNLS} are non-singular and can exhibit some interesting dynamics such as the amplitude-changing interactions~\cite{JKY,Wu}. In addition, Ref.~\cite{Chen2020} derived the bright multi-soliton solutions in the determinant form by the Hirota's bilinear method, and Ref.~\cite{WX2024} obtained the multi-parametric $N$th-order rogue wave solution in terms of Schur polynomials via the Darboux transformation.

Up to now, there has been  no research on the defocusing case of Eq.~\eref{NNLS}, which is expected to admit the dark soliton solutions.
In this paper, we are devoted to developing the IST theory for the defocusing case of Eq.~\eref{NNLS} under non-zero boundary condition (NZBC), which is given as follows:
\begin{align}
	&\displaystyle\lim_{x \to \pm \infty} q(x,t) = q_{\pm} e^{-2 \mi (|q_+|^2+|q_-|^2) t},
\end{align}
where $q_{\pm} \neq 0$ are  independent of $x$ and $t$.
By referring to the method for dealing with the defocusing Manakov system in Refs.~\cite{DJK,Biondini2015,BP}, we introduce an adjoint Lax pair and
auxiliary eigenfunctions, provide a rigorous proof for the analyticity and symmetries of the Jost eigenfunctions, auxiliary Jost eigenfunctions and scattering coefficients.
Different from the Manakov system, the parity-symmetric constraint~\eref{PS} introduces an additional symmetry (see the second symmetry in Section~\ref{section2.4}), which results in stricter constraints on the scattering coefficients, eigenfunctions and norming constants. Moreover, we formulate a matrix RHP for the inverse problem and derive the reconstruction formulas
when the scattering coefficients admit an arbitrary number of simple zeros.
On the other side, we present the N-soliton solutions in the determinant form for the reflectionless case, and discuss the soliton dynamical behavior based on the distribution of discrete eigenvalues and the asymptotic behavior of solutions. It turns out that the defocusing Eq.~\eref{NNLS} supports two types of fundamental solitons: the dark soliton and beating soliton, which are respectively linked to the discrete eigenvalue pairs lying on and off the circle on the spectrum plane. We stress that it is a nontrivial task to derive the N-soliton solutions despite they are a subset of Manakov solitons. One may notice that  Eq.~\eref{NNLSS1} is invariant under the transformation $q(x,t)\to q(-x,t)$, but this invariance does not imply that the admissible solutions of Eq.~\eref{NNLSS1}  must be even- or odd-symmetric with respect to $x$. For example, the beating one-soliton solution~\eref{Eq512} does not keep such symmetry in $x$. It should be noted that in the zero-background case of focusing Eq.~\eref{NNLS}, only the multi-soliton solutions with $N>2$ would exhibit asymmetry~\cite{JKY,Wu}.



The structure of this paper  is organized as follows: Section~\ref{Sec2} is dedicated to the study of direct scattering, including the analyticity and symmetries of the scattering coefficients, Jost eigenfunctions and auxiliary Jost eigenfunctions.
In Section~\ref{sec3}, we analyze the distribution and constraints of discrete eigenvalues, determine the asymptotic behaviors of analytic scattering coefficients and the analytic components of eigenfunctions, and establish the corresponding trace formulas. In Section~\ref{sec4}, we construct the RHP, residue conditions and reconstruction formulas. In Section 5, we present the explicit determinant representation of N-soliton solutions, and discuss the soliton dynamical properties based on  the distribution of discrete eigenvalues and the asymptotic behavior of solutions.
Finally, Section~\ref{sec6} addresses the conclusions and discussions of this paper.

\section{Direct scattering under NZBC}
\label{Sec2}

In this section, we study the direct scattering problem of Eq.~\eref{NNLS} under NZBC, including the Jost eigenfunctions, auxiliary Jost eigenfunctions and the associated scattering coefficients, along with their analyticity and symmetries.


\subsection{Lax pair}
The Lax pair of the defocusing Eq.~\eref{NNLS} can be expressed as:
\begin{subequations}
\label{Laxpair1}
	\begin{align}
		& \bm{\psi}_x(x,t,k)
		= \mathbf{X}(x,t,k) \bm{\psi}(x,t,k),\quad \mathbf{X}(x,t,k) = -\mi k \bm{\sigma}_{3} + \mathbf{Q}(x,t),   \\
		& \bm{\psi}_t(x,t,k)
		= \mathbf{T}(x,t,k) \bm{\psi}(x,t,k), \quad  \mathbf{T}(x,t,k) = 2 k \mathbf{X}(x,t,z)  + \mi \bm{\sigma}_{3} \mathbf{Q}_x(x,t) - \mi  \bm{\sigma}_{3}{\mathbf{Q}}^2(x,t),
	\end{align}
\end{subequations}
with
\begin{equation}
	\label{Laxpair_trainmatrix}
	\begin{aligned}
		& \bm{\sigma}_{3}
		= \left(
		\begin{array}{ccc}
			1 & 0 & 0 \\
			0 & 1 & 0 \\
			0 & 0 & -1 \\
		\end{array}
		\right),
		& \mathbf{Q}(x,t)
		=  \left(
		\begin{array}{ccc}
			0 & 0 & q(x,t) \\
			0 & 0 & q(-x,t) \\
			q^*(x,t) & q^*(-x,t) & 0 \\
		\end{array}
		\right),
	\end{aligned}
\end{equation}
where $\bm{\psi}(x,t,k)$ is the $3 \times 3$ eigenfunction matrix, $k$ is the spectral parameter in $\mathbb{C}$, and the asterisk represents complex conjugate.
Eq.~\eref{NNLS} has equivalence in nature to the zero-curvature condition $\mathbf{X}_t(x,t,k) - \mathbf{T}_x(x,t,k) + [\mathbf{X}(x,t,k), \mathbf{T}(x,t,k)] = 0$ (also known as the compatibility condition).

To make the boundary conditions independent of $t$, we make the transformation $q(x,t) \to \hat{q}(x,t) e^{-2 \mi  (|q_+|^2+|q_-|^2) t}$ for Eq.~\eref{NNLS}, obtaining that
\begin{align}
	\label{NNLS1}
	\mi  q_t(x,t) + q_{xx}(x,t) - 2  q(x,t) \big( |q(x,t)|^2  + |q(-x,t)|^2 - 2 q^2_0 \big) = 0,
\end{align}
where $q_0^2: = (|q_+|^2+|q_-|^2)/2$, $q_0>0$ and the hat has been dropped for simplicity. Thus, the corresponding NZBC becomes
\begin{align}
	\label{xcx6}
	\displaystyle\lim_{x \to \pm \infty} q(x,t) = q_{\pm}.
\end{align}
Obviously, if $q(x,t)$ is a solution to Eq.~\eref{NNLS1}, then $q(x,t) e^{-4 \mi  q^2_0 t}$ must exactly satisfy Eq.~\eref{NNLS}.
Meanwhile, the Lax pair of Eq.~\eref{NNLS1} becomes
\begin{subequations}
	\label{Laxpairjvzhenxingshi}
	\begin{align}
		&  \bm{\phi}_x(x,t,k)
		= \mathbf{X}(x,t,k) \bm{\phi}(x,t,k),\quad \mathbf{X}(x,t,k) = -\mi  k \bm{\sigma}_{3} + \mathbf{Q}(x,t),   \label{Laxpairjvzhenxingshia}  \\
		& \bm{\phi}_t(x,t,k)
		= \mathbf{T}(x,t,k) \bm{\phi}(x,t,k), \quad  \mathbf{T}(x,t,k) = 2 k \mathbf{X}(x,t,k)  + \mi  \bm{\sigma}_{3} \mathbf{Q}_x(x,t) - \mi  \bm{\sigma}_{3}{\mathbf{Q}}^2(x,t) - \mathbf{V_0},  \label{Laxpairjvzhenxingshib}
	\end{align}
\end{subequations}
where  $\bm{\phi}(x,t,k) =e^{- \mathbf{V}_0 t} \bm{\psi}(x,t,k)  $ with $\mathbf{V}_0 = \diag(-2 \mi  q^2_0, -2 \mi  q^2_0, 2 \mi  q^2_0)$.

Keeping in mind~\eref{xcx6}, calculation of the limits of Eqs.~\eref{Laxpairjvzhenxingshi} as $x\to \pm \infty$ yields the asymptotic scattering problem:
\begin{equation}
	\label{sg5}
	\begin{aligned}
		&\bm{\phi}_{\pm,x}(x,t,k) = \mathbf{X}_{\pm}(k) \bm{\phi}_{\pm}(x,t,k), \quad  \bm{\phi}_{\pm,t}(x,t,k) = \mathbf{T}_{\pm}(k) \bm{\phi}_{\pm}(x,t,k),                                     \\
	\end{aligned}
\end{equation}
where
\begin{subequations}
	\begin{align}
		&\mathbf{X}_{\pm}(k) = \left(
		\begin{array}{ccc}
			-\mi  k & 0 & q_{\pm} \\
			0 & -\mi  k & q_{\mp} \\
			q^{*}_{\pm} & q^{*}_{\mp} & \mi  k \\
		\end{array}
		\right),  \quad\\
		&\mathbf{T}_{\pm}(k) = \left(
		\begin{array}{ccc}
			-\mi  |q_{\pm}|^2  -2 \mi  k^2 + 2 \mi  q_0^2 & -\mi  q_{\pm} q^{*}_{\mp} & 2 k q_{\pm} \\
			-\mi  q_{\mp} q^{*}_{\pm} & -\mi  |q_{\mp}|^2 -2 \mi  k^2 + 2 \mi  q_0^2 & 2 k q_{\mp} \\
			2 k q^{*}_{\pm} & 2 k q^{*}_{\mp} &  2 \mi  k^2  \\
		\end{array}
		\right).\label{2.7}
	\end{align}
\end{subequations}
Since the matrices $\mathbf{X}_{\pm}(k)$ and  $\mathbf{T}_{\pm}(k)$ are independent of $x$ and $t$, the compatibility condition of Eqs.~\eref{sg5} becomes $[\mathbf{X}_{\pm}(k), \mathbf{T}_{\pm}(k)] = 0$.
So, $\mathbf{X}_{\pm}(k)$ and $\mathbf{T}_{\pm}(k)$ share the common eigenfunctions, and their eigenvalues can be, respectively, given by $\{-\mi k$, $\pm \mi  \lambda\}$ and $\{-2\mi k^2 + 2 \mi  q^2_0$, $\pm 2\mi k\lambda\}$, where
\begin{equation}
	\label{Riemann surface}
	\begin{aligned}
		\lambda(k) =  \sqrt{k^2 - 2 q^2_0}.
	\end{aligned}
\end{equation}
Noticing that the eigenvalues have the branching,
we introduce the two-sheeted Riemann surface to make $\lambda(k)$ become single-valued on each individual sheet.
Since the branch points occur at $k= \pm  \sqrt{2} q_0$, thus  the branch cut can be taken as the interval $(-\infty, - \sqrt{2} q_0] \cup [\sqrt{2} q_0, \infty)$. In the following, we denote the regions with $\Im (\lambda) > 0$ and $\Im (\lambda) < 0$ as the first and second sheets, respectively.

\begin{figure}[H]
	\centering
	\includegraphics[width=2.7in]{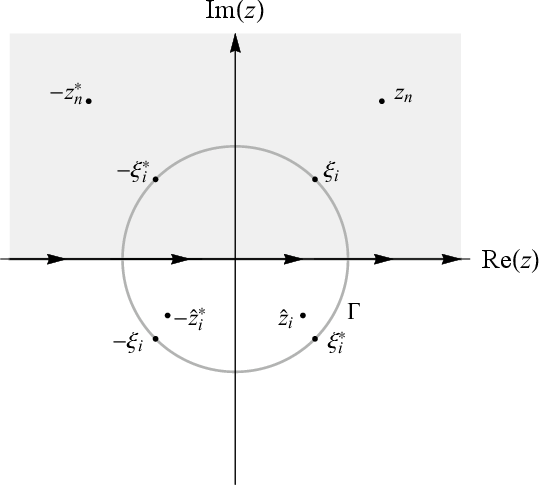}\hfill
	\caption{\small  The regions $C^{+}$ with $\Im (z) > 0$ (gray) and $C^{-}$ with $\Im (z) < 0$ (white) in the complex $z$-plane, where $\Gamma$ represents the circle with radius $\sqrt{2}|q_0|$. The orientations of the contours for the RHP and  the symmetries of  discrete eigenvalues are also illustrated.  \label{f0} }
\end{figure}

In order to avoid the direct treatment of the  two-sheeted Riemann surface, the following uniformization variable is introduced
\begin{equation}
	\label{sl2}
\begin{aligned}
z = k + \lambda(k),
\end{aligned}
\end{equation}
which allows us to work in the complex $z$-plane. Meanwhile,  the inverse transformation can be obtained by
\begin{equation}
	\label{hlj11}
	\begin{aligned}
		k = \frac{1}{2}(z + \frac{2q^2_0}{z}),\quad \lambda = \frac{1}{2}(z - \frac{2q^2_0}{z}).
	\end{aligned}
\end{equation}
As a result, the transformation~\eref{sl2} can map the branch cut $(-\infty, - \sqrt{2} q_0] \cup [\sqrt{2} q_0, \infty)$ onto the real axis of $z$-plane, and the first and second sheets in the $k$-plane onto the regions $C^+$ ($\Im (z) > 0$ ) and $C^-$ ($\Im (z) < 0$), respectively.



\subsection{Jost eigenfunctions}
\label{sub2}
In this section, over the continuous spectrum $\Sigma_z: \mathbb{R}$ in which  $\lambda(k)$ is a real number, we present the Jost eigenfunctions. The eigenvector matrices of Eq.~\eref{sg5} can be formulated as
\begin{equation}
	\label{xcx17}
	\begin{aligned}
		\mathbf{Y}_{\pm}(z)
		=\left(
		\begin{array}{ccc}
			\frac{q_{\pm}}{q_0} & -\frac{q^{*}_{\mp}}{\sqrt{2} q_0}  & \frac{-\mi  \sqrt{2}q_{\pm}}{z}  \\
			\frac{q_{\mp}}{q_0} &  \frac{q^{*}_{\pm}}{\sqrt{2}q_0} & \frac{-\mi  \sqrt{2}q_{\mp}}{z}   \\
			\frac{ 2 \mi  q_0 }{z} & 0 & \sqrt{2} \\
		\end{array}
		\right),\\
	\end{aligned}
\end{equation}
which can diagonalize $\mathbf{X}_{\pm}$ and $\mathbf{T}_{\pm}$ as
\begin{equation}
	\label{sm99}
	\begin{aligned}
		&\mathbf{Y}^{-1}_{\pm}(z) \mathbf{X}_{\pm}(z) \mathbf{Y}_{\pm}(z) = \mi  \mathbf{J}(z) =  \diag(-\mi  \lambda, -\mi  k, \mi  \lambda), \\
		&\mathbf{Y}^{-1}_{\pm}(z) \mathbf{T}_{\pm}(z) \mathbf{Y}_{\pm}(z) = \mi  \bm{\Omega}(z) =  \diag(- 2 \mi  k \lambda, -2 \mi  k^2 + 2 \mi  q^2_0, 2 \mi  k \lambda).
	\end{aligned}
\end{equation}

The Jost eigenfunctions $\bm{\phi}_{\pm}(x,t,z)$ are the solutions of the Lax pair~\eref{Laxpairjvzhenxingshi} with  $z \in \Sigma_z$, and satisfy the asymptotic behavior
\begin{equation}
\label{4}
\bm{\phi}_{\pm}(x, t, z) \sim \mathbf{Y}_{\pm}(z)  \mathbf{E}(x, t, z) + o(1), \quad x \to \pm \infty,
\end{equation}
where $\mathbf{E}(x, t, z) = e^{\mi  \bm{\Theta}(x,t,z)}$ and $\bm{\Theta}(x,t,z)$ is the $3 \times 3$ diagonal matrix
\begin{equation}
	\label{xcx20}
	\begin{aligned}
		&\bm{\Theta}(x,t,z) =   \mathbf{J}(z) x +  \bm{\Omega}(z)  t = \diag[\theta_{1}(x,t,z),\theta_{2}(x,t,z), - \theta_{1}(x,t,z) ],
	\end{aligned}
\end{equation}
with $\theta_{1}(x,t,z) = -\lambda x -  2  k \lambda t, \, \theta_{2}(x,t,z) = -kx - 2(k^2 -  q^2_0)t$.


Based on the modified Lax pair~\eref{Laxpairjvzhenxingshi},  the following total differential form can be obtained:
\begin{equation}
	\label{2.15}
	\begin{aligned}
		d \big(  \mathbf{E}^{-1}(x,t,z) \mathbf{Y}^{-1}_{\pm}(z) \bm{\phi}_{\pm}(x,t,z)  \big) =
		&\big(\mathbf{E}^{-1}(x,t,z) \mathbf{Y}^{-1}_{\pm}(z)
		\bm{\Delta}\!\mathbf{Q}_{\pm}(x,t) \bm{\phi}_{\pm} (x,t,z)
		 \big) dx \\
		&+
		\big(\mathbf{E}^{-1}(x,t,z) \mathbf{Y}^{-1}_{\pm}(z)
		\bm{\Delta}\!\mathbf{T}_{\pm}(x,t,z) \bm{\phi}_{\pm}(x,t,z)
		 \big) dt,
	\end{aligned}
\end{equation}
where $\bm{\Delta}\!\mathbf{Q}_{\pm}(x,t): = \mathbf{Q}(x,t) - \mathbf{Q}_{\pm}$ and $\bm{\Delta}\!\mathbf{T}_{\pm}(x,t,z) := \mathbf{T}(x,t,z) - \mathbf{T}_{\pm}(x,t,z) = 2k \bm{\Delta}\!\mathbf{Q}_{\pm}(x,t) + \mi  \mathbf{Q}_x(x,t) \bm{\sigma_{3}} +\mi  \bm{\sigma}_{3}\mathbf{Q}^2(x,t) - \mi  \bm{\sigma_{3}} \mathbf{Q}^2_{\pm}$.
Then, integrating both the sides of Eq.~\eref{2.15} along the  paths $(-\infty, t) \to (x, t)$ and $(x, t) \to (\infty, t)$ yields
\begin{subequations}
	\begin{align}
		&\bm{\phi}_{-}(x,t,z) = \mathbf{Y}_{-}(z)\mathbf{E}(x,t,z) + \int_{-\infty}^{x} \mathbf{Y}_{-}(z) e^{\mi  \mathbf{J}(z) (x-y)} \mathbf{Y}^{-1}_{-}(z) \bm{\Delta}\!\mathbf{Q}_{-}(y,t) \bm{\phi}_{-}(y,t,z) dy, \label{7}\\
		&\bm{\phi}_{+}(x,t,z) = \mathbf{Y}_{+}(z)\mathbf{E}(x,t,z) - \int_{x}^{\infty} \mathbf{Y}_{+}(z)  e^{\mi  \mathbf{J}(z) (x-y)} \mathbf{Y}^{-1}_{+}(z) \bm{\Delta}\!\mathbf{Q}_{+}(y,t) \bm{\phi}_{+}(y,t,z) dy. \label{8}
	\end{align}
\end{subequations}
To factorize the exponential oscillations of $\bm{\phi}_{\pm}(x,t,z)$, we make the following transformation:
\begin{equation}
	\label{s16}
\begin{aligned}
\bm{\mu}_{\pm}(x,t,z) = \bm{\phi}_{\pm}(x,t,z) \mathbf{E}^{-1}(x,t,z),
\end{aligned}
\end{equation}
where $\bm{\mu}_{\pm}$ are referred to as the modified Jost eigenfunctions and satisfy the asymptotic behavior $\displaystyle\lim_{x \to \pm \infty} \bm{\mu}_{\pm}(x,t,z) = \mathbf{Y}_{\pm}(z)$.
Then, we can obtain the analyticity of modified Jost eigenfunctions $\bm{\mu}_{\pm}$ (see Appendix~\ref{appendixA1}):
\begin{theorem}
	\label{th1}
Suppose that $||\bm{\Delta}\!\mathbf{Q}_{\pm}(\cdot,t)|| \in L^1(\mathbb{R})$, the analyticity of the following
columns of $\bm{\mu}_{+}$ and $\bm{\mu}_{-}$ can be extended into  their respective regions of the $z$-plane:
\begin{equation}
		\begin{aligned}
			\mu_{-,1}(x, t, z), \mu_{+,3}(x, t, z): C^+;\quad \mu_{+,1}(x, t, z), \mu_{-,3}(x, t, z): C^-,
		\end{aligned}
	\end{equation}
where the subscripts denote matrix columns, i.e., $\bm{\mu}_{\pm}(x,t,z) = (\mu_{\pm,1}(x,t,z), \mu_{\pm,2}(x,t,z), \mu_{\pm,3}(x,t,z))$.
\end{theorem}
In view of Eq.~\eref{s16},  the same analyticity and boundedness properties are also possessed by the corresponding columns of $\bm{\phi}_{\pm}(x,t,z)$. However, $\mu_{\pm,2}(x,t,z)$ (or $ \phi_{\pm,2}(x,t,z))$ are generally nowhere analytic.


For convenience, we define that $\gamma(z):= \det (\mathbf{Y}_{\pm}(z)) =2 - 4 q^2_0/z^2$. Since $\bm{\phi}(x,t,z)$ satisfies~\eref{Laxpairjvzhenxingshi},  the derivatives of $e^{- \mi  \bm{\Theta}(x,t,z) } \bm{\phi}(x,t,z)$ with respect to $x$ and $t$ can be respectively obtained by
\begin{equation}
	\begin{aligned}
			&\frac{\partial}{\partial x} \big[ e^{- \mi  \bm{\Theta}(x,t,z) } \bm{\phi}(x,t,z) \big] =\big[-\mi\mathbf{J}(z) e^{- \mi  \bm{\Theta}(x,t,z) } + e^{- \mi  \bm{\Theta}(x,t,z) } \mathbf{X}(x,t,z) \big] \bm{\phi}(x,t,z), \\
			&\frac{\partial}{\partial t} \big[ e^{- \mi  \bm{\Theta}(x,t,z) } \bm{\phi}(x,t,z) \big] =\big[-\mi\bm{\Omega}(z) e^{- \mi  \bm{\Theta}(x,t,z) } + e^{- \mi  \bm{\Theta}(x,t,z) } \mathbf{T}(x,t,z) \big] \bm{\phi}(x,t,z).
		\end{aligned}
\end{equation}
Noticing that
\begin{equation}
\begin{aligned}
& \tr\big[-\mi\mathbf{J}(z) e^{- \mi  \bm{\Theta}(x,t,z) } + e^{- \mi  \bm{\Theta}(x,t,z) } \mathbf{X}(x,t,z)\big] = \tr\big[-\mi \bm{\Omega}(z) e^{- \mi  \bm{\Theta}(x,t,z) } + e^{- \mi  \bm{\Theta}(x,t,z) } \mathbf{T}(x,t,z) \big]= 0,
\end{aligned}
\end{equation}
it follows from Abel's theorem that $\det\big[e^{- \mi  \bm{\Theta}(x,t,z) } \bm{\phi}(x,t,z) \big] = \det\big[ \bm{\phi}(x,t,z) \big]e^{- \mi  \theta_{2}(x,t,z)  }$  is independent of $x$ and $t$. By virtue of Eq.~\eref{4}, one can obtain that
\begin{equation}
	\label{xcx2.21}
	\begin{aligned}
		\det (\bm{\phi}_{\pm}(x,t,z) )  = \gamma(z) e^{\mi  \theta_{2}(x,t,z)  } \quad \mathrm{for\, all}\,\, z \in \Sigma_o:\, \mathbb{R} \setminus \{\pm \sqrt{2} q_0 \}.
	\end{aligned}
\end{equation}
%

Recalling that both $\bm{\phi}_{+}  $  and  $\bm{\phi}_{-}$ satisfy Eq.~\eref{sg5}, thus there exist the scattering matrices $\mathbf{A}(z)$ and $\mathbf{B}(z)$ such that
\begin{subequations}
	\begin{align}
		&\bm{\phi}_{-}(x,t,z)  = \bm{\phi}_{+}(x,t,z)  \mathbf{A}(z), \quad  \mathbf{A}(z) = (a_{ij}(z)) \quad  z \in \Sigma_o,\label{sm1}\\
		&\bm{\phi}_{+}(x,t,z)  = \bm{\phi}_{-}(x,t,z)  \mathbf{B}(z), \quad  \mathbf{B}(z) = (b_{ij}(z)) \quad  z \in \Sigma_o,
	\end{align}
\end{subequations}
where $\mathbf{B}(z) = \mathbf{A}^{-1}(z)$ and $\det (\mathbf{A}(z)) = \det (\mathbf{B}(z)) = 1$ owing to Eq.~\eref{xcx2.21}.
Meanwhile, the scattering coefficients $a_{ij}$ and $b_{ij}$ can be represented as
\begin{equation}
	\begin{aligned}
		&a_{ij}(z) = [\bm{\mu}^{-1}_{+}(x,t,z) ]^{i}  \mu_{-,j}(x,t,z) , \quad b_{ij}(z) = [\bm{\mu}^{-1}_{-}(x,t,z) ]^{i} \mu_{+,j}(x,t,z) , \\
	\end{aligned}
\end{equation}
where $[\bm{\mu}^{-1}_{\pm}(x,t,z) ]^{i}$ are the $i$th rows of  $\bm{\mu}^{-1}_{\pm}(x,t,z) $.
Then, we can prove the analyticity of
scattering coefficients (see the details in Appendix~\ref{appendixA2}):
\begin{theorem}
	\label{th2}
	Suppose that $||\bm{\Delta}\!\mathbf{Q}_{\pm}(\cdot,t)|| \in L^1(\mathbb{R})$, the analyticity of the following scattering coefficients can
be  extended into  their respective regions of the $z$-plane:
	\begin{equation}
		\begin{aligned}
			a_{11}(z),\, b_{33}(z):\,  C^+,\quad a_{33}(z),\, b_{11}(z):\,  C^-.
		\end{aligned}
	\end{equation}
\end{theorem}
Note that the remaining scattering coefficients in $\mathbf{A}(z)$ and $\mathbf{B}(z)$ are generally nowhere analytic.

\subsection{Adjoint problem and auxiliary eigenfunctions}
It is known that a complete set of analytic eigenfunctions is necessary for solving the inverse problem.
However, $\phi_{\pm,2}(x,t,z)$ are generally not analytic in any specific region.
To address this issue,  following the approach in Refs.~\cite{DJK,Biondini2015,BP}, we introduce the adjoint Lax pair:
	\begin{align}
		& \tilde{\bm{\phi}}_x(x,t,k)
		= \tilde{\mathbf{X}}(x,t,k)  \bm{\phi}(x,t,k) ,
		\quad \tilde{\bm{\phi}}_t(x,t,k)
		= \tilde{\mathbf{T}}(x,t,k)  \bm{\phi}(x,t,k) ,  \label{Laxpairjvzhenxingshi2}
	\end{align}
with
\begin{subequations}
	\begin{align}
		&\tilde{\mathbf{X}}(x,t,k)  = \mathbf{X}^*(x,t,k^*) = \mi  k \bm{\sigma}_{3} + \mathbf{Q}^*(x,t),   \\
		&\tilde{\mathbf{T}}(x,t,k)  = \mathbf{T}^*(x,t,k^*)  = 2 k \tilde{\mathbf{X}}(x,t,k)  - \mi  \bm{\sigma}_{3} \mathbf{Q}^*_x(x,t) + \mi  \bm{\sigma}_{3}({\mathbf{Q}^*})^2(x,t) + \mathbf{V_0}.
	\end{align}
\end{subequations}
It is easy to check that Eq.~\eref{NNLS1} can also be obtained from the zero-curvature condition $\tilde{\mathbf{X}}_x(x,t,k) - \tilde{\mathbf{T}}_t(x,t,k) + [\tilde{\mathbf{X}}(x,t,k), \tilde{\mathbf{T}}(x,t,k)] = 0$. With the same procedure in section~\ref{sub2}, we obtain the eigenvector matrix of  $\tilde{\mathbf{X}}_{\pm}(z)$ and $\tilde{\mathbf{T}}_{\pm}(z)$:
\begin{equation}
	\begin{aligned}
		\tilde{\mathbf{Y}}_{\pm}(z) = \mathbf{Y}^*_{\pm}(z^*)
		=\left(
		\begin{array}{ccc}
			\frac{q^{*}_{\pm}}{q_0} & -\frac{q_{\mp}}{\sqrt{2} q_0}  & \frac{\mi  \sqrt{2}q^{*}_{\pm}}{z}  \\
			\frac{q^{*}_{\mp}}{q_0} &  \frac{q_{\pm}}{\sqrt{2}q_0} & \frac{\mi  \sqrt{2}q^{*}_{\mp}}{z}   \\
			\frac{- 2 \mi  q_0 }{z} & 0 & \sqrt{2} \\
		\end{array}
		\right),\\
	\end{aligned}
\end{equation}
which has $\det (\tilde{\mathbf{Y}}_{\pm}(z)) = \gamma(z)$ and can make $\tilde{\mathbf{Y}}^{-1}_{\pm}(z) \tilde{\mathbf{X}}_{\pm}(z) \tilde{\mathbf{Y}}_{\pm}(z) = -\mi  \mathbf{J}(z)$ and $\tilde{\mathbf{Y}}^{-1}_{\pm}(z) \tilde{\mathbf{T}}_{\pm}(z) \tilde{\mathbf{Y}}_{\pm}(z) = -\mi  \bm{\Omega}(z)$.

Similarly, the adjoint Jost eigenfunctions $\tilde{\bm{\phi}}_{\pm}$ are the solutions of Eq.~\eref{Laxpairjvzhenxingshi2} satisfying the asymptotic behavior:
\begin{equation}
	\label{xcx35}
	\tilde{\bm{\phi}}_{\pm}(x,t,z) \sim \tilde{\mathbf{Y}}_{\pm}(z)  \mathbf{E}^{-1}(x, t, z) + o(1), \quad x \to \pm \infty,
\end{equation}
and the modified adjoint Jost eigenfunctions $\tilde{\bm{\mu}}_{\pm}(x,t,z) = \tilde{\bm{\phi}}_{\pm}(x,t,z) \mathbf{E}(x, t, z)$ are also introduced.
Again, there exist the adjoint scattering matrices $\tilde{\mathbf{A}}(z)$ and $\tilde{\mathbf{B}}(z)$ such that
\begin{align}
& \tilde{\bm{\phi}}_{-} = \tilde{\bm{\phi}}_{+} \tilde{\mathbf{A}}(z), \quad \tilde{\bm{\phi}}_{+} = \tilde{\bm{\phi}}_{-} \tilde{\mathbf{B}}(z),
 \quad z \in \Sigma_o,\label{xcx38}
\end{align}
where $\tilde{\mathbf{A}}(z) = (\tilde{a}(z))_{ij}$ and $ \tilde{\mathbf{B}}(z)= \tilde{\mathbf{A}}^{-1}(z) = (\tilde{b}(z))_{ij}$.

Repeating the same procedure in proving Theorems~\ref{th1} and~\ref{th2},
the analyticity of some columns of $\tilde{\bm{\mu}}_{\pm}$ and adjoint scattering coefficients of $\tilde{\mathbf{A}}(z)$ and $\tilde{\mathbf{B}}(z)$ can be obtained as follows:
\begin{theorem}
\label{th3}
Suppose that $||\bm{\Delta}\!\mathbf{Q}_{\pm}(\cdot,t)|| \in L^1(\mathbb{R})$, the analyticity of the columns $\tilde{\bm{\mu}}_{\pm,1}(x,t,z)$,
$\tilde{\bm{\mu}}_{\pm,3}(x,t,z)$ and the following adjoint scattering coefficients can be extended into their respective  regions of $z$-plane:
\begin{equation}
\begin{aligned}
\tilde{\mu}_{-,1}(x, t, z),\, \tilde{\mu}_{+,3}(x, t, z), \, \tilde{a}_{11}(z),\,  \tilde{b}_{33}(z):\quad C^-;  \\
\tilde{\mu}_{+,1}(x, t, z),\,  \tilde{\mu}_{-,3}(x, t, z), \, \tilde{a}_{33}(z), \, \tilde{b}_{11}(z):\quad C^+.
\end{aligned}
\end{equation}
\end{theorem}


Referring to~\cite{DJK,BP,Biondini2015}, we can present the following useful result (see Appendix~\ref{appendixA4}):
\begin{proposition}
\label{pro1}
Given two arbitrary column vector solutions $u(x,t,z)$ and $v(x,t,z)$ of the adjoint Lax pair~\eref{Laxpairjvzhenxingshi2}, the function
\begin{equation}
	\label{sm40}
	\begin{aligned}
		&r(x,t,z) = e^{\mi \theta_{2}(x,t,z)} \bm{\sigma}_{3} \big[u(x,t,z) \times v(x,t,z) \big],
		\end{aligned}
	\end{equation}
satisfies the original Lax pair~\eref{Laxpairjvzhenxingshi}, where the symbol `$\times$' represents the standard vector cross product.
\end{proposition}

Using Proposition~\ref{pro1}, two new analytic solutions of the original Lax pair~\eref{Laxpairjvzhenxingshi} can be constructed as follows:
\begin{equation}
	\label{xcx41}
	\begin{aligned}
		&w(x, t, z) = e^{\mi  \theta_{2}} \bm{\sigma}_{3} \big[\tilde{\phi}_{-,3}(x,t,z) \times \tilde{\phi}_{+,1}(x,t,z) \big] /\gamma(z), \\
		&\overline{w}(x, t, z) = e^{\mi  \theta_{2}} \bm{\sigma}_{3} \big[\tilde{\phi}_{-,1}(x,t,z) \times \tilde{\phi}_{+,3}(x,t,z) \big] /\gamma(z),
	\end{aligned}
\end{equation}
which are called the \textit{auxiliary eigenfunctions}. It follows from Theorem~\ref{th3} that $w(x, t, z)$ and $\overline{w}(x, t, z)$ are analytic in $C^{+}$ and $C^{-}$, respectively.
Similarly, to eliminate the exponential oscillations, we introduce the modified auxiliary eigenfunctions:
\begin{equation}
	\label{xcx47}
	\begin{aligned}
		&m(x, t, z)            = w(x, t, z)            e^{-\mi  \theta_{2}(x, t, z)}, \quad
		\overline{m}(x, t, z) = \overline{w}(x, t, z) e^{-\mi  \theta_{2}(x, t, z)},
	\end{aligned}
\end{equation}
which are analytic in $C^{+}$ and $C^{-}$, respectively.

Furthermore, we establish the relationship between the columns of $\phi_{\pm}$ and $\tilde{\phi}_{\pm}$, as well as the connection between the scattering matrices $\mathbf{A}(z)$ and $\tilde{\mathbf{A}}(z)$  (see Appendix~\ref{appendixA4}):
\begin{lemma}
	\label{th2.7}
	Considering the cyclic indices $j$, $l$ and $m$, we have
	\begin{subequations}
		\label{xcx42}
		\begin{align}
			&\phi_{\pm,j}(x,t,z) = e^{\mi  \theta_{2}(x,t,z)} \bm{\sigma}_{3} \big[\tilde{\phi}_{\pm,l}(x,t,z) \times \tilde{\phi}_{\pm,m}(x,t,z) \big] /r_j(z), \quad z \in \Sigma_o, \label{xcx42a} \\
			&\tilde{\phi}_{\pm,j}(x,t,z) = e^{-\mi  \theta_{2}(x,t,z)} \bm{\sigma}_{3} \big[\phi_{\pm,l}(x,t,z) \times \phi_{\pm,m}(x,t,z) \big] /r_j(z), \quad z \in \Sigma_o,  \label{xcx42b}
		\end{align}
	\end{subequations}
    with
    \begin{equation}
	    \begin{aligned}
		r_1(z) = 1,\quad r_2(z) =  \gamma(z),\quad r_3(z) = -1.
	    \end{aligned}
     \end{equation}
\end{lemma}

\begin{lemma}
\label{co2.8}
The scattering matrix $\mathbf{A}(z)$ and adjoint scattering matrix $\tilde{\mathbf{A}}(z)$ obey the relation
\begin{equation}
		\begin{aligned}
			\bm{\Xi}(z) (\mathbf{A}^{-1}(z))^T \bm{\Xi}^{-1}(z)  & =   \tilde{\mathbf{A}}(z), \,\,  z \in \Sigma_o,
			\quad
			\bm{\Xi}(z)
			= \left(
			\begin{array}{ccc}
				1 & 0 & 0 \\
				0 & \gamma(z) & 0 \\
				0 & 0 & -1 \\
			\end{array}
			\right).
		\end{aligned}
\end{equation}
\end{lemma}

Besides, based on Lemmas~\ref{th2.7} and~\ref{co2.8} together with Eq.~\eref{xcx41},  there are the following decompositions for $\phi_{\pm,2}$  (see Appendix~\ref{appendixA4}):
\begin{proposition}
	\label{co2.9}
The non-analytic Jost eigenfunctions $\phi_{\pm,2}$ admit the following decompositions:
	\begin{subequations}
		\label{sm48}
		\begin{align}
			\phi_{+,2}(x,t,z) &= \frac{1}{b_{33}(z)} \big[ b_{32}(z) \phi_{+,3}(x,t,z)  + w(x,t,z) \big]  \notag \\
			&= \frac{1}{b_{11}(z)} \big[ b_{12}(z) \phi_{+,1}(x,t,z)  - \overline{w}(x,t,z) \big], \quad z \in \mathbb{R}, \label{sm48a} \\
			\phi_{-,2}(x,t,z) &= \frac{1}{a_{33}(z)} \big[ a_{32}(z) \phi_{-,3}(x,t,z)  - \overline{w}(x,t,z) \big]  \notag \\
			&= \frac{1}{a_{11}(z)} \big[ a_{12}(z) \phi_{-,1}(x,t,z)   + w(x,t,z) \big], \quad z \in \mathbb{R}. \label{sm48b}
		\end{align}
	\end{subequations}
\end{proposition}

In Sections~\ref{section2.4} and~\ref{section3.1}, by virtue of Proposition~\ref{co2.9}, we will give the symmetry relations of Jost eigenfunctions and their properties at the discrete eigenvalues.


\subsection{Symmetries}
\label{section2.4}

As shown in Ref.~\cite{JKY}, the scattering problem of Eq.~\eref{NNLS} under ZBC admits two symmetries $k \mapsto k^*$ and $k \mapsto -k$.
For the NZBC case, other symmetries should be taken into account with the presence of a Riemann surface.
In fact, there are three symmetries due to the introduction of nonlocal reduction, in contrast that the Manakov system  under NZBC has only two symmetries (see, e.g., Ref.~\cite{Biondini2015}). In this section, the related proofs for our results are provided in Appendix~\ref{appendixA5}.

\subsubsection{The first symmetry}
\label{section2.4.1}

We begin by considering the complex-conjugate mapping $z \mapsto z^{*}$, which induces the corresponding transformation $(k, \lambda)  \mapsto (k^{*}, \lambda^{*})$.
\begin{lemma}
	\label{le2.12}
If $\bm{\phi}(x,t,z)$ solves the Lax pair~\eref{Laxpairjvzhenxingshi}, then so does $\bm{\sigma}_{3} (\phi^{\dagger}(x,t,z^*))^{-1}$.
Moreover, the Jost eigenfunctions $\bm{\phi}_{\pm}(x,t,z)$ meet the symmetry
	\begin{equation}
		\label{xcx49}
		\begin{aligned}
			&\bm{\phi}_{\pm}(x,t,z) = \bm{\sigma}_{3}(\bm{\phi}^{\dagger}_{\pm} (x,t,z^*))^{-1} \mathbf{C}(z),
			\quad
			\mathbf{C}(z)
			= \left(
			\begin{array}{ccc}
				\gamma(z) & 0 & 0 \\
				0 & 1 & 0 \\
				0 & 0 & -\gamma(z) \\
			\end{array}
			\right).
		\end{aligned}
	\end{equation}
\end{lemma}

Then,  combining~\eref{sm48} and~\eref{xcx49}, we can obtain the result:
\begin{lemma}
	\label{le2.13}
	For the columns $\phi_{\pm, 1}$ and $\phi_{\pm, 3}$ of Jost eigenfunctions, the following symmetry relations hold:
\begin{subequations}
	\label{xcx491}
	\begin{align}
		&b_{33}(z) \phi^*_{+, 1}(x,t,z^*) =    \bm{\sigma}_{3} [w \times \phi_{+,3}](x,t,z) e^{-\mi  \theta_{2}(x,t,z)},   \label{xcx49a}\\
		&a_{33}(z) \phi^*_{-, 1}(x,t,z^*) = -  \bm{\sigma}_{3} [\overline{w} \times \phi_{-,3}](x,t,z) e^{-\mi  \theta_{2}(x,t,z)}, \label{xcx49b}  \\
		&b_{11}(z) \phi^*_{+, 3}(x,t,z^*) = -  \bm{\sigma}_{3} [\overline{w} \times \phi_{+,1}](x,t,z) e^{-\mi  \theta_{2}(x,t,z)},  \label{xcx49c} \\
		&a_{11}(z) \phi^*_{-, 3}(x,t,z^*) =    \bm{\sigma}_{3} [w \times \phi_{-,1}](x,t,z) e^{-\mi  \theta_{2}(x,t,z)}.   \label{xcx49d}
	\end{align}
\end{subequations}
\end{lemma}
\begin{lemma}
	\label{le2.14}
The matrix $\mathbf{A}(z)$ is linked to its inverse $\mathbf{B}(z)$ through the following symmetry relation:
	\begin{equation}
		\label{xcx50}
		\begin{aligned}
			\mathbf{A}(z) =  \mathbf{C}^{-1}(z)    \mathbf{B}^{\dagger}(z^*) \mathbf{C}(z), \quad z \in \Sigma_o.
		\end{aligned}
	\end{equation}
\end{lemma}

It can be seen from Lemma~\ref{le2.14} that $a_{11}(z) = b^*_{11}(z^*), \, a_{33}(z) = b^*_{33}(z^*)$ for $z \in \Sigma_o$.
Meanwhile, based on Theorem~\ref{th2}, the Schwarz reflection principle implies that
\begin{equation}
 	\label{xcx2.41}
	\begin{aligned}
		a_{11}(z) = b^*_{11}(z^*), \quad z \in C^{+}; \quad  a_{33}(z) = b^*_{33}(z^*), \quad  z \in C^{-}.
	\end{aligned}
\end{equation}
Notably, both $\phi^*_{\pm}(x,t,z^*)$ and $\tilde{\phi}_{\pm}(x,t,z)$ satisfy Eq.~\eref{Laxpairjvzhenxingshi2} and possess the same asymptotic behaviors. Therefore, we infer that
\begin{equation}
	\label{xcx501}
	\begin{aligned}
		&\tilde{\phi}_{\pm, 1}(x,t,z) =  \phi^*_{\pm, 1}(x,t,z^*), \quad z \in C^{\pm},\\
		&\tilde{\phi}_{\pm, 2}(x,t,z) =  \phi^*_{\pm, 2}(x,t,z^*), \quad z \in \Sigma_o, \\
		&\tilde{\phi}_{\pm, 3}(x,t,z) =  \phi^*_{\pm, 3}(x,t,z^*), \quad z \in C^{\mp}.
	\end{aligned}
\end{equation}
Then, substitution of Eqs.~\eref{xcx501} into Eqs.~\eref{xcx41} and~\eref{xcx42a} yields the relations:
\begin{subequations}
\label{xcx54}
\begin{align}
&w(x, t, z) = e^{\mi  \theta_{2}(x,t,z)} \bm{\sigma}_{3} \big[\phi^*_{-, 3}(x,t,z^*) \times \phi^*_{+, 1}(x,t,z^*) \big] /\gamma(z), \quad z\in C^+, \label{xcx54a} \\
&\overline{w}(x, t, z) = e^{\mi  \theta_{2}(x,t,z)} \bm{\sigma}_{3} \big[\phi^*_{-,1}(x,t,z^*) \times \phi^*_{+,3} (x,t,z^*) \big] /\gamma(z), \quad z\in C^-,
\label{xcx54b}
	\end{align}
\end{subequations}
and
\begin{equation}
	\begin{aligned}
		&\phi_{\pm,j}(x,t,z) = e^{\mi  \theta_{2}(x,t,z)} \bm{\sigma}_{3} \big[\phi^*_{\pm,l} (x,t,z^*) \times \phi^*_{\pm,m}(x,t,z^*) \big] /r_j(z), \quad z\in \Sigma_o, \\
	\end{aligned}
\end{equation}
with $j$, $l$ and $m$ representing the cyclic indices.


\subsubsection{The second and third symmetries}
Next, we consider the mappings $z \mapsto -z$ and $z \mapsto \hat{z}: = \frac{2 q^2_0}{z}$, which correspond to the transformations $(k, \lambda) \mapsto (-k,  -\lambda)$ and $(k, \lambda) \mapsto (k,  -\lambda)$, respectively.
Similar to Lemma~\ref{le2.12}, we can derive that
\begin{lemma}
	\label{le2.17}
If $\bm{\phi}(x,t,z)$ solves the Lax pair~\eref{Laxpairjvzhenxingshi}, both $\bm{\delta}_1 \bm{\phi}(-x,t,-z)$ and $\bm{\phi}(x,t,\hat{z})$ are also the solutions of Eq.~\eref{Laxpairjvzhenxingshi}. In addition, the Jost eigenfunctions $\bm{\phi}_{\pm}$ possess the  symmetries as follows:
\begin{subequations}
		\label{xcx56a}
		\begin{align}
			&\bm{\phi}_{\pm}(x,t,z)  = \bm{\delta}_1 \bm{\phi}_{\mp}(-x,t,-z) \bm{\sigma}_{4},\quad z \in \Sigma_o,  \label{xcx56} \\
			&\bm{\phi}_{\pm}(x,t,z) = \bm{\phi}_{\pm}(x,t,\hat{z}) \bm{\Pi}(z),  \quad z \in \Sigma_o, \label{xcx62}
		\end{align}
	\end{subequations}
where
\begin{align}
\bm{\sigma}_{4}= \begin{pmatrix}
	1  & 0  &  0 \\
	0  & -1  &  0 \\
	0  & 0  & -1 \\
\end{pmatrix}, \quad \bm{\delta}_1 = \begin{pmatrix}
	0  & 1  &  0 \\
	1  & 0  &  0 \\
	0  & 0  & -1 \\
\end{pmatrix}, \quad \bm{\Pi}(z) = \begin{pmatrix}
0  & 0  &  -\mi \frac{\sqrt{2} q_{0} }{z} \\
0  & 1  &  0 \\
\mi \frac{\sqrt{2} q_{0} }{z}  & 0  & 0 \\
\end{pmatrix}. \label{eq246}
\end{align}
\end{lemma}

Then, combining the symmetries~\eref{xcx56a} and Eq.~\eref{sm1} gives the following results:

\begin{lemma}
	\label{le2.18}
The matrix $\mathbf{A}(z)$ is related to $\mathbf{B}(-z)$ and $\mathbf{A}(\hat{z})$ by the following symmetry relations:
	\begin{subequations}
		\begin{align}
			&\mathbf{A}(z) =  \bm{\sigma}_{4} \mathbf{B}(-z) \bm{\sigma}_{4}, \quad z \in \Sigma_o, \label{xcx58} \\
			&\mathbf{A}(z) =  \bm{\Pi}(z)^{-1} \mathbf{A}(\hat{z}) \bm{\Pi}(z), \quad z \in \Sigma_o. \label{xcx64}
		\end{align}
	\end{subequations}
\end{lemma}

It follows from Lemma~\ref{le2.18} that $a_{11}(z) = b_{11}(-z) = a_{33}(\hat{z})$ and $a_{33}(z) = b_{33}(-z)$ for $z \in \Sigma_o$.
Again, using the Schwarz reflection principle, one can show that
\begin{equation}
	\begin{aligned}
		a_{11}(z) = b_{11}(-z) = a_{33}(\hat{z}), \quad z \in C^{+};  \quad  a_{33}(z) = b_{33}(-z),\quad z \in C^{-}. \label{Eq248}
	\end{aligned}
\end{equation}
For the auxiliary eigenfunctions, by using the relation~\eref{xcx54} together with the symmetry~\eref{xcx56}, we can obtain that
\begin{lemma}
\label{le2.19}
The auxiliary eigenfunctions possess the symmetries:
\begin{equation}
	\label{2.49}
w(x, t, z)  =  \bm{\delta}_{1}  \overline{w}(-x,t,-z), \quad w(x, t, z)  = - \overline{w}(x,t,\hat{z}), \quad z \in C^{+}.
\end{equation}
\end{lemma}

\subsubsection{Reflection coefficients}
\label{Combined symmetry and reflection coefficients}

Similarly to the Manakov system~\cite{DJK,Biondini2015,BP}, we define the reflection coefficients as
\begin{equation}
\rho_{1}(z) = \frac{b_{13}(z)}{b_{11}(z)}, \quad \rho_{2}(z) = \frac{a_{21}(z)}{a_{11}(z)}, \label{Eq250}
\end{equation}
based on which the symmetries of scattering coefficients can induce the corresponding symmetries of the reflection coefficients.
In fact, applying the  symmetry in Lemma~\ref{le2.14} to the scattering coefficients yields
\begin{equation}
	\label{xcx71a}
	\begin{aligned}
		\rho_{1}(z) = - \frac{a^*_{31}(z^*)}{a^*_{11}(z^*)}, \quad\, \rho_{2}(z) = \gamma(z) \frac{b^*_{12}(z^*)}{b^*_{11}(z^*)}, \quad z\in \Sigma_o;
	\end{aligned}
\end{equation}
whereas using the  symmetries in Lemma~\ref{le2.18} gives
\begin{subequations}
	\label{xcx71}
\begin{align}
		& \rho_{1}(z)  = - \frac{a_{13}(-z)}{a_{11}(-z)} = \frac{b^*_{31}(-z^*)}{b^*_{11}(-z^*)}, \quad \rho_{2}(z) =  - \frac{b_{21}(-z)}{b_{11}(-z)} = -\gamma(z) \frac{a^*_{12}(-z^*)}{a^*_{11}(-z^*)}, \quad z\in \Sigma_o,\\
		&\rho_{1}(z) = - \frac{b_{31}( \hat{z} )}{b_{33}( \hat{z} ) } = \frac{a^*_{13}(\hat{z}^*)}{a^*_{33}(\hat{z}^*)},  \quad
		\rho_{2}(z)  =   \frac{\mi \sqrt{2} q_0 }{ z } \frac{a_{23}( \hat{z})}{a_{33}(\hat{z}) } =  \frac{\mi z \gamma(z)}{ \sqrt{2} q_0} \frac{b^*_{23}(\hat{z}^*)}{b^*_{33}(\hat{z}^*)}, \quad z\in \Sigma_o.
	\label{xcx71b}
		\end{align}
	\end{subequations}

\section{Discrete spectrum, asymptotic behaviors and trace formulas}
\label{sec3}
In this section, we analyze the distribution and constraints of discrete eigenvalues,
derive the asymptotic behaviors of the eigenfunctions and scattering coefficients, and obtain the corresponding
trace formulas.
Considering the  three symmetries in Section~\ref{section2.4}, the constraints of discrete eigenvalues in Eq.~\eref{NNLS} under NZBC are stricter than those for the defocusing Eq.~\eref{Manakov} under NZBC~\cite{Biondini2015}.

\subsection{Discrete spectrum}
\label{section3.1}
In order to study the properties of discrete eigenvalues, we introduce the following analytic matrices:
\begin{equation}
	\begin{aligned}
		\mathbf{\Phi}^+(x, t, z) & = \big(\phi_{-,1}(x,t,z), w(x,t,z),\phi_{+,3}(x,t,z)\big), \quad z \in C^+, \\
		\mathbf{\Phi}^-(x, t, z) & = \big(\phi_{+,1}(x,t,z), -\overline{w}(x,t,z),\phi_{-,3}(x,t,z)\big), \quad z \in C^-.
	\end{aligned}
\end{equation}
Based on the scattering relation~\eref{sm1} and the decompositions of $\phi_{\pm,2}$ in Eq.~\eref{sm48a}, we can obtain  the determinants (see Appendix~\ref{appendixA8}):
\begin{equation}
	\label{xcx73}
	\begin{aligned}
		\det(\mathbf{\Phi}^+(x, t, z)) & = \gamma(z) a_{11}(z) b_{33}(z) e^{\mi\theta_{2}(x, t, z)}, \quad z \in C^+,  \\
		\det(\mathbf{\Phi}^-(x, t, z)) & = \gamma(z) a_{33}(z) b_{11}(z) e^{\mi\theta_{2}(x, t, z)}, \quad z \in C^-.
	\end{aligned}
\end{equation}
Hence, the columns of $\mathbf{\Phi}^+(x, t, z)$ exhibit linear dependence at the zeros of $a_{11}(z)$ and $b_{33}(z)$, whereas the columns of $\mathbf{\Phi}^-(x, t, z)$ show linear dependence at the zeros of $a_{33}(z)$ and $b_{11}(z)$. Moreover, the symmetries of scattering coefficients imply that these zeros are interrelated, as suggested by Lemmas~\ref{le2.14} and~\ref{le2.18}.

One should recall that $a_{11}(z)$ is analytic in $C^{+}$ and satisfies $a_{11}(z)=b_{11}(-z), a_{11}(z)=b^*_{11}(z^*)$. Thus, we can suppose that $a_{11}(z)$ has zeros $\{\xi_i,-\xi_i^*\}^{n_1}_{i=1}$  and
$\{\xi_{0,j} \}^{m_1}_{j=1}$ (where $\Re(\xi_i) \neq 0$, $\Re(\xi_{0,j}) = 0$ and $m_1 =0$ or $1$) on the upper part of $\Gamma$ (denoted as $\Gamma^+_1$), and has zeros $\{z_i,-z_i^*\}^{n_2}_{i=1}$ and $\{z_{0,j}\}^{m_2}_{j=1}$ (where $\Re(z_i) \neq 0$ and $\Re(z_{0,j}) = 0$) in $C^{+}$ but off $\Gamma^+_1$. Besides, Lemmas~\ref{le2.14} and~\ref{le2.18} imply that $a_{11}(z)=a_{33}(\hat{z}), a_{33}(z)=b_{33}(-z), a_{33}(z)=b^*_{33}(z^*)$. Thus, we  obtain that  the scattering coefficients $a_{33}(z), b_{11}(z), b_{33}(z)$ have zeros as follows:

\begin{equation}
	\label{3.3}
	\begin{aligned}
		&a_{33}(-\xi_i) =0\, \Leftrightarrow \, a_{33}(\xi_i^*) = 0,\quad a_{33}(-\xi_{0,j})  = 0, \\
		&b_{11}(\xi_i^*) = 0 \,\Leftrightarrow \,b_{11}(-\xi_i) = 0, \quad  b_{11}(-\xi_{0,j}) = 0, \\
		&b_{33}(-\xi_i^*) =0 \, \Leftrightarrow\, b_{33}(\xi_i) = 0, \quad  b_{33}(\xi_{0,j})=0, \\
	\end{aligned}
\end{equation}
and 	
\begin{equation}
	\label{3.4}
	\begin{aligned}
		&a_{33}( \hat{z}_i ) = 0\,   \Leftrightarrow \, a_{33}(- \hat{z}_i^*) = 0, \quad a_{33}(\hat{z}_{0,j})  = 0,\\
		&b_{11}(z^*_i) = 0\,  \Leftrightarrow \,  b_{11}(-z_i) = 0, \quad   b_{11}(z^*_{0,j}) = 0, \\
		&b_{33}( \hat{z}_i^*) = 0\, \Leftrightarrow\, b_{33}(- \hat{z}_i ) = 0, \quad b_{33}(\hat{z}_{0,j}^*)  = 0.
	\end{aligned}
\end{equation}

In Appendix~\ref{appendixA8}, we prove that the Jost eigenfunctions at the discrete eigenvalues have the following results:
\begin{lemma}
\label{le3.3}
For the discrete eigenvalues $\xi_i$ ($\Re(\xi_i) \neq 0$,  $1 \leq i \leq n_1$) and $\xi_{0,j}$ ($\Re(\xi_{0,j}) = 0$, $1 \leq j \leq m_1=0$ or $1$)
on the circle $\Gamma$,  the following relations hold
\begin{equation}
\label{xcx76}
\begin{aligned}
	& w(x,t, \pm \xi_i) = \overline{w}(x,t, \pm \xi^*_i) = 0, \quad w(x,t, \xi_{0,j}) = \overline{w}(x,t, - \xi_{0,j}) = 0.
\end{aligned}
\end{equation}	
Meanwhile, the Jost eigenfunctions at $z=\xi_i$ and $z=\xi_{0,j}$, respectively, satisfy
\begin{equation}
		\label{xcx77}
		\begin{aligned}
		&\phi_{-, 1}(x,t,\xi_i)    =  c_{i}   \phi_{+, 3}(x,t,\xi_i), \quad  \phi_{+, 1}(x,t,\xi_i^*)  =  \overline{c}_{i} \phi_{-, 3}(x,t,\xi_i^*),  \\
		&\phi_{+, 1}(x,t,-\xi_i)   = -c_{i} \phi_{-, 3}(x,t,-\xi_i), \quad \phi_{-, 1}(x,t,-\xi_i^*) = -\overline{c}_{i} \phi_{+, 3}(x,t,-\xi_i^*),
	\end{aligned}
\end{equation}
and
\begin{equation}
\label{xcx77b}
\begin{aligned}
&\phi_{-, 1}(x,t,\xi_{0,j}) = c_{0,j} \phi_{+, 3}(x,t,\xi_{0,j}), \quad \phi_{+, 1}(x,t,-\xi_{0,j}) = -c_{0,j} \phi_{-,3}(x,t,-\xi_{0,j}),
\end{aligned}
\end{equation}
with the norming constants $c_{i}, \overline{c}_{i}$ and $c_{0,j}$ obeying the constraints:
\begin{align}
	\label{C38}
c_{i} \overline{c}_{i} = -1, \quad \overline{c}_{i} c_{i}^* \frac{ a'_{33}(\xi_i^*)}{ b'_{11}(\xi_i^*)} = 1, \quad c_{0,j}^2 = 1,
\end{align}
where the prime denotes the differentiation with respect to $z$.
\end{lemma}

\begin{lemma}
\label{le3.5}
At the discrete eigenvalues $z_i$ ($\Re(z_i)\neq 0$,  $1 \leq i \leq n_2$) off the circle $\Gamma$, the following relations hold:
\begin{equation}
	\label{xcx81}
	\begin{aligned}
		&w(x,t,z_i) = d_{i} \phi_{-,1}(x,t,z_i), \quad w(x,t,-z^*_i) = \breve{d}_{i} \phi_{-,1}(x,t,-z^*_i),        \\
		&w(x,t,- \hat{z}_{i}) = \hat{d}_{i} \phi_{+,3} (x,t,- \hat{z}_{i}), \quad w(x,t,\hat{z}^*_{i}) =  \overline{d}_{i} \phi_{+,3}(x,t,\hat{z}^*_{i}),
	\end{aligned}
\end{equation}
and
\begin{equation}
	\label{xcx81a}
	\begin{aligned}
		&\overline{w}(x,t,-z_i) = d_{i} \phi_{+,1} (x,t,-z_i), \quad  \overline{w}(x,t,z^*_i) = \breve{d}_{i}  \phi_{+,1} (x,t,z^*_i),       \\
		&\overline{w}(x,t,\hat{z}_{i})  = -\hat{d}_{i} \phi_{-,3}(x,t,\hat{z}_{i}), \quad  \overline{w}(x,t,-\hat{z}^*_{i})  = - \overline{d}_{i} \phi_{-,3}(x,t,\hat{z}^*_{i}),
	\end{aligned}
\end{equation}
where the norming constants $d_i$, $\breve{d}_{i}$, $\hat{d}_i$ and $\overline{d}_{i}$ obey the constraints:
\begin{equation}
	\label{xcx81c}
	\begin{aligned}
		&\breve{d}_{i} = \frac{b^*_{33}(z_i)}{\gamma(z_i^*) d_i^*}, \quad \hat{d}_i= \frac{\mi  \sqrt{2} q_0}{z_i} d_i, \quad \overline{d}_{i} = - \frac{\mi  \sqrt{2} q_0}{z^*_i}  \frac{b^*_{33}(z_i)}{\gamma(z_i^*) d_i^*}.
	\end{aligned}
\end{equation}
Particularly for the pure imaginary eigenvalues $z_{0,j}$ ($\Re(z_{0,j}) = 0$,  $1 \leq j \leq m_2$), the Jost eigenfunctions satisfy
\begin{equation}
	\label{xcx81d}
	\begin{aligned}
&w(x,t,z_{0,j}) = d_{0,j} \phi_{-,1}(x,t,z_{0,j}), \quad w(x,t,- \hat{z}_{0,j}) = \hat{d}_{0,j} \phi_{+,3} (x,t,- \hat{z}_{0,j}), \\
&\overline{w}(x,t,-z_{0,j}) = d_{0,j} \phi_{+,1} (x,t,-z_{0,j}), \quad \overline{w}(x,t,\hat{z}_{0,j})  = -\hat{d}_{0,j} \phi_{-,3}(x,t,\hat{z}_{0,j}), \\
	\end{aligned}
\end{equation}
with  the constraints:
\begin{equation}
	\label{xcx81e}
	\begin{aligned}
		&\hat{d}_{0,j}= \frac{\mi \sqrt{2} q_0}{z_{0,j}} d_{0,j}, \quad |d_{0,j}|^2 = \frac{b^*_{33}(z_{0,j})}{\gamma(z_{0,j}^*)}.
	\end{aligned}
\end{equation}
\end{lemma}

One should note that a discrete eigenvalue $z=\xi_i$ on the circle $\Gamma$ can be regarded as the special case for the coalescence of two discrete eigenvalues $z_i, \hat{z}^*_i$ off $\Gamma$. If letting $z_i,\hat{z}_i^* \to \xi_i$, it follows from Eq.~\eref{xcx81} that $\displaystyle\lim_{z_i \to \xi_i} d_i =\displaystyle\lim_{z_i \to \xi_i} \overline{d}_i=0 $. Based on the first relation in Eq.~\eref{xcx77} and the first and fourth relations in Eq.~\eref{xcx81},
we have
\begin{equation}
	\begin{aligned}
		&\displaystyle\lim_{z_i \to \xi_i} \frac{w(x,t,z_i)}{d_i}
		= \displaystyle\lim_{z_i \to \xi_i} \phi_{-,1}(x,t,z_i)
		=  c_{i}   \phi_{+, 3}(x,t,\xi_i)
		= \displaystyle\lim_{z_i \to \xi_i} c_{i}   \frac{w(x,t,\hat{z}^*_i)}{\overline{d}_i},
	\end{aligned}
\end{equation}
which implies the constraint between the related norming
constants $c_i$ and $\{d_i, \overline{d}_i\}$:
\begin{equation}
	\label{315}
	\begin{aligned}
\displaystyle\lim_{z_i \to \xi_i} \frac{c_{i} d_i}{ \overline{d}_i}   = 1.
	\end{aligned}
\end{equation}
Substituting $\overline{d}_i$ in Eq.~\eref{xcx81c} into Eq.~\eref{315} yields
\begin{equation}
	\label{3156}
	\begin{aligned}
		\displaystyle\lim_{z_i \to \xi_i}  \frac{c_{i} |d_i|^2}{b^*_{33}(z_i)} = - \frac{\mi  \sqrt{2} q_0}{\xi^*_i}  \gamma(\xi_i^*).
	\end{aligned}
\end{equation}

\subsection{Asymptotic behaviors}


In this section, we analyze the asymptotic behaviors of scattering coefficients and modified Jost eigenfunctions as $z \to \infty$ and $z \to 0$,  and then regularize the RHP which will be given in section~\ref{sectionRHP}. To do so, we expand the functions $\mu_{\pm}$:
\begin{equation}
	\label{xcx82}
	\begin{aligned}
\mu_{\pm}(x,t,z) =  \mu^{\pm}_{0}(x,t) +  \frac{1}{z} \mu^{\pm}_{-1}(x,t)  + \frac{1}{z^2} \mu^{\pm}_{-2}(x,t) + \cdots, \quad z \to \infty,
	\end{aligned}
\end{equation}
where $\mu^{\pm}_{i}(x,t)$ are independent of $z$. Then, substituting Eq.~\eref{s16} into Lax pair~\eref{Laxpairjvzhenxingshi} yields
\begin{subequations}
	\label{xcx83}
	\begin{align}
		& \big( \mathbf{Y}^{-1}_{\pm} \mu_{\pm} \big)_x + \mi \big[ \mathbf{Y}^{-1}_{\pm} \mu_{\pm},  \mathbf{J} \big] = \mathbf{Y}^{-1}_{\pm} \bm{\Delta}\!\mathbf{Q}_{\pm} \mu_{\pm}, \\
		& \big( \mathbf{Y}^{-1}_{\pm} \mu_{\pm} \big)_t + \mi \big[ \mathbf{Y}^{-1}_{\pm} \mu_{\pm}, \bm{\Omega} \big] = \mathbf{Y}^{-1}_{\pm} \bm{\Delta}\!\mathbf{T}_{\pm} \mu_{\pm}.
	\end{align}
\end{subequations}

\begin{lemma}
	\label{co3.6}
	The modified Jost eigenfunctions $\mu_{\pm,1}$ and $\mu_{\pm,3}$ satisfy the asymptotic behaviors:
\begin{subequations}
	\begin{align}
		&\mu_{\pm,1} =\Big(
			\frac{q_{\pm}}{q_0},
			\frac{q_{\mp}}{q_0},
			\frac{\mi  \left(q_{\mp} q^{*}(-x,t)+q_{\pm} q^{*}(x,t)\right)}{q_0 z}
		\Big)^T+ O\Big(\frac{1}{z^2}\Big),
		\quad\quad\quad \,\,\,\,\, z \to \infty,  \label{xcx84}\\
		&\mu_{\pm,3} =\Big(
			 -\frac{\mi  \sqrt{2} q(x,t)}{z},
			 -\frac{\mi  \sqrt{2} q(-x,t)}{z},
			 \sqrt{2}
		\Big)^T+ O\Big(\frac{1}{z^2}\Big),
		\quad\quad\quad\quad\quad\,\, z \to \infty, \label{xcx84b} \\
		&\mu_{\pm,1} =\Big(
		\frac{q(x,t)}{q_0},
		\frac{q(-x,t)}{q_0},
		\frac{2 \mi q_0}{z}
		\Big)^T+ O(z),
		\quad\quad\quad\quad\quad\quad\quad\quad\quad\quad\quad \, z \to 0, \label{xcx85} \\
		&\mu_{\pm,3} =\Big(
		-\frac{\mi  \sqrt{2} q_{\pm}}{z},
		-\frac{\mi  \sqrt{2} q_{\mp}}{z},
		\frac{q_{\mp} q^{*}(-x,t)+q_{\pm} q^{*}(x,t)}{\sqrt{2} q_0^2}
		\Big)^T + O(z), \quad z \to 0.  \label{xcx85b}
	\end{align}
\end{subequations}
\end{lemma}
A detailed proof for Lemma~\ref{co3.6} can refer to Appendix~\ref{appendixA9}.
Importantly, from  Eqs.~\eref{xcx84b} and~\eref{xcx85}, it is straightforward to derive
\begin{subequations}
\label{xcx86}
\begin{align}
& q(x,t) = \frac{\mi  \sqrt{2}}{2}  \displaystyle\lim_{z \to \infty}( z \mu_{\pm})_{13}, \quad q(-x,t)=\frac{\mi  \sqrt{2}}{2}  \displaystyle\lim_{z \to \infty}( z \mu_{\pm})_{23}, \\
& q(x,t) = q_0 \displaystyle\lim_{z \to 0}(\mu_{\pm})_{11}, \quad \quad \,\,\,\,\, q(-x,t)= q_0   \displaystyle\lim_{z \to 0}( \mu_{\pm})_{21}.
\end{align}
\end{subequations}
The above four relations are consistent owing to the symmetries in Eqs.~\eref{xcx56} and~\eref{xcx62},
thus we will use the first one to reconstruct the potential later.


Further, the asymptotic behaviors of the modified auxiliary Jost eigenfunctions and scattering coefficients can be obtained by
\begin{corollary}
	\label{co3.7}
	The modified auxiliary eigenfunctions $m $ and $\overline{m}$ satisfy the asymptotic behaviors:
\begin{subequations}
	\begin{align}
		&m= \Big(-\frac{q^{*}_-}{\sqrt{2} q_0},\frac{q^{*}_+}{\sqrt{2} q_0},\frac{\mi  \left(q^{*}_+ q^*(-x,t)-q^{*}_- q^{*}(x,t)\right)}{\sqrt{2} q_0 z}\Big)^T + O\Big(\frac{1}{z^2}\Big),
		\quad z \to \infty, \\
		&\overline{m}= \Big(\frac{q^{*}_+}{\sqrt{2} q_0},-\frac{q^{*}_-}{\sqrt{2} q_0},\frac{\mi  \left(q^{*}_+ q^{*}(x,t)-q^{*}_- q^{*}(-x,t)\right)}{\sqrt{2} q_0 z}\Big)^T + O\Big(\frac{1}{z^2}\Big),
		\quad \, \, z \to \infty, \\
		&m = \Big(-\frac{q^{*}_+}{\sqrt{2} q_0},\frac{q^{*}_-}{\sqrt{2} q_0},0 \Big)^T+ O(z),
		\,\, \, \quad\quad\quad\quad\quad\quad\quad\quad \quad\quad\quad\quad\quad z \to 0, \\
		&\overline{m} = \Big(\frac{q^{*}_-}{\sqrt{2} q_0},-\frac{q^{*}_+}{\sqrt{2} q_0},0\Big)^T+ O(z),
		\quad \quad\quad\quad\quad\quad\quad\quad\quad \quad\quad\quad\quad\quad z \to 0.
      \end{align}
\end{subequations}
\end{corollary}

\begin{corollary}
	\label{co3.8}The scattering coefficients satisfy the asymptotic behaviors:
	\begin{subequations}
		\label{3.19}
		\begin{align}
			&a_{11} = \frac{q_+ q^*_-+q_- q^*_+}{2 q_0^2} + O\Big(\frac{1}{z}\Big), \quad b_{11} = \frac{q_+ q^*_-+q_- q^*_+}{2 q_0^2} + O\Big(\frac{1}{z}\Big), \,\quad 	z \to \infty, \\
			&a_{33} = 1+ O\Big(\frac{1}{z}\Big), \quad b_{33} = 1+ O\Big(\frac{1}{z}\Big),
			\,\quad\quad \quad\quad\quad \quad\quad\quad\quad\quad\quad\quad 	z \to \infty, \\
		&a_{11} = 1 + O(z), \quad b_{11} = 1 + O(z),
		\,\, \, \quad\quad\quad\quad \quad\quad\quad\quad\quad\quad \quad\quad\quad z \to 0, \\
		&a_{33} = \frac{q_+ q^*_-+q_- q^*_+}{2 q_0^2}+ O(z), \quad b_{33} = \frac{q_+ q^*_-+q_- q^*_+}{2 q_0^2}+ O(z),
		\quad \quad \,\,\, z \to 0.  	
	\end{align}
\end{subequations}
\end{corollary}

\subsection{Trace formulas}

The following Lemma  is proved in Appendix~\ref{appendixC3}.
\begin{lemma}
	\label{le4.4}
	The trace formulas can be explicitly given by
	\begin{subequations}
		\label{xcx108}
		\begin{align}
			a_{33} = &\exp \bigg[- \frac{1}{2 \pi \mi }  \int_{\Sigma_o}^{ }  \frac{\log \Big( 1 -  \rho_{1}( \hat{z}) \rho^*_{1}( \hat{z}^*)   - \frac{\hat{z}^2 \gamma^*( \hat{z}^*) }{2 q^2_{0}}\rho_{2}( \hat{z})  \rho^*_{2}( \hat{z}^*) \Big) }{s-z}ds  \bigg]  \notag \\
			& \times \displaystyle\prod_{i = 1}^{n_1}\left( \frac{ z - \xi^*_i  }{z- \xi_i }   \frac{ z + \xi_i }{z + \xi^*_i }\right)
			\displaystyle\prod_{i = 1}^{m_1} \frac{ z + \xi_{0,j} }{z - \xi_{0,j} }
			\displaystyle\prod_{i = 1}^{n_2} \left(\frac{  z + \hat{z}^*_i }{z -\hat{z}^*_i }   \frac{ z -\hat{z}_i  }{z+ \hat{z}_i }\right)
			\displaystyle\prod_{i = 1}^{m_2}  \frac{ z -\hat{z}_{0,j} }{z +\hat{z}_{0,j} }, \label{xcx108a} \\
			b_{33} = &\exp \bigg[   \frac{1}{2 \pi \mi }  \int_{\Sigma_o}^{ }  \frac{\log \Big( 1 -  \rho_{1}( \hat{z}) \rho^*_{1}( \hat{z}^*)   - \frac{\hat{z}^2 \gamma^*( \hat{z}^*) }{2 q^2_{0}}\rho_{2}( \hat{z})  \rho^*_{2}( \hat{z}^*) \Big) }{s-z}ds  \bigg] \notag \\
			& \times \displaystyle\prod_{i = 1}^{n_1} \left( \frac{ z - \xi_i  }{z- \xi^*_i }   \frac{ z + \xi^*_i }{z + \xi_i }\right)
			\displaystyle\prod_{i = 1}^{m_1} \frac{ z - \xi_{0,j} }{z + \xi_{0,j} }
			\displaystyle\prod_{i = 1}^{n_2} \left(\frac{ z - \hat{z}^*_i  }{z + \hat{z}^*_i}   \frac{ z +\hat{z}_i }{z -\hat{z}_i }\right)
			\displaystyle\prod_{i = 1}^{m_2}  \frac{ z +\hat{z}_{0,j} }{z -\hat{z}_{0,j} }. \label{xcx108b}
		\end{align}
	\end{subequations}
\end{lemma}

\begin{remark}
Note that $a_{33}(z)$ and $b_{33}(z)$ should satisfy the asymptotic behaviors in~\eref{3.19} as $z\to 0$. Thus, we have $|a_{33}(0)| =|b_{33}(0)| = 1$, i.e., $(q_+  q^*_- + q^*_+  q_-)^2 = (2 q^2_0)^2$, which means that there must be $q_{+}  = \pm q_{-}$.
\end{remark}

\section{Inverse problem}
\label{sec4}
In this section, we  formulate an appropriate matrix RHP to relate the eigenfunctions $\mu_{-,1}(z), m(z), \mu_{+,3}(z)$ (meromorphic in $C^+$) to the eigenfunctions $\mu_{+,1}(z), \overline{m}(z), \mu_{-,3}(z)$ (meromorphic in $C^-$), and then present the reconstruction formula for the solutions of Eq.~\eref{NNLS1} under NZBC~\eref{xcx6}.

\subsection{Riemann-Hilbert problem}
\label{sectionRHP}
The properties of eigenfunctions and scattering coefficients in Lemmas~\ref{th1} and~\ref{th2}  enable us to introduce the sectionally meromorphic matrices $\mathbf{M}^{\pm}$:
\begin{subequations}
	\label{xcx90}
	\begin{align}
		\mathbf{M}^{+}(x,t,z) & = \Big(\frac{\mu_{-,1}(x,t,z)}{a_{11}(z)}, \frac{m(x,t,z) }{b_{33}(z)}, \mu_{+,3}(x,t,z) \Big) \notag \\
  &= \mathbf{\Phi}_{+}(x,t,z) e^{-\mi \bm{\Theta}(x,t,z)} \diag\Big(\frac{1}{a_{11}(z)}, \frac{1 }{b_{33}(z)}, 1  \Big), \\
		\mathbf{M}^{-}(x,t,z)& = \Big(\mu_{+,1}(x,t,z), - \frac{\overline{m}(x,t,z) }{b_{11}(z)}, \frac{\mu_{-,3}(x,t,z)}{a_{33}(z)}   \Big)  \notag \\
& = \mathbf{\Phi}_{-}(x,t,z)  e^{-\mi \bm{\Theta}(x,t,z)} \diag\Big(1, - \frac{1}{b_{11}(z)}, \frac{1}{a_{33}(z)}   \Big).
	\end{align}
\end{subequations}
To formulate RHP, it is essential to analyze the jump conditions, asymptotic behaviors of $\mathbf{M}^{\pm}$, and the pole contributions at the zeros of $a_{11}(z)$, $a_{33}(z)$, $b_{11}(z)$ and $b_{33}(z)$.

As proved in Appendix~\ref{appendixC1}, we can give the jump condition between $\mathbf{M}^{+}$ and $\mathbf{M}^{-}$.
\begin{lemma}
	\label{le4.1}
	The matrices $\mathbf{M}^{+}$ and $\mathbf{M}^{-}$ satisfy the jump condition:
	\begin{equation}
		\label{xcx91}
		\begin{aligned}
			&\mathbf{M}^{+}(x,t,z) = \mathbf{M}^{-}(x,t,z)  e^{\mi  \bm{\Theta}(x,t,z)} \mathbf{G}(z)  e^{-i\bm{\Theta}(x,t,z)},
		\end{aligned}
	\end{equation}
     where $\mathbf{G}(z) = (G_{ij}(z))$ is called the jump matrix with
     \begin{equation}
     	\begin{aligned}
     		&G_{11} = 1, \quad G_{12} = - \frac{\rho^*_{2}(z^*)}{\gamma(z)}, \quad G_{13} = \rho^*_{1}(\hat{z}^*), \\
     		&G_{21} = - \rho_{2}(z), \quad G_{22} = 1+ \frac{\rho^*_{2}(z^*)\rho_{2}(z)}{\gamma(z)}, \quad
     		G_{23} = \frac{\rho_{2}(\hat{z})\hat{z}}{\mi  \sqrt{2} q_{0}} - \rho^*_{1}(\hat{z}^*)  \rho_{2}(z),\\
     		&G_{31} = \rho^*_{1}(z^*)	- \frac{\mi \sqrt{2} q_{0}}{\hat{z}} \frac{\rho_{2}^*(\hat{z}^*) \rho_{2}(z)}{\gamma(\hat{z})}, \quad
     		G_{32} =  \frac{\mi \sqrt{2} q_{0}}{\hat{z}} \frac{\rho_{2}^*(\hat{z}^*)}{\gamma(\hat{z})} \Big(1 + \frac{\rho_{2}(z)  \rho^*_{2}(z^*)}{\gamma(z)}\Big)
     		- \frac{\rho^*_{1}(z^*) \rho_{2}^*(\hat{z}^*)}{\gamma(\hat{z}^*)}, \\
     		&G_{33}(z) = 1+\rho^*_{1}(\hat{z}^*) \rho_{1}^*(z^*) +  \frac{\mi \sqrt{2} q_{0}}{\hat{z}} \frac{\rho_{2}^*(\hat{z}^*)}{\gamma(\hat{z})} \Big(\frac{\rho_{2}(\hat{z})\hat{z}}{\mi  \sqrt{2} q_{0}}- \rho^*_1(\hat{z}^*) \rho_{2}(z)\Big) .
     	\end{aligned}
     \end{equation}
\end{lemma}

Based on Lemma~\ref{co3.6} and Corollaries~\ref{co3.7},~\ref{co3.8}, the asymptotic behaviors of $\mathbf{M}^{\pm}$ can be derived:
\begin{equation}
	\begin{aligned}
		&\mathbf{M}^{\pm}(x, t, z) = \mathbf{M}_ {\infty} +O(\frac{1}{z}),\quad   z\to \infty,  \\
		&\mathbf{M}^{\pm}(x, t, z) = \mathbf{M}_ {0} +O(1),\quad        z\to 0,       \\
	\end{aligned}
\end{equation}
where
\begin{equation}
	\begin{aligned}
		&\mathbf{M}_{\infty}   = \left(
		\begin{array}{ccc}
			\frac{q_+}{q_0} & -\frac{q^*_-}{\sqrt{2} q_0} & 0 \\
			\frac{q_-}{q_0} & \frac{q^*_+}{\sqrt{2} q_0} & 0 \\
			0 & 0 & \sqrt{2} \\
		\end{array}
		\right), \quad
		&\mathbf{M}_ {0}   =
		\left(
		\begin{array}{ccc}
			0 & 0 & -\frac{\mi  \sqrt{2} q_+}{z} \\
			0 & 0 & -\frac{\mi  \sqrt{2} q_-}{z} \\
			\frac{2 \mi  q_0}{z} & 0 & 0 \\
		\end{array}
		\right),
	\end{aligned}
\end{equation}
with $\mathbf{M}_ {\infty} + \mathbf{M}_ {0} = \mathbf{Y}_{+}$.

On the other hand, combining Lemmas~\ref{le3.3} and~\ref{le3.5}, we derive the residue conditions for the RHP:

\begin{lemma}
	\label{le4.2}
	For the discrete eigenvalues $\xi_i$ ($\Re(\xi_i) \neq 0$,  $1 \leq i \leq n_1$) and $\xi_{0,j}$ ($\Re(\xi_{0,j}) = 0$, $1 \leq j \leq m_1=0$ or $1$)
	on the circle $\Gamma$,
	the matrices $\mathbf{M}^{+}$ and $\mathbf{M}^{-}$ satisfy the residue conditions:
\begin{equation}
	\label{xcx95}
	\begin{aligned}
		&\underset{z = \xi_i}{\Res}(\mathbf{M}^{+})     = C_{i} ( M^{+}_{3} (x,t,\xi_i), \mb{0}, \mb{0}), \quad \underset{z = - \xi_i^*}{\Res}(\mathbf{M}^{+}) = \overline{C}_{i} ( M^{+}_{3} (x,t,-\xi_i^*), \mb{0}, \mb{0}),  \\
		&\underset{z = \xi_{0,j}}{\Res}(\mathbf{M}^{+})     = C_{0,j} ( M^{+}_{3} (x,t,\xi_{0,j}), \mb{0}, \mb{0}), \quad \underset{z = \xi^*_i}{\Res}(\mathbf{M}^{-})   = D_{i} ( \mb{0}, \mb{0},  M^{-}_{1} (x,t,\xi^*_i)), \\
		&\underset{z = -\xi_i}{\Res}(\mathbf{M}^{-})    = \overline{D}_{i} ( \mb{0}, \mb{0},  M^{-}_{1} ( x,t,-\xi_i)),  \quad \underset{z = -\xi_{0,j}}{\Res}(\mathbf{M}^{-})    = D_{0,j} ( \mb{0}, \mb{0},  M^{-}_{1} ( x,t,-\xi_{0,j})),
	\end{aligned}
\end{equation}
where  $M^{\pm}_{j}$ represent the $j$th columns of $\mathbf{M}^{\pm}$, and
\begin{equation}
	\begin{aligned}
		& C_{i} =  \frac{  c_{i}  e^{-2 i \theta_{1}(x,t,\xi_i) }   }{a'_{11}(\xi_i)}, \quad   \overline{C}_{i}  = - \frac{\overline{c}_{i} e^{-2 i \theta_{1} (x,t,-\xi_i^*) }}{ a'_{11}(-\xi^*_i)},
		\quad C_{0,j} =  \frac{  c_{0,j}  e^{-2 \mi  \theta_{1}(x,t,\xi_{0,j}) }   }{a'_{11}(\xi_{0,j})}, \\
		& D_{i}  = \frac{ e^{ 2 i \theta_{1} (x,t,\xi^*_i)  } }{\overline{c}_{i} a'_{33}(\xi^*_i) }, \quad   \overline{D}_{i} =  - \frac{e^{2 i \theta_{1} (x,t,-\xi_i)}}{ c_i a'_{33}(-\xi_i) },
		\quad D_{0,j}  = - \frac{e^{2 \mi  \theta_{1} (x,t,-\xi_{0,j})}}{ c_{0,j} a'_{33}(-\xi_{0,j}) }.
	\end{aligned}
\end{equation}
\end{lemma}

\begin{lemma}
	\label{le4.2b}
	For the discrete eigenvalues $z_i$ ($\Re(z_i) \neq 0$,  $1 \leq i \leq n_2$) and $z_{0,j}$ ($\Re(z_{0,j}) = 0$, $1 \leq j \leq m_2$) off the circle $\Gamma$,
	the matrices $\mathbf{M}^{+}$ and $\mathbf{M}^{-}$ satisfy the residue conditions:
	\begin{equation}
		\label{xcx96}
		\begin{aligned}
			&\underset{z = z_i}{\Res}(\mathbf{M}^{+}) =             E_{i}  ( M^{+}_{2}(x,t,z_i) , \mb{0}, \mb{0}),\quad
			\underset{z = -z^*_i}{\Res}(\mathbf{M}^{+})   =   \breve{E}_{i}  ( M^{+}_{2}(x,t,-z^*_i)  , \mb{0}, \mb{0}), \\
			&\underset{z = - \hat{z}_i }{\Res}(\mathbf{M}^{+})   =  \hat{E}_{i}  (\mb{0},  M^{+}_{3} (x,t,- \hat{z}_i), \mb{0}), \quad
			\underset{z = \hat{z}^*_i}{\Res}(\mathbf{M}^{+})   =   \overline{E}_{i}   (\mb{0},   M^{+}_{3} (x,t,\hat{z}^*_i), \mb{0}),          \\
			&\underset{z = z_{0,j}}{\Res}(\mathbf{M}^{+}) =             E_{0,j}  ( M^{+}_{2}(x,t,z_{0,j}) , \mb{0}, \mb{0}),\quad
			\underset{z = - \hat{z}_{0,j} }{\Res}(\mathbf{M}^{+})   =  \hat{E}_{0,j}  (\mb{0},  M^{+}_{3} (x,t,- \hat{z}_{0,j}), \mb{0}), \\
			&\underset{z = -z_i}{\Res}(\mathbf{M}^{-})            =  F_{i}  (\mb{0}, M^{-}_{1}(x,t,-z_i), \mb{0}), \quad
			\underset{z = z^*_i}{\Res}(\mathbf{M}^{-})           =  \breve{F}_{i}  (\mb{0}, M^{-}_{1}(x,t,z^*_i), \mb{0}),  \\
			&\underset{z = \hat{z}_i}{\Res}(\mathbf{M}^{-})       =  \hat{F}_{i}  (\mb{0},   \mb{0}, M^{-}_{2} (x,t,\hat{z}_i)), \quad
			\underset{z = -\hat{z}^*_i }{\Res}(\mathbf{M}^{-})   =   \overline{F}_{i}  (\mb{0},   \mb{0}, M^{-}_{2} (x,t,-\hat{z}^*_i)), \\
			&\underset{z = -z_{0,j}}{\Res}(\mathbf{M}^{-})            =  F_{0,j}  (\mb{0}, M^{-}_{1}(x,t,-z_{0,j}), \mb{0}), \quad
			\underset{z = \hat{z}_{0,j}}{\Res}(\mathbf{M}^{-})       = \hat{F}_{0,j}  (\mb{0},   \mb{0}, M^{-}_{2} (x,t,\hat{z}_{0,j})), \, \\
		\end{aligned}
	\end{equation}
	where
	\begin{equation}
		\begin{aligned}
			&E_{i} = \frac{b_{33}(z_i) e^{-\mi  \theta_{1}(x,t,z_i) + \mi  \theta_{2}(x,t,z_i)}}{d_{i} a'_{11}(z_i) }, \quad
			\breve{E}_{i} = \frac{b_{33}(-z^*_i) e^{-\mi  \theta_{1}(x,t,-z^*_i) + \mi  \theta_{2}(x,t,-z^*_i)}}{\breve{d}_{i} a'_{11}(-z^*_i) } , \\
			&\hat{E}_{i} = \frac{ \hat{d}_{i}   e^{ -\mi  \theta_{1}(x,t,  - \hat{z}_i  )  - \mi  \theta_{2}(x,t,  - \hat{z}_i  )  }    }{b'_{33}( - \hat{z}_i ) }, \quad
			\overline{E}_{i} = \frac{ \overline{d}_{i}   e^{ -\mi  \theta_{1}(x,t, \hat{z}^*_i )  - \mi  \theta_{2}(x,t, \hat{z}^*_i )  }    }{b'_{33}(\hat{z}^*_i) },   \\
			&E_{0,j} = \frac{b_{33}(z_{0,j}) e^{-\mi  \theta_{1}(x,t,z_{0,j}) + \mi  \theta_{2}(x,t,z_{0,j})}}{d_{0,j} a'_{11}(z_{0,j}) }, \quad
			\hat{E}_{0,j} = \frac{ \hat{d}_{0,j}   e^{ -\mi  \theta_{1}(x,t,  - \hat{z}_{0,j}  )  - \mi  \theta_{2}(x,t,  - \hat{z}_{0,j}  )  }    }{b'_{33}( - \hat{z}_{0,j} ) }, \\
			&F_{i} = - \frac{d_{i} e^{\mi  \theta_{1}(x,t,-z_i)   - \mi  \theta_{2}(x,t,-z_i)} }{b'_{11}(-z_i) }, \quad
			\breve{F}_{i} = - \frac{\breve{d}_{i} e^{\mi  \theta_{1}(x,t, z^*_i )   - \mi  \theta_{2}(x,t, z^*_i )}  }{b'_{11}(z^*_i) }, \\
			&\hat{F}_{i} =   \frac{b_{11}(\hat{z}_i) e^{\mi  \theta_{1}(x,t,\hat{z}_i) + \mi  \theta_{2}(x,t,\hat{z}_i)}}{\hat{d}_{i} a'_{33}(\hat{z}_i) }, \quad
			\overline{F}_{i} =   \frac{b_{11}(-\hat{z}^*_i) e^{\mi  \theta_{1}(x,t,-\hat{z}^*_i) + \mi  \theta_{2}(x,t,-\hat{z}^*_i)}}{\overline{d}_{i} a'_{33}(-\hat{z}^*_i) }, \\
			&F_{0,j} = - \frac{d_{0,j} e^{\mi  \theta_{1}(x,t,-z_{0,j})      - \mi  \theta_{2}(x,t,-z_{0,j})} }{b'_{11}(-z_{0,j}) }, \quad
			\hat{F}_{0,j} =   \frac{b_{11}(\hat{z}_{0,j}) e^{\mi  \theta_{1}(x,t,\hat{z}_{0,j}) + \mi  \theta_{2}(x,t,\hat{z}_{0,j})}}{\hat{d}_{0,j} a'_{33}(\hat{z}_{0,j}) }.  \\
		\end{aligned}
	\end{equation}
\end{lemma}

\subsection{Reconstruction formula}
The RHP~\eref{xcx91} can be regularized by removing  the asymptotic behaviors of $\mathbf{M}^{\pm}$ and the contributions at the zeros of scattering coefficients.
By virtue of Plemelj's formula, we have
\begin{equation}
	\label{xcx98}
	\begin{aligned}
		\mathbf{M}(x,t,z) =& \mathbf{Y}_{+} +
		\displaystyle\sum_{i = 1}^{n_1}
		\Bigg(
		\frac{\underset{z =  \xi_i}{\text{Res}}(\mathbf{M}^{+})}{z - \xi_i}+
		\frac{\underset{z =  -\xi^*_i}{\text{Res}}(\mathbf{M}^{+})}{z + \xi^*_i}+
		\frac{\underset{z =  \xi^*_i}{\text{Res}}(\mathbf{M}^{-})}{z - \xi^*_i}+
		\frac{\underset{z = -\xi_i}{\text{Res}}(\mathbf{M}^{-})}{z + \xi_i}
		\Bigg)
		\\
		&+\displaystyle\sum_{i = 1}^{n_2}
		\Bigg(
		\frac{\underset{z =  z_i}{\text{Res}}(\mathbf{M}^{+})}{z - z_i}+
		\frac{\underset{z = -z^*_i}{\text{Res}}(\mathbf{M}^{+})}{z + z^*_i}+
		\frac{\underset{z =  \hat{z}^*_i}{\text{Res}}(\mathbf{M}^{+})}{z - \hat{z}^*_i }+
		\frac{\underset{z = -\hat{z}_i }{\text{Res}}(\mathbf{M}^{+})}{z + \hat{z}_i }
		\Bigg)\\
		&+\displaystyle\sum_{i= 1}^{n_2}
		\Bigg(
		\frac{\underset{z =  z^*_i}{\text{Res}}(\mathbf{M}^{-})}{z - z^*_i}+
		\frac{\underset{z = -z_i}{\text{Res}}(\mathbf{M}^{-})}{z + z_i}+
		\frac{\underset{z =  \hat{z}_i }{\text{Res}}(\mathbf{M}^{-})}{z - \hat{z}_i}+
		\frac{\underset{z = -\hat{z}^*_i }{\text{Res}}(\mathbf{M}^{-})}{z + \hat{z}^*_i }
		\Big) \\
		&+\displaystyle\sum_{j = 1}^{m_1}
		\Bigg(
		\frac{\underset{z =  \xi_{0,j}}{\text{Res}}(\mathbf{M}^{+})}{z - \xi_{0,j}}+
		\frac{\underset{z =  \xi^*_{0,j}}{\text{Res}}(\mathbf{M}^{-})}{z - \xi^*_{0,j}}
		\Bigg)
		+\displaystyle\sum_{j = 1}^{m_2}
		\Bigg(
		\frac{\underset{z =  z_{0,j}}{\text{Res}}(\mathbf{M}^{+})}{z - z_{0,j}}+
		\frac{\underset{z =  \hat{z}^*_{0,j}}{\text{Res}}(\mathbf{M}^{+})}{z - \hat{z}^*_{0,j} }
		\Bigg)\\
		&+\displaystyle\sum_{j = 1}^{m_2}
		\Bigg(
		\frac{\underset{z =  z^*_{0,j}}{\text{Res}}(\mathbf{M}^{-})}{z - z^*_{0,j} }+
		\frac{\underset{z =  \hat{z}_{0,j}}{\text{Res}}(\mathbf{M}^{-})}{z - \hat{z}_{0,j} }
		\Bigg)  +
		\frac{1}{2 \pi \mi } \int_{\Sigma}^{ }{  \frac{\mathbf{M}^{-}(s) \big(e^{\mi \bm{\Theta}} \mathbf{G}(x,t,s)  e^{-\mi \bm{\Theta}} - \mathbf{I}\big)}{s-z}ds}, \quad z \in \mathbb{C} \setminus \Sigma,
	\end{aligned}
\end{equation}
and thus obtain the representation for the solution of Eq.~\eref{NNLS1} under NZBC~\eref{xcx6}:
		\begin{align}
q(x,t)=& q_{+} +
			\mi  \frac{\sqrt{2}}{2}
			\bigg[ \displaystyle\sum_{i = 1}^{n_1}
			\big(
			D_{i} M^{-}_{11} (x,t,\xi^*_i)+
			\overline{D}_{i} M^{-}_{11} (x,t,-\xi_i)
			\big) \notag
			 \\
&	+ 	\displaystyle\sum_{i = 1}^{n_2}
			\big(
			\hat{F}_{i} M^{-}_{12} (x,t,\hat{z}_i)+
			\overline{F}_{i} M^{-}_{12} (x,t,-\hat{z}^*_i)
			\big) + \displaystyle\sum_{j =1}^{m_1}
			D_{0,j} M^{-}_{11} (x,t,\xi^*_{0,j})  	\notag   \\
&	+ \displaystyle\sum_{j =  1}^{m_2}
			\hat{F}_{0,j} M^{-}_{12} (x,t,\hat{z}_{0,j})
			+  \frac{1}{2 \pi \mi } \int_{\Sigma}^{ }{ \big[\mathbf{M}^{-}(x,t,s) \big(e^{\mi \bm{\Theta}(x,t,s)} \mathbf{G}(s)e^{-\mi \bm{\Theta}(x,t,s)} - \mathbf{I}\big) \big]_{13}ds }
			\bigg]. 		\label{xcx107}
		\end{align}

\section{N-soliton solutions in the reflectionless case}
\label{sec5}

In this section,  considering the case of reflectionless potentials
(that is,  $\rho_{1}(z)$ and $\rho_{2}(z)$ are equal to $0$ for $z \in \mathbb{R}$),  we derive the N-soliton solutions  in the  compact form for Eq.~\eref{NNLS1} under NZBC~\eref{xcx6} (see Appendix~\ref{appendixC4}), and then  study the dynamical behavior of the obtained soliton solutions.

\subsection{Determinant representation of N-soliton solutions}
\begin{theorem}
	\label{Thm5.1}
Suppose that $||\bm{\Delta}\!\mathbf{Q}_{\pm}(\cdot,t)|| \in L^1(\mathbb{R})$,
$a_{11}(z)$ has simple zeros $\{\xi_i,-\xi_i^*\}^{n_1}_{i=1}$ and $\{ \xi_{0,j} \}^{m_1}_{j=1}$ on the half upper circle $\Gamma^+$, and simple zeros $\{z_i,-z_i^*\}^{n_2}_{i=1}$, $\{z_{0,j}\}^{m_2}_{j=1}$ in $C^{+}$ but off $\Gamma^+$, where $\Re(\xi_i), \Re(z_i) \neq 0$, $\Re(\xi_{0,j}) = \Re(z_{0,j}) = 0$. Then, in the reflectionless case, the N-soliton solutions of Eq.~\eref{NNLS1} under NZBC~\eref{xcx6} can be represented by
	\begin{equation}
		\label{Nfold2}
		\begin{aligned}
&q =q_{+}  + \frac{\det(\mathbf{S}^{\rm{aug}})}{\det(\mathbf{S})}, \quad \mathbf{S} = \begin{pmatrix}
		\mathbf{S}^{(1)} & \mathbf{S}^{(2)}   & \mathrm{\mathbf{0}} \\
		\mathrm{\mathbf{0}}       & \mathrm{I }         & \mathbf{S}^{(3)} \\
		\mathbf{S}^{(4)} & \mathbf{S}^{(5)}   & \mathrm{I} \\
	\end{pmatrix}, \quad \mathbf{S}^{\rm{aug}} = \begin{pmatrix}
		0 & \mathbf{H}  \\
		\mathbf{K} & \mathbf{S}  \\
	\end{pmatrix},
		\end{aligned}
	\end{equation}
with
	\begin{equation}
	\mathbf{S}^{(1)}_{k,l} =
			\left\{
			\begin{aligned}
				& \frac{C_{l}}{\xi_k^*-\xi_l} \frac{\mi  \sqrt{2} q_{0} }{\xi_l} + \delta_{kl}, \quad k \in [1,n_1], \,\, l \in  [1,n_1], \\
				& \frac{\overline{C}_{l-n_1} }{\xi_k^*+\xi^*_{l - n_1}}   \frac{\mi  \sqrt{2} q_{0} }{-\xi^*_{l - n_1}}, \quad k \in [1,n_1], \,\, l \in  [n_1+1, 2n_1], \\
				& \frac{C_{0,l-2n_1}}{\xi_k^*-\xi_{0,l-2n_1}} \frac{\mi  \sqrt{2} q_{0} }{\xi_{0,l-2n_1}}, \quad k \in [1,n_1],\,\, l \in  [2n_1+1, N_1], \\
				& \frac{C_{l}}{-\xi_{k-n_1}-\xi_l} \frac{\mi  \sqrt{2} q_{0} }{\xi_l}, \quad
				k \in [n_1+1,2n_1],\,\, l \in  [1,n_1], \\
				& \frac{\overline{C}_{l-n_1} }{-\xi_{k-n_1}+\xi^*_{l - n_1}}   \frac{\mi  \sqrt{2} q_{0} }{-\xi^*_{l - n_1}}  + \delta_{kl}, \quad
				k \in [n_1+1,2n_1],\,\, l \in  [n_1+1, 2n_1], \\
				& \frac{C_{0,l-2n_1}}{-\xi_{k-n_1}-\xi_{0,l-2n_1}} \frac{\mi  \sqrt{2} q_{0} }{\xi_{0,l-2n_1}}, \quad
				k \in [n_1+1,2n_1],\,\, l \in  [2n_1+1, N_1], \\
				& \frac{C_{l}}{-\xi_{0,k-2n_1}-\xi_{l}} \frac{\mi  \sqrt{2} q_{0} }{\xi_l}, \quad
				k \in [2n_1+1,N_1],\,\, l \in  [1,n_1], \\
				& \frac{\overline{C}_{l-n_1} }{-\xi_{0,k-2n_1}+\xi^*_{l - n_1}}   \frac{\mi  \sqrt{2} q_{0} }{-\xi^*_{l - n_1}}, \quad
				k \in [2n_1+1,N_1],\,\, l \in  [n_1+1, 2n_1], \\
				& \frac{C_{0,l-2n_1}}{-\xi_{0,k-2n_1}-\xi_{0,l-2n_1}} \frac{\mi  \sqrt{2} q_{0} }{\xi_{0,l-2n_1}}  + \delta_{kl}, \quad
				k \in [2n_1+1,N_1],\,\, l \in  [2n_1+1, N_1], \\
			\end{aligned}
			\right.  \notag
	\end{equation}
	\begin{equation}
\hspace{-4.16cm} \mathbf{S}^{(2)}_{k,l} =
			\left\{
			\begin{aligned}
				& -\frac{ E_{l}}{\xi_k^* - z_l}, \quad
				k \in [1,n_1],\,\, l \in  [1,n_2], \\
				& -\frac{  \breve{E}_{l - n_2} }{\xi_k^* + z^*_{l - n_2}}, \quad
				k \in [1,n_1],\,\, l \in  [n_2+1, 2n_2], \\
				& -\frac{ E_{0,l-2n_2} }{\xi_{k}^* - z_{0,l-2n_2}}, \quad
				k \in [1,n_1],\,\, l \in  [2n_2+1, N_2], \\
				& -\frac{ E_{l}}{-\xi_{k-n_1} - z_{l}}, \quad
				k \in [n_1+1, 2n_1],\,\, l \in  [1,n_2], \\
				& -\frac{  \breve{E}_{l - n_2} }{-\xi_{k-n_1} + z^*_{l - n_2}}, \quad
				k \in [n_1+1, 2n_1],\,\, l \in  [n_2+1, 2n_2], \\
				& -\frac{ E_{0,l-2n_2} }{-\xi_{k-n_1} - z_{0,l-2n_2}}, \quad
				k \in [n_1+1, 2n_1],\,\, l \in  [2n_2+1, N_2], \\
				& -\frac{ E_{l}}{-\xi_{0,k-2n_1} - z_{l}}, \quad
				k \in [2n_1+1, N_1],\,\, l \in  [1,n_2], \\
				& -\frac{  \breve{E}_{l - n_2} }{-\xi_{0,k-2n_1} + z^*_{l - n_2}}, \quad
				k \in [2n_1+1, N_1],\,\, l \in  [n_2+1, 2n_2], \\
				& -\frac{ E_{0,l-2n_2} }{-\xi_{0,k-2n_1} - z_{0,l-2n_2}}, \quad
				k \in [2n_1+1, N_1],\,\, l \in  [2n_2+1, N_2], \\
			\end{aligned}
			\right.  \notag
	\end{equation}
\begin{equation}
	\mathbf{S}^{(3)}_{k,l} =
		\left\{
		\begin{aligned}
			& \frac{  \overline{E}_{l} }{\hat{z}_{k}-\hat{z}^*_l}  \frac{\mi  \sqrt{2} q_{0} }{\hat{z}^*_l } - \frac{ \breve{F}_{l} }{\hat{z}_{k} - z^*_l} , \quad
			k \in [1,n_2],\,\, l \in  [1 ,n_2], \\
			& \frac{\hat{E}_{l-n_2}}{\hat{z}_{k}+\hat{z}_{l-n_2}} \frac{\mi  \sqrt{2} q_{0} }{-\hat{z}_{l-n_2}} -   \frac{ F_{l-n_2}}{\hat{z}_{k} + z_{l-n_2}}, \quad
			k \in [1,n_2],\,\, l \in  [n_2+1, 2n_2], \\
			& \frac{  \hat{E}_{0,l-2n_2} }{\hat{z}_{k}+\hat{z}_{0,l-2n_2}}  \frac{\mi  \sqrt{2} q_{0} }{-\hat{z}_{0,l-2n_2} } - \frac{ F_{0,l-2n_2} }{\hat{z}_{k} +z_{0,l-2n_2}}  , \quad
			k \in [1,n_2],\,\, l \in  [2n_2+1, N_2], \\
			& \frac{  \overline{E}_{l} }{-\hat{z}^*_{k-n_2}-\hat{z}^*_l}  \frac{\mi  \sqrt{2} q_{0} }{\hat{z}^*_l } - \frac{ \breve{F}_{l} }{-\hat{z}^*_{k-n_2} - z^*_l} , \quad
			k \in [n_2+1,2n_2],\,\, l \in  [1 ,n_2], \\
			& \frac{\hat{E}_{l-n_2}}{-\hat{z}^*_{k-n_2}+\hat{z}_{l-n_2}} \frac{\mi  \sqrt{2} q_{0} }{-\hat{z}_{l-n_2}} -   \frac{ F_{l-n_2}}{-\hat{z}^*_{k-n_2} + z_{l-n_2}}, \quad
			k \in [n_2+1,2n_2],\,\, l \in  [n_2+1, 2n_2], \\
			& \frac{  \hat{E}_{0,l-2n_2} }{-\hat{z}^*_{k-n_2}+\hat{z}_{0,l-2n_2}}  \frac{\mi  \sqrt{2} q_{0} }{- \hat{z}_{0,l-2n_2} } - \frac{ F_{0,l-2n_2} }{-\hat{z}^*_{k-n_2} +z_{0,l-2n_2}}  , \quad
			k \in [n_2+1,2n_2],\,\, l \in  [2n_2+1, N_2], \\
			& \frac{  \overline{E}_{l} }{\hat{z}_{0,k-2n_2}-\hat{z}^*_l}  \frac{\mi  \sqrt{2} q_{0} }{\hat{z}^*_l } - \frac{ \breve{F}_{l} }{\hat{z}_{0,k-2n_2} - z^*_l} , \quad
			k \in [2n_2+1,N_2],\,\, l \in  [1 ,n_2], \\
			& \frac{\hat{E}_{l-n_2}}{\hat{z}_{0,k-2n_2}+\hat{z}_{l-n_2}} \frac{\mi  \sqrt{2} q_{0} }{-\hat{z}_{l-n_2}} -   \frac{ F_{l-n_2}}{\hat{z}_{0,k-2n_2} + z_{l-n_2}}, \quad
			k \in [2n_2+1,N_2],\,\, l \in  [n_2+1, 2n_2], \\
			& \frac{  \hat{E}_{0,l-2n_2} }{\hat{z}_{0,k-2n_2}+\hat{z}_{0,l-2n_2}}  \frac{\mi  \sqrt{2} q_{0} }{-\hat{z}_{0,l-2n_2} } - \frac{ F_{0,l-2n_2} }{\hat{z}_{0,k-2n_2} +z_{0,l-2n_2}}  , \quad
			k \in [2n_2+1,N_2],\,\, l \in  [2n_2+1, N_2], \\
		\end{aligned}
		\right. \notag
\end{equation}
\begin{equation}
	\hspace{0.8cm} \mathbf{S}^{(4)}_{k,l} =
		\left\{
		\begin{aligned}
			& \frac{C_{l}}{z_{k}^*-\xi_l} \frac{\mi  \sqrt{2} q_{0} }{\xi_l}, \quad
			k \in [1,n_2],\,\, l \in  [1,n_1], \\
			& \frac{\overline{C}_{l-n_1} }{z_{k}^*+\xi^*_{l - n_1}}   \frac{\mi  \sqrt{2} q_{0} }{-\xi^*_{l - n_1}}, \quad
			k \in [1,n_2],\,\, l \in  [n_1+1, 2n_1], \\
			& \frac{C_{0,l-2n_1}}{z_{k}^*-\xi_{0,l-2n_1}} \frac{\mi  \sqrt{2} q_{0} }{\xi_{0,l-2n_1}}, \quad
			k \in [1,n_2],\,\, l \in  [2n_1+1, N_1], \\
			& \frac{C_{l}}{-z_{k-n_2}-\xi_l} \frac{\mi  \sqrt{2} q_{0} }{\xi_l}, \quad
			k \in [n_2+1,2n_2],\,\, l \in  [1,n_1], \\
			& \frac{\overline{C}_{l-n_1} }{-z_{k-n_2}+\xi^*_{l - n_1}}   \frac{\mi  \sqrt{2} q_{0} }{-\xi^*_{l - n_1}}, \quad
			k \in [n_2+1,2n_2],\,\, l \in  [n_1+1, 2n_1], \\
			& \frac{C_{0,l-2n_1}}{-z_{k-n_2}-\xi_{0,l-2n_1}} \frac{\mi  \sqrt{2} q_{0} }{\xi_{0,l-2n_1}}, \quad
			k \in [n_2+1,2n_2],\,\, l \in  [2n_1+1, N_1], \\
			& \frac{C_{l}}{-z_{0,k-2n_2}-\xi_l} \frac{\mi  \sqrt{2} q_{0} }{\xi_l}, \quad
			k \in [2n_2+1,N_2],\,\, l \in  [1,n_1], \\
			& \frac{\overline{C}_{l-n_1} }{- z_{0,k-2n_2}+\xi^*_{l - n_1}}   \frac{\mi  \sqrt{2} q_{0} }{-\xi^*_{l - n_1}}  , \quad
			k \in [2n_2+1,N_2],\,\, l \in  [n_1+1, 2n_1], \\
			& \frac{C_{0,l-2n_1}}{-z_{0,k-2n_2}-\xi_{0,l-2n_1}} \frac{\mi  \sqrt{2} q_{0} }{\xi_{0,l-2n_1}}, \quad
			k \in [2n_2+1,N_2],\,\, l \in  [2n_1+1, N_1], \\
		\end{aligned}
		\right. \notag
\end{equation}
\begin{equation}
		\hspace{0cm} \mathbf{S}^{(5)}_{k,l} =
		\left\{
		\begin{aligned}
			& -\frac{ E_{l}}{z_{k}^* - z_l}, \quad
			k \in [1,n_2],\,\, l \in  [1 ,n_2], \\
			& -\frac{  \breve{E}_{l - n_2} }{z_{k}^* + z^*_{l - n_2}}, \quad
			k \in [1,n_2],\,\, l \in  [n_2+1, 2n_2], \\
			& -\frac{ E_{0,l-2n_2}}{z_{k}^* - z_{0,l-2n_2}}, \quad
			k \in [1,n_2],\,\, l \in  [2n_2+1, N_2], \\
			& -\frac{ E_{l}}{-z_{k-n_2} - z_l}, \quad
			k \in [n_2+1,2n_2],\,\, l \in  [1 ,n_2], \\
			& -\frac{  \breve{E}_{l - n_2} }{-z_{k-n_2} + z^*_{l - n_2}}, \quad
			k \in [n_2+1,2n_2],\,\, l \in  [n_2+1, 2n_2], \\
			& -\frac{ E_{0,l-2n_2}}{-z_{k-n_2} - z_{0,l-2n_2}}, \quad
			k \in [n_2+1,2n_2],\,\, l \in  [2n_2+1, N_2], \\
			& -\frac{ E_{l}}{-z_{0,k-2n_2} - z_l}, \quad
			k \in [2n_2+1,N_2],\,\, l \in  [1 ,n_2], \\
			& -\frac{  \breve{E}_{l - n_2} }{-z_{0,k-2n_2} + z^*_{l - n_2}}, \quad
			k \in [2n_2+1,N_2],\,\, l \in  [n_2+1, 2n_2], \\
			& -\frac{ E_{0,l-2n_2}}{-z_{0,k-2n_2} - z_{0,l-2n_2}}, \quad
			k \in [2n_2+1,N_2],\,\, l \in  [2n_2+1, N_2], \\
		\end{aligned}
		\right. \notag
\end{equation}
    where $N_1 = 2n_1+m_1$, $N_2 = 2n_2+m_2$, and the vectors $\mathbf{H} =  - \mi  \frac{\sqrt{2}}{2} (H_{1},...,H_{N_1+2N_2})$ and $\mathbf{K} = (K_{1},...,K_{ N_1+2N_2})^{T}$ are, respectively, given by
	\begin{equation}
\label{Eq57}
		\begin{aligned}
		\quad H_{k} &=
		\left\{
		\begin{aligned}
			& D_{k}, \quad
			k \in [1, n_1], \\
			& \overline{D}_{k-n_1}, \quad
			k \in [n_1+1, 2n_1], \\
			& D_{0,k-2n_1}, \quad
			k \in [2n_1+1, N_1], \\
			& \hat{F}_{k-N_1}, \quad
			k \in [N_1+1,N_1+n_2], \\
			& \overline{F}_{k-N_1-n_2}, \quad
			k \in [N_1+n_2+1,N_1+2n_2], \\
			& \hat{F}_{0,k-N_1-2n_2}, \quad
			k \in [N_1+2n_2+1,N_1+N_2], \\
			&0, \quad
			k \in[N_1+N_2+1, N_1+2N_2], \\
		\end{aligned}
		\right.
		\end{aligned}
	\end{equation}
 	\begin{equation}
		\begin{aligned}
\hspace{-1.3cm}		K_k =
		\left\{
		\begin{aligned}
			&\frac{q_{+}}{q_0}, \quad
			k \in [1,N_1], \\
			&- \frac{q^{*}_{-}}{\sqrt{2} q_0}, \quad
			k \in [N_1+1,N_1+N_2], \\
			&\frac{q_{+}}{q_0}, \quad
			k \in[N_1+N_2+1, N_1+2N_2]. \\
		\end{aligned}
		\right.
	\end{aligned}
	\end{equation}
\end{theorem}

\begin{remark}
\label{remark}
	In the reflectionless case, the trace formulas~\eref{xcx108} have the following expressions:
	\begin{subequations}
	\begin{align}
		a_{33}(z) =
		&\displaystyle\prod_{i = 1}^{n_1} \left(\frac{ z - \xi^*_i  }{z- \xi_i }   \frac{ z + \xi_i }{z + \xi^*_i }\right)
		\displaystyle\prod_{i = 1}^{m_1} \frac{ z + \xi_{0,j} }{z - \xi_{0,j} }
		\displaystyle\prod_{i = 1}^{n_2} \left(\frac{  z + \hat{z}^*_i }{z -\hat{z}^*_i }   \frac{ z -\hat{z}_i  }{z+ \hat{z}_i }\right)
		\displaystyle\prod_{i = 1}^{m_2}  \frac{ z -\hat{z}_{0,j} }{z +\hat{z}_{0,j} },\\
		b_{33}(z) = &
		\displaystyle\prod_{i = 1}^{n_1} \left(\frac{ z - \xi_i  }{z- \xi^*_i }   \frac{ z + \xi^*_i }{z + \xi_i }\right)
		\displaystyle\prod_{i = 1}^{m_1} \frac{ z - \xi_{0,j} }{z + \xi_{0,j} }
		\displaystyle\prod_{i = 1}^{n_2} \left(\frac{ z - \hat{z}^*_i  }{z + \hat{z}^*_i}   \frac{ z +\hat{z}_i }{z -\hat{z}_i }\right)
		\displaystyle\prod_{i = 1}^{m_2}  \frac{ z +\hat{z}_{0,j} }{z -\hat{z}_{0,j} }.
	\end{align}
\end{subequations}
Based on the relation~\eref{Eq248}, one can also obtain the expressions of $a_{11}(z)$ and $b_{11}(z)$. In view of the asymptotic limits of $a_{33}(z)$ (or $b_{33}(z)$) as $z \to 0$ in Eq.~\eref{3.19}, the following constraint should be satisfied:
\begin{equation}
\begin{aligned}
	\label{xcx111}
	(-1)^{m_1 + m_2}
	= \frac{q_+  q^*_- + q^*_+  q_-}{2 q^{2}_0},
\end{aligned}
\end{equation}
which implies that if $q_{+}=-q_{-}$, the index $m_1 + m_2$ must be odd; whereas for $q_{+}=q_{-}$, $m_1 + m_2$ must be even.
\end{remark}

\subsection{Dynamical behavior of soliton solutions}

In the following, we study the dynamical behavior of soliton solutions with $N_1 + N_2 = 1$ and $N_1 + N_2 = 2$ as illustrative examples.
According to Remark~\ref{remark}, there must be $q_{+}=-q_{-}$  for $N_1 + N_2 = 1$ and $q_{+}=q_{-}$ for $N_1 + N_2 = 2$, in which the  heteroclinic and homoclinic soliton solutions can be derived respectively.

\begin{figure}[H]
	\centering
	\subfigure[]{
		\label{f1aa}
		\includegraphics[width=2.7in]{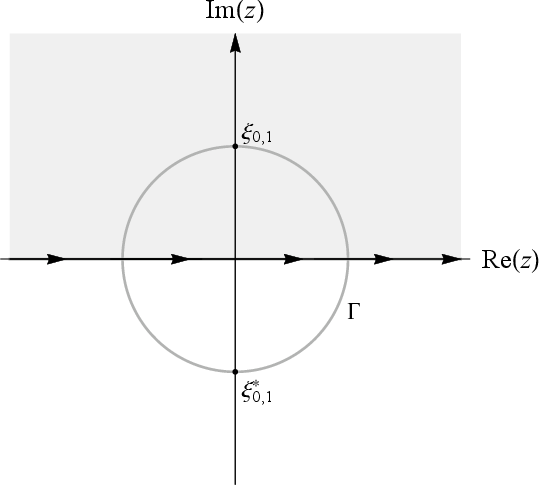}}\hfill
	\subfigure[]{
		\label{f1bb}
		\includegraphics[width=2.7in]{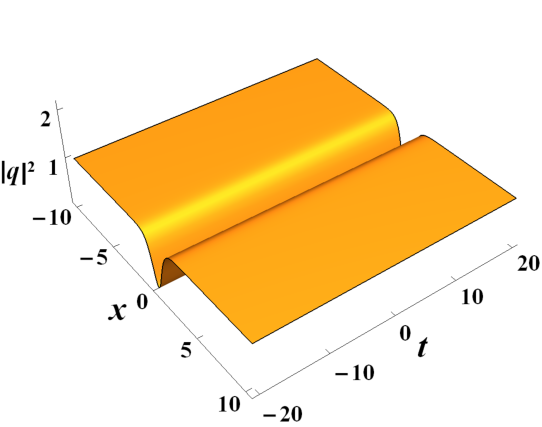}}\hfill
	\caption{\small
		(a) Distribution of discrete eigenvalues $\{\xi_{0,1}, \xi^*_{0,1}\}$;
		(b) A non-propagating dark soliton with $q_+ = 1$, $\xi_{0,1} =  \sqrt{2} \mi $ and $c_{0,1} = 1$.
		\label{f1} }
\end{figure}

\textbf{Case 5.1} $N_1 + N_2 = 1$ ($n_{1,2}=0$, $m_{1}=1$ and $m_2=0$).
In this case,  there is just one pair of pure imaginary eigenvalues $\xi_{0,1} = \mi  \sqrt{2} q_0$ and $\xi^*_{0,1} =-\mi  \sqrt{2} q_0$ on the circle $\Gamma$ (see Fig.~\ref{f1aa}) and the norming constant satisfies $c_{0,1}^2 = 1$.
If $c_{0,1} = 1$, we have the dark one-soliton solution:
\begin{equation}
	\label{511}
	\begin{aligned}
		q(x,t)
		& =  q_{+}  \tanh (\sqrt{2} q_0 x) ,
	\end{aligned}
\end{equation}
which represents a non-propagating black soliton (see Fig.~\ref{f1bb}). For  $c_{0,1} = - 1$, the solution is $q(x,t) = q_{+}\coth (\sqrt{2} q_0 x)$, which has always the singularity at $x = 0$.

\begin{figure}[H]
	\centering
	\subfigure[]{
		\label{f1a}
		\includegraphics[width=2.7in]{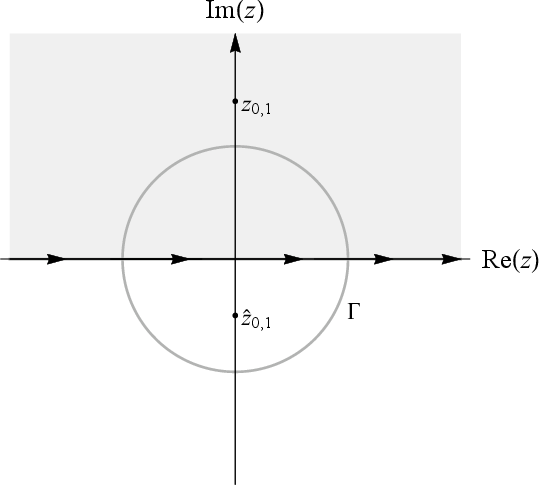}}\hfill
	\subfigure[]{
		\label{f1b}
		\includegraphics[width=2.7in]{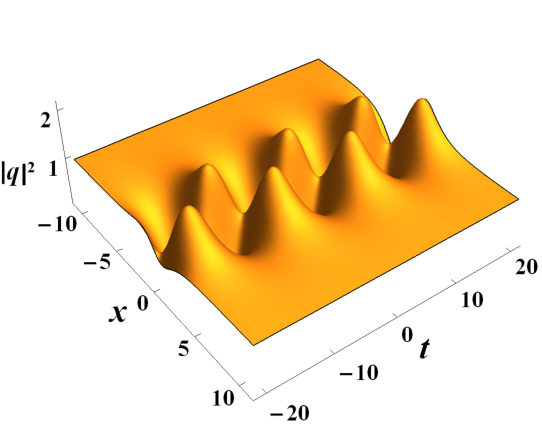}}\hfill
	\caption{\small
		(a) Distribution of discrete eigenvalues $\{z_{0,1}, \hat{z}_{0,1}\}$;
		(b) A non-propagating beating soliton with $q_+ = e^{2\mi }$, $z_{0,1} = 2 \sqrt{2} \mi $ and $d_{0,1} = \frac{\sqrt{6}}{5} e^{3 \mi}$.
		\label{f2} }
\end{figure}

\textbf{Case 5.2} $N_1 + N_2 = 1$ ($n_{1,2}=0$, $m_{1}=0$ and $m_2=1$).
In such case, there is one pair of pure imaginary eigenvalues $z_{0,1}, \hat{z}_{0,1}$ ($|z_{0,1}|>\sqrt{2}q_0$) lying off the circle $\Gamma$ (see Fig.~\ref{f1a}), so we can take $z_{0,1} = \mi  \kappa \sqrt{2} q_0$ with $\kappa> 1$. In view of the second  constraint of~\eref{xcx81e}, the constant $d_{0,1}$  can be expressed as $d_{0,1} =   - \frac{\kappa \sqrt{\kappa^2 - 1}}{\sqrt{2} \left(\kappa^2+1\right)} e^{\mi \omega_1}$, where $\omega_1 = \mathrm{arg} (d_{0,1})$.
Thus,
the dark-bright one-soliton solution can be presented as follows:
\begin{equation}
	\label{Eq512}
	\begin{aligned}
		q(x,t)
		& =  q_{+} \Big[\tanh \Big(\frac{\sqrt{2} q_0}{\kappa} x \Big) - \frac{\sqrt{\kappa^{2} - 1}}{\kappa} \sech \Big(\frac{\sqrt{2} q_0}{\kappa} x \Big)e^{\mi   \big(\frac{2 q_0^{2}} {\kappa^{2}} t  - \omega_1 - 2 \theta_{+} \big) } \Big],
	\end{aligned}
\end{equation}
which reduces to the dark one-soliton solution~\eref{511} at $\kappa \to 1$.
As shown in Fig.~\ref{f1b}, the solution is heteroclinic in $x$ and periodic in $t$, so the dark and bright solitons beat each other with the period $\frac{\pi \kappa^{2}}{ q_0^{2}}$ as $t$ evolves. Via the extreme value analysis, we find that $|q(x,t)|^2$  attains its maximum value $2 - \frac{1}{\kappa^{2}}$ periodically at the points $\Big(\frac{\kappa}{\sqrt{2} q_0} \ln \Big(\frac{\sqrt{ 2\kappa^{2} -1}+ \kappa}{\sqrt{ \kappa^{2} -1} }\Big), \frac{\kappa^{2} (2 n \pi  + \omega_1 + 2 \theta_{+} )}{ 2 q_0^{2} }\Big)$ and $\Big(\frac{\kappa}{\sqrt{2} q_0} \ln \Big(\frac{\sqrt{2 \kappa^{2} -1}- \kappa}{\sqrt{ \kappa^{2} -1} }\Big), \frac{\kappa^{2} [ (2 n+1)\pi  + \omega_1 + 2 \theta_{+} ]}{ 2 q_0^{2} }\Big)$ ($n \in \mathbb{Z}$),  and it periodically drops to its minimum value $0$ at the points $\Big(\frac{\kappa}{\sqrt{2} q_0} \ln \Big(\frac{\sqrt{2 \kappa^{2} -1} + \sqrt{\kappa^{2} -1}}{\sqrt{2} q_0} \Big), \frac{\kappa^{2} (2 n \pi  + \omega_1 + 2 \theta_{+} )}{ 2 q_0^{2} }\Big)$ and $\Big(\frac{\kappa}{\sqrt{2} q_0} \ln \Big(\frac{\sqrt{2 \kappa^{2} -1} - \sqrt{\kappa^{2} -1}}{\sqrt{2} q_0}\Big)$, $\frac{\kappa^{2} [ (2 n+1)\pi  + \omega_1 + 2 \theta_{+} ]}{ 2 q_0^{2} }\Big)$ ($n \in \mathbb{Z}$).

Recall that Eq.~\eref{NNLS} arises from a parity-symmetric reduction of the Manakov system~\eref{Manakov}, where $q(x,t)$ and $q(-x,t)$ are two components of system~\eref{Manakov}. Thus, it is not surprising that the intensities of individual components exhibit the beating behavior~\cite{QHP}, but their superposed intensity shows a non-propagating dark soliton, namely,
\begin{equation}
	\begin{aligned}
		|q(x,t)|^2 + |q(-x,t)|^2
		& =  2 q^2_0 \Big(1 - \frac{1}{\kappa^2}\sech^2\Big(\frac{\sqrt{2} q_0}{\kappa} x \Big) \Big).
	\end{aligned}
\end{equation}
So, solution~\eref{Eq512} is also called the beating one-soliton~\cite{LCZ} and it is neither even- or odd-symmetric with respect to $x$.

\begin{figure}[H]
	\centering
	\subfigure[]{\label{f3a}
		\includegraphics[width=2.7in]{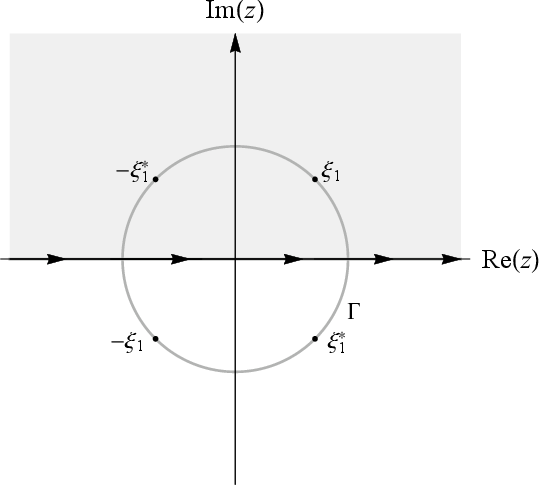}}\hfill
	\subfigure[]{\label{f2a}
		\includegraphics[width=2.7in]{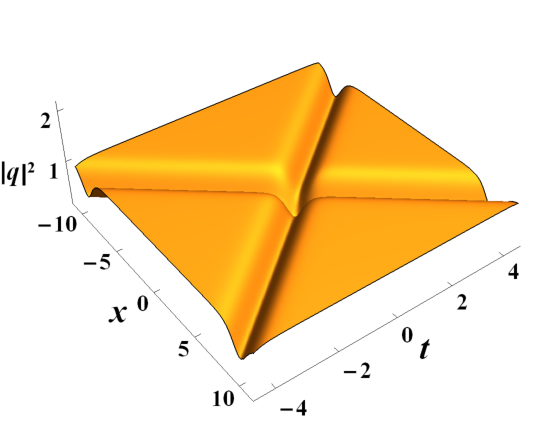}}\hfill
	\caption{\small
		(a) Distribution of discrete eigenvalues $\{\xi_1, -\xi_1, \xi^*_1, -\xi^*_1\}$;
		(b) An interaction between two propagating dark solitons with $q_+ = 1$, $\xi_{1} = \sqrt{2} e^{ \frac{\pi}{4} \mi}$ and $c_{1} = 2\mi$.
		\label{Fig461} }
\end{figure}

\textbf{Case 5.3} $N_1 + N_2 = 2$ ($n_{1}=1$ and $n_{2}=m_{1,2}=0$).
For this case, the discrete eigenvalues  appear in symmetric quartet $\{\xi_1, -\xi_1, \xi^*_1, -\xi^*_1\}$ on the circle $\Gamma$,  as shown in Fig.~\ref{f3a}. By  the relations in Eq.~\eref{C38}, the norming constants  satisfy $c_{1} \overline{c}_1 = - 1$ and $\overline{c}_1  c^*_{1} = 1$, which means $c_{1}$ is a  pure imaginary number. Then, by setting $\xi_{1} =  \sqrt{2} q_0 e^{\mi \varsigma_1 }$ $(\varsigma_1 \in (0, \frac{\pi}{2}))$, we obtain the two-soliton solution:
\begin{equation}
\begin{aligned}
&q(x,t) = q_+\Bigg(1-2 \tan(\varsigma_1)\frac{c_{1} e^{- \mi \varsigma_1} e^{-2 \mi \theta_1(\xi_{1})} + c^{-1}_{1} e^{\mi \varsigma_1} e^{-2 \mi \theta_1(-\xi^*_{1})} }{1 + e^{-2 \mi \theta_1(\xi_{1})-2 \mi \theta_1(-\xi^*_{1})}  - \mi \sec(\varsigma_1)
		\big(c_{1} e^{-2 \mi \theta_1(\xi_{1})} - c^{-1}_{1}  e^{-2 \mi \theta_1(-\xi^*_{1})}\big)}\Bigg),
\end{aligned}
\end{equation}
where
\begin{equation}
	\label{xcx88}
	\begin{aligned}
		&\theta_1(\xi_{1}) = - \mi  \sqrt{2} q_0 \sin(\varsigma_1)(x + 2 \sqrt{2} q_0 \cos(\varsigma_1) t), \quad
		\theta_1(-\xi^*_{1}) =  - \mi  \sqrt{2} q_0 \sin(\varsigma_1)(x - 2 \sqrt{2} q_0 \cos(\varsigma_1) t).
	\end{aligned}
\end{equation}

By respectively assuming that $\Im[\theta_1(\xi_{1})]= O(1)$ and  $\Im[\theta_1(-\xi^*_{1})] = O(1)$, we make asymptotic analysis of the solution and obtain two asymptotic solitons as $t \to \pm \infty$:
\begin{subequations}
\label{xcx92}
\begin{align}
&q^{\pm}_{\rm{I}} =  q_+ e^{\pm \mi  \varsigma_1} \big[\cos(\varsigma_1) - \mi  \sin(\varsigma_1)  \tanh( \eta_{\rm{I}}^{\pm} ) \big], \label{xcx92a}\\
&q^{\pm}_{\rm{I\!I}} = q_+ e^{\pm \mi  \varsigma_1} \big[\cos(\varsigma_1) + \mi  \sin(\varsigma_1)  \tanh( \eta_{\rm{I\!I}}^{\pm} ) \big], \label{xcx92b}
	\end{align}
\end{subequations}
where
\begin{equation}
	\begin{aligned}
		&\eta^{\pm}_{\rm{I}}=    \sqrt{2} q_0 \sin(\varsigma_1)[x - 2 \sqrt{2} q_0 \cos(\varsigma_1) t]
		\mp \frac{1}{2}   \ln \Bigg(\frac{|c_{1}|^{\mp 1}}{\cos(\varsigma_1)}\Bigg), \\
		&\eta^{\pm}_{\rm{I\!I}}=  \sqrt{2} q_0 \sin(\varsigma_1)[x + 2 \sqrt{2} q_0 \cos(\varsigma_1) t]
		\pm \frac{1}{2}   \ln \Bigg(\frac{|c_{1}|^{\mp 1}}{\cos(\varsigma_1)}\Bigg). \nonumber
	\end{aligned}
\end{equation}
Apparently, both the asymptotic expressions $q^{\pm}_{\rm{I}}$ and $q^{\pm}_{\rm{I\!I}}$ have no singularity, and they
can describe two gray solitons with the velocities $v^{\pm}_{\rm{I}} = - v^{\pm}_{\rm{I\!I}}= 2 \sqrt{2} q_0 \cos(\varsigma_1)$ (see Fig.~\ref{f2a}). Based on the expressions of asymptotic solitons, the extreme values of $|q^{\pm}_{\rm{I}}|^2$ and $|q^{\pm}_{\rm{I\!I}}|^2$ can be obtained as
\begin{equation}
	\begin{aligned}
		&A_{\rm{I}}^{\pm} = A_{\rm{I\!I}}^{\pm} = q_0^2 \cos^2 (\varsigma_1 ),  \nonumber
	\end{aligned}
\end{equation}
where  $\sin (\varsigma_1)$ characterizes the darkness of soliton.
The soliton position shifts between $q_{\rm{I}, \rm{I\!I}}^{+}$ and $q_{\rm{I}, \rm{I\!I}}^{-}$ are, respectively, given by  $\Delta x_{0,\rm{I}}= \frac{\ln(\cos(\varsigma_1))}{\sqrt{2} q_0 \sin(\varsigma_1)},  \Delta x_{0,\rm{I\!I}} = -\frac{\ln(\cos(\varsigma_1))}{\sqrt{2} q_0 \sin(\varsigma_1)}$, and their phase shift are $\Delta \phi_{\rm{I}}=\Delta  \phi_{\rm{I\!I}}=2 \varsigma_1$.

\begin{figure}[H]
	\centering
	\subfigure[]{\label{f3aa}
		\includegraphics[width=2.7in]{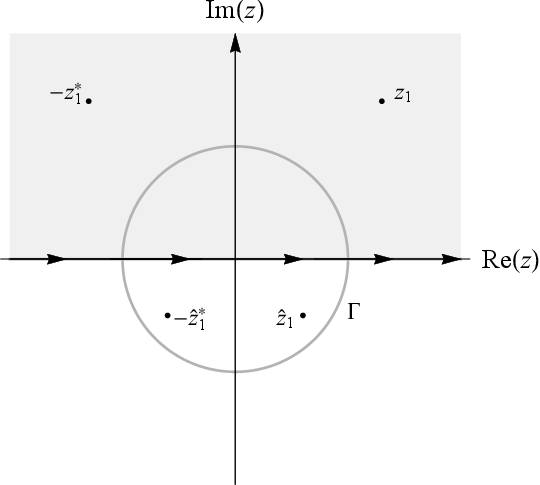}}\hfill
	\subfigure[]{\label{f2aa}
		\includegraphics[width=2.7in]{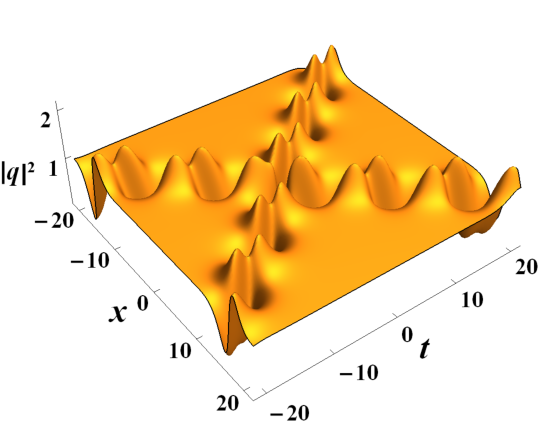}}\hfill
	\caption{\small
		(a) Distribution of discrete eigenvalues $\{z_1, \hat{z}_1, -z^*_1, -\hat{z}^*_1\}$;
		(b) An interaction between two propagating beating solitons with $q_+ = 1$,  $z_1 =\sqrt{5} e^{\arctan(2) \mi}$ and $d_{1} = 1$.
		\label{Fig4611} }
\end{figure}

\textbf{Case 5.4} $N_1 + N_2 = 2$ ($n_{2}=1$ and $n_{1}=m_{1,2}=0$).
In this case, the discrete eigenvalues  appear in symmetric quartet $\{z_1, -z^*_1, \hat{z}_1, -\hat{z}^*_1\}$ off the circle $\Gamma$, as shown in Fig.~\ref{f3aa}.
Correspondingly, the norming constants satisfy the constraints in~\eref{xcx81c} and the eigenvalue $z_1$ can be
taken as $z_{1} =  \sqrt{2} \kappa q_0 e^{\mi \varsigma_1 }$, where $\varsigma_1 \in (0, \frac{\pi}{2})$ and $\kappa>1$ (Note that for $0<\kappa < 1$
the denominator always crosses zeros, which implies that the solutions have singularities).
At the same time,  one should notice that there exist the following relations:
\begin{subequations}
	\begin{align}
& \theta_2(z) -\theta_1(z)  = \theta_1(\hat{z}) + \theta_2(\hat{z}), \\
		&\theta_2(z_{1}) -\theta_1(z_{1})  =
		\frac{ q_0 }{\kappa^2}
		\Big[-\big(\sqrt{2} \kappa \cos (\varsigma_1) x + 2 q_0 \cos (2 \varsigma_1) t \big) + \mi \sin (\varsigma_1)  \big(\sqrt{2} \kappa x + 4 q_0 \cos (\varsigma_1) t\big)\Big], \\
		&\theta_2(-z^*_{1}) -\theta_1(-z^*_{1})  =
		\frac{ q_0 }{\kappa^2}
		\Big[\big(\sqrt{2} \kappa \cos (\varsigma_1) x - 2 q_0 \cos (2 \varsigma_1) t \big) + \mi \sin (\varsigma_1)  \big(\sqrt{2} \kappa x - 4 q_0 \cos (\varsigma_1) t\big)\Big].
	\end{align}
\end{subequations}
By respectively assuming that $\Im[\theta_2(z_{1})-\theta_1(z_{1})] = O(1)$ and  $\Im[\theta_2(-z^*_{1})-\theta_1(-z^*_{1}) ] = O(1)$,
 the asymptotic analysis shows that there are two asymptotic solitons as $t \to \pm \infty$:
\begin{subequations}
	\label{xcx921}
\begin{align}
		&q^{\pm}_{\rm{I}} =  q_+  e^{\pm \mi  \varsigma_1}
		\Big[\cos(\varsigma_1) - \mi  \sin(\varsigma_1)
		\tanh (\eta^{\pm}_{\rm{I}})
		- \tau_1^{\pm}
		\sech(\eta^{\pm}_{\rm{I}})
		e^{\mi \varpi^{\pm}_{\rm{I}}}
		\Big], \label{xcx92aa}\\
		&q^{\pm}_{\rm{I\!I}} = q_+ e^{\pm \mi  \varsigma_1}
		\Big[
		\cos(\varsigma_1) + \mi  \sin(\varsigma_1) \tanh (\eta^{\pm}_{\rm{I\!I}})
		 + \tau_1^{\pm}
		\sech(\eta^{\pm}_{\rm{I\!I}} )
		e^{\mi \varpi^{\pm}_{\rm{I\!I}}}
		\Big], \label{xcx92bb}
\end{align}
\end{subequations}
with
\begin{equation}
	\begin{aligned}
		&\eta^{\pm}_{\rm{I}}=  \frac{q_0 \sin(\varsigma_1)}{\kappa^2} [\sqrt{2} \kappa x - 4 q_0 \cos(\varsigma_1) t] + \frac{1}{2}\ln(\nu_1^{\pm}),
		\quad
		\varpi^{\pm}_{\rm{I}}=  \frac{ q_0}{\kappa^2} [\sqrt{2} \kappa  \cos(\varsigma_1)  x - 2q_0 \cos(2 \varsigma_1) t] \mp \varsigma_1 \\
		&\eta^{\pm}_{\rm{I\!I}}= \frac{q_0 \sin(\varsigma_1)}{\kappa^2} [\sqrt{2} \kappa x + 4 q_0 \cos(\varsigma_1) t] - \frac{1}{2}\ln(\nu_1^{\pm}),
		\quad
		\varpi^{\pm}_{\rm{I\!I}}=  -\frac{ q_0}{\kappa^2} [\sqrt{2} \kappa  \cos(\varsigma_1)  x + 2 q_0 \cos(2 \varsigma_1) t] \mp \varsigma_1,  \\
		&\nu_1^{+} = \frac{\kappa^2 \cos^2(\varsigma_1)}
		{\kappa^4 - 2 \kappa^2 \cos(2 \varsigma_1) + 1 }
		\frac{\kappa^2 - 1}{2 |d_1|^2},
		\quad
		\nu_1^{-} =  \frac{ \kappa^4 + 2 \kappa^2 \cos(2 \varsigma_1) + 1 }
		{\cos^4(\varsigma_1) (\kappa^2 + 1)^2 }\nu_1^{+}, \\
		&\tau_1^{+} =  \frac{d^*_{1} }{|d_{1}|}\frac{\sqrt{\kappa^2-1} }{ \kappa} \frac{\mi e^{- 2 \mi \theta_{+}} \sin(\varsigma_1) (e^{2\mi \varsigma_1} - \kappa^2)}{ \sqrt{\kappa^4 - 2 \kappa^2 \cos(2\varsigma_1) + 1}}
		,
		\quad
		\tau_1^{-} =
		\frac{\sqrt{\kappa^4 + 2 \kappa^2 \cos(2\varsigma_1) + 1}}{\kappa^2 + e^{2 \mi \varsigma_1}}
		\tau_1^{+}.  \nonumber
	\end{aligned}
\end{equation}
It can be seen from Eqs.~\eref{xcx92aa} and~\eref{xcx92bb} that both the asymptotic expressions $q^{\pm}_{\rm{I}}$ and $q^{\pm}_{\rm{I\!I}}$ are non-singular, and they can
describe two propagating beating solitons with the same $t$-period $\frac{\pi \kappa^2}{q^2_0  \cos(2 \varsigma_1)}$ and same $x$-period $\frac{\sqrt{2} \pi \kappa}{q_0  \cos(\varsigma_1)}$ (see Fig.~\ref{f2aa}).

Besides, for the special case that $z_1, \hat{z}_1$ merges into an eigenvalue $\xi_1=\sqrt{2} q_0 e^{\mi \varsigma_1}$ on $\Gamma$,
Eq.~\eref{3156} says that $\displaystyle\lim_{\kappa \to 1} \frac{\kappa^2 - 1}{|d_1|^2} = 8 |c_1| \sin(\varsigma_1) \tan(\varsigma_1)$,
$\displaystyle\lim_{\kappa \to 1}\tau_1^{\pm} = 0$ and $\displaystyle\lim_{\kappa \to 1}\nu_1^{\pm} = |c_1| \cos^{\pm 1}(\varsigma_1)$,
where $c_1$ is the norming constant corresponding to $\xi_1$.
Hence,  $q^{\pm}_{\rm{I}}$ and $q^{\pm}_{\rm{I\!I}}$ can be reduced to
the asymptotic solitons~\eref{xcx92a} and~\eref{xcx92b} at $\kappa\to 1$.

\begin{figure}[H]
	\centering
	\subfigure[]{\label{f3d}
		\includegraphics[width=2.7in]{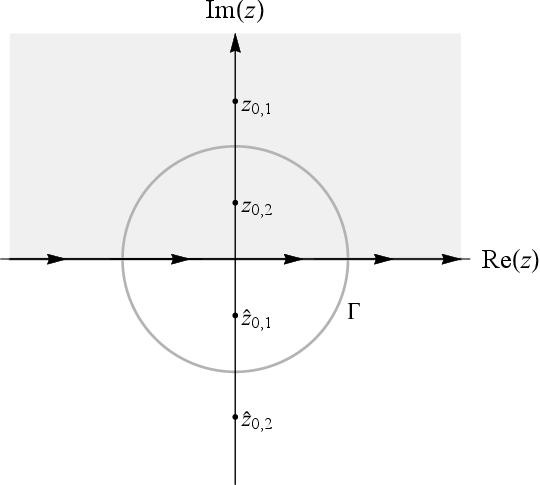}}\hfill
	\subfigure[]{\label{f2d}
		\includegraphics[width=2.7in]{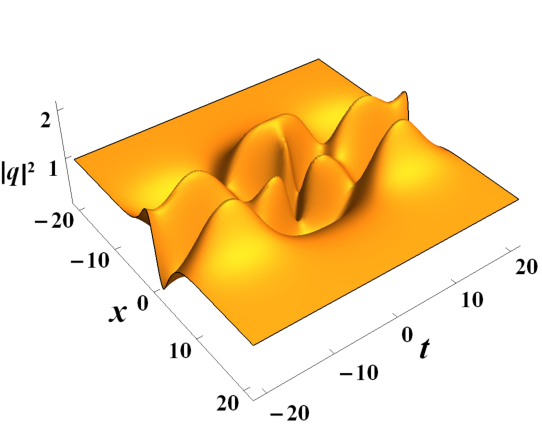}}\hfill
	\subfigure[]{\label{f3c}
		\includegraphics[width=2.7in]{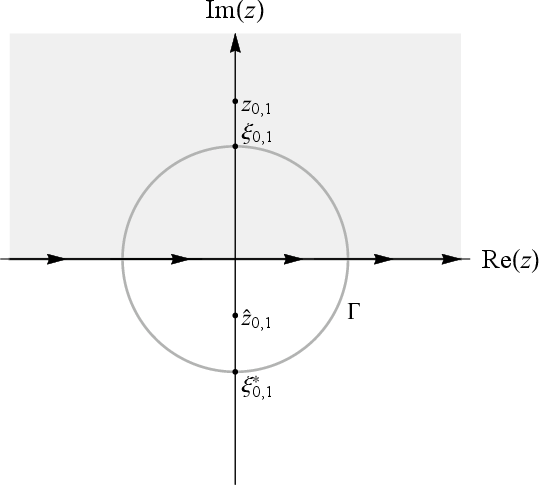}}\hfill
	\subfigure[]{\label{f2c}
		\includegraphics[width=2.7in]{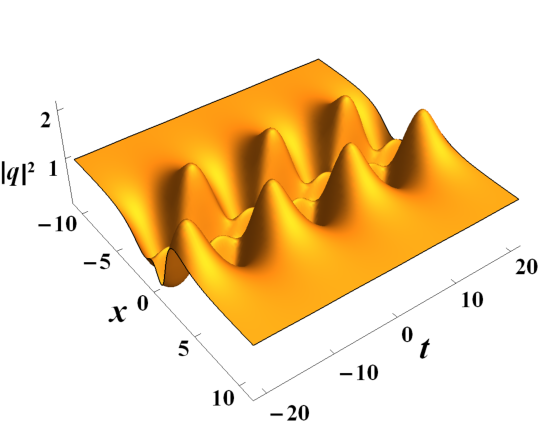}}\hfill
	\caption{\small
		(a) Distribution of discrete eigenvalues $\{z_{0,1}, z_{0,2}, \hat{z}_{0,1}, \hat{z}_{0,2}\}$;
		(b) Superposition of two non-propagating beating solitons with $q_+ = 1$, $z_{0,1} = 2 \sqrt{2} \mi$, $z_{0,2} = 3 \sqrt{2} \mi$, $d_{0,1} = \sqrt{\frac{6}{35}}$ and $d_{0,2} = \frac{\sqrt{3}}{5}$;
		(c) Distribution of discrete eigenvalues $\{\xi_{0,1}, \xi^*_{0,1}, z_{0,1},  \hat{z}_{0,1}\}$;
		(d) Superposition of non-propagating beating soliton and  dark soliton with $q_+ = 1$, $\xi_{0,1} = \sqrt{2} \mi$, $z_{0,1} = 2 \sqrt{2} \mi$, $c_{0,1} = 1$ and $d_{0,1} = \frac{\sqrt{2}}{5}$.
		\label{Fig46} }
\end{figure}

\textbf{Case 5.5} $N_1 + N_2 = 2$ ($m_{2}=2, \,n_{1,2}=m_{1}=0$ and $m_{1,2}=1,\, n_{1,2}=0$).
In both the two subcases, all the four discrete eigenvalues are  pure imaginary numbers.
(i) For the first subcase, the distribution of two pairs of eigenvalues $\{z_{0,1},  \hat{z}_{0,1}\}$ and $\{z_{0,2},
\hat{z}_{0,2}\}$ is shown in Fig.~\ref{f3d}. Because of their being off the circle $\Gamma$, the eigenvalue pairs lead to
two beating solitons which are both homoclinic in $x$ but have different periods in $t$. Such two beating solitons are non-propagating and are
linearly superposed around $x=O(1)$, as seen in Fig.~\ref{f2d}.
(ii) For the second subcase, there are one pair of imaginary eigenvalues
$\{z_{0,1}, \hat{z}_{0,1}\}$  off the circle $\Gamma$ and another pair of imaginary eigenvalues $\{\xi_{0,1}, \xi^*_{0,1}\}$ on $\Gamma$ (see Fig.~\ref{f3c}).
Fig.~\ref{f2c} shows the  superimposition of one  beating  soliton and one  dark soliton around $x=O(1)$.


\section{Conclusions}
\label{sec6}
For the defocusing NLS equation with local and nonlocal nonlinearities under NZBC, we have developed the theory of IST by using the Riemann-Hilbert method.
Similar to the Manakov system, we have introduced the adjoint Lax pair and auxiliary eigenfunctions for the direct scattering problem,
and studied the analyticity, symmetries and asymptotic behaviors of the Jost eigenfunctions, auxiliary Jost eigenfunctions and scattering coefficients. Then, we have regularized a RHP for the inverse problem,  analyzed the  residue conditions at the discrete  eigenvalues, and obtained the integral
representation for the solutions. It is worth noting that the Manakov system~\eref{Manakov} under NZBC admits two symmetries,
while Eq.~\eref{NNLS} under NZBC has one more symmetry due to the presence of the nonlocality. This additional symmetry leads to some stricter
constraints on the scattering coefficients, eigenfunctions and norming constants.
In particular, we have studied the properties of the Jost eigenfunctions and auxiliary Jost eigenfunctions in cases  when the discrete eigenvalues lie either on or off the circle,
and explicitly obtained the associated constraints on norming constants. We point that  it is also essential to introduce an adjoint Lax pair for the IST theory of the focusing Eq.~\eref{NNLS} under NZBC, and the analyticity must be examined in four domains, analogous to the focusing Manakov system studied in Ref.~\cite{DK}.

On the other hand, we have presented the N-soliton solutions in the determinant form for the reflectionless case. If $ \theta_{+} - \theta_{-} = \pi$ ($q_+ = -q_-$), the number of pure imaginary eigenvalues must be odd; whereas if $\theta_{+} - \theta_{-} = 0$ ($q_+ = q_-$) the number of pure imaginary eigenvalues must be even. Also, we have discussed the soliton dynamical behavior based on the distribution of discrete eigenvalues and the asymptotic behavior of solutions. It has been found that the discrete eigenvalue pairs lying on and off the circle $\Gamma$, respectively, lead to the dark and beating solitons. Moreover, our analysis has demonstrated  that the two-soliton solutions can  manifest diverse interaction patterns, including the interactions between two dark solitons or two beating solitons, as well as the superpositions of two beating solitons or one beating soliton and one dark soliton. This rich variety of soliton interactions stems from the possibility of different distribution of discrete eigenvalue pairs in the spectrum plane. In addition, we emphasize that the admissible solutions of Eq.~\eref{NNLSS1} are not necessarily even- or odd-symmetric with respect to $x$ (e.g., the beating one-soliton solution~\eref{Eq512}), despite the invariance of this equation under the transformation $q(x,t)\to q(-x,t)$.

\section*{Acknowledgments}

This work was supported by the National Natural Science Foundation of China (Grant Nos. 12475003 and 11705284), and by the Beijing Natural Science Foundation (Grant Nos. 1252016, 1232022 and 1212007).

\section*{Authors Contributions}
Chuanxin Xu: Conceptualization; Formal analysis; Software; Writing original draft. Tao Xu: Conceptualization; Formal analysis; Validation; Writing original draft; Funding acquisition. Min Li: Formal analysis; Validation; Writing review \& editing; Funding acquisition.


\begin{appendix}
\renewcommand{\thesection}{Appendix~\Alph{section}}
\renewcommand{\thesubsection}{\Alph{section}.\arabic{subsection}}

\section{}
\label{appendixA}
\renewcommand{\theequation}{A.\arabic{equation}}
\setcounter{equation}{0}

\subsection{Proof of Theorem~\ref{th1}}
\label{appendixA1}
In this part, the dependence on $t$ is omitted for brevity.
Based on the relation~\eref{s16}, Eq.~\eref{7} can be rewritten as
\begin{equation}
	\begin{aligned}
		\label{a1}
		&\bm{\mu}_{-}(x,z) = \mathbf{Y}_{-}(z) \bigg[\mathrm{I} + \int_{-\infty}^{x}  e^{\mi \mathbf{J}(z) (x-y)} \mathbf{Y}^{-1}_{-}(z) \bm{\Delta}\!\mathbf{Q}_{-}(y) \bm{\mu}_{-}(y,z) e^{\mi \mathbf{J}(z) (y-x)} dy \bigg].
	\end{aligned}
\end{equation}
By denoting the first column of $\mathbf{Y}^{-1}_{-}(z) \bm{\mu}_{-}(x,z)$ as $\chi(x, z)$, we have
\begin{equation}
	\begin{aligned}
		\label{a3}
		&\chi(x, z) =
		\left(
		\begin{array}{ccc}
			1 \\
			0 \\
			0 \\
		\end{array}
		\right)
	 + \int_{-\infty}^{x}  \mathbf{G}(x-y,z) \bm{\Delta}\!\mathbf{Q}_{-}(y) \mathbf{Y}_{-}(z) \chi(y,z) dy,
	\end{aligned}
\end{equation}
with
\begin{equation}
	\begin{aligned}
		\label{a4}
		&\mathbf{G}(x-y,z) = \diag\big(1,e^{\mi(\lambda - k) (x-y)}, e^{2\mi \lambda (x-y)}\big) \mathbf{Y}^{-1}_{-}(z).
	\end{aligned}
\end{equation}
Then, we represent $\chi(x, z)$ in the Neumann series:
\begin{equation}
	\begin{aligned}
		\label{a5}
		&\chi(x,z) =\sum_{n=0}^{\infty} \chi^{(n)}(x,z),
	\end{aligned}
\end{equation}
with
\begin{equation}
	\begin{aligned}
		&\chi^{(0)}(x,z) =
		\left(
		\begin{array}{ccc}
			1 \\
			0 \\
			0 \\
		\end{array}
		\right), \quad
		\chi^{(n+1)}(x,z) = \int_{-\infty}^{x}  \mathbf{\Upsilon}(x, y, z) \chi^{(n)}(y,z) dy,
	\end{aligned}
\end{equation}
where $\mathbf{\Upsilon}(x, y, z) = \mathbf{G}(x-y,z) \bm{\Delta}\!\mathbf{Q}_{-}(y) \mathbf{Y}_{-}(z)$.

Given the $L^{1}$ vector norm $||\chi(x,z)||=|\chi_{1}|+|\chi_{2}|+|\chi_{3}|$ and its associated subordinate matrix norm $||\mathbf{\Upsilon}(x, y, z)||$, we have
\begin{equation}
	\label{a6}
	\begin{aligned}
\parallel\chi^{(n+1)}(x,z)\parallel \leq \int_{-\infty}^{x}  \parallel \mathbf{\Upsilon}(x, y, z)\parallel \, \parallel  \chi^{(n)}(y,z)\parallel  dy.
	\end{aligned}
\end{equation}
Meanwhile, it is easy to know that $\parallel \mathbf{Y}_{-}(z)\parallel < 2  +  \frac{2 \sqrt{2} q_{0}}{|z|}  $ and $\parallel \mathbf{Y}^{-1}_{-}(z)\parallel  <
\big(\sqrt{2} +  \frac{2 q_{0} }{|z|}\big)/|\gamma(z)|  + \frac{\sqrt{2}}{2}$ with $\gamma(z)=2-\frac{4q^2_0}{z^2}$.
Then, we can obtain the inequality for $||\mathbf{\Upsilon}(x, y, z)||$:
\begin{align}
		\parallel \mathbf{\Upsilon}(x, y, z)\parallel  & \leq \parallel \diag(1,e^{\mi(\lambda - k) (x-y)}, e^{2\mi \lambda (x-y)})\parallel \, \parallel \mathbf{Y}^{-1}_{-}(z)\parallel  \,  \parallel  \bm{\Delta}\!\mathbf{Q}_{-}(y)\parallel  \, \parallel \mathbf{Y}_{-}(z)\parallel  \notag \\
		&\leq c(z) (1 + e^{-(\lambda_{\mathrm{I}} - k_{\mathrm{I}} ) (x-y)} + e^{-2 \lambda_{\mathrm{I}} (x-y)})  \parallel  \bm{\Delta}\!\mathbf{Q}_{-}(y)\parallel,
	\end{align}
where $\lambda_{\mathrm{I}} = \Im (\lambda)$, $k_{\mathrm{I}} = \Im (k)$, and $c(z) = \frac{\sqrt{2} (|z|^2 + \sqrt{2} q_0)^2}{|z^2-2q^2_0|} + \sqrt{2}  +  \frac{2 q_{0}}{|z|} $.  On the other hand, $c(z) \to \infty$ as $z \to  0$ or $z\to\pm\sqrt{2} q_{0}$.
Thus, for any given $\epsilon>0$,  we consider the domain $(C^{+})_{\epsilon} = C^{+} \setminus \big(B_{\epsilon}( 0) \cup  B_{\epsilon}( \sqrt{2} q_{0}) \cup B_{\epsilon}(-  \sqrt{2} q_{0})\big)$, where $B_{\epsilon}(z_{0}) = \{z \in \mathbb{C}: |z - z_{0}| < \epsilon q_{0} \}$. Obviously, one can check that $c_{\epsilon} = \displaystyle\max_{z \in (C^{+})_{\epsilon}} c(z) < \infty$.

Next, we use the inductive method to prove that the following inequality holds:
\begin{equation}
	\label{a8}
	\begin{aligned}
		&\parallel \chi^{(n)}(x,z)\parallel  \leq \frac{M^{n}(x)}{n!}, \quad z \in (C^{+})_{\epsilon}\, \text{and}\,\, n \in \mathbb{N},
	\end{aligned}
\end{equation}
where
\begin{equation}
	\begin{aligned}
		M(x) = 3 c_{\epsilon} \int_{-\infty}^{x}  \parallel  \bm{\Delta}\!\mathbf{Q}_{-}(y)\parallel   dy.
	\end{aligned}
\end{equation}
For $n=0$, Eq.~\eref{a8} is satisfied naturally. Observing that $\parallel \diag(1,e^{\mi(\lambda - k) (x-y)}, e^{2\mi \lambda (x-y)})\parallel \leq 3$ for $z\in (C^{+})_{\epsilon},\,y \leq x$ and
assuming that the inequality~\eref{a8} holds at $n = j$, we deduce  from Eq.~\eref{a6} that
\begin{equation}
\parallel \chi^{(j+1)}(x,z)\parallel  \leq \frac{3 c_{\epsilon}}{j!} \int_{-\infty}^{x}  \parallel  \bm{\Delta}\!\mathbf{Q}_{-}(y)\parallel   M^{j}(y) dy = \frac{ M^{j+1}(x)}{(j+1)!},
\end{equation}
which proves the induction step. Therefore, if $\parallel  \bm{\Delta}\!\mathbf{Q}_{-}(x)\parallel  \in L^1 (\mathbb{R})$,
the Neumann series converges absolutely for $z \in (C^{+})_{\epsilon}$, which immediately implies that $\mu_{-,1}$  is analytic for $z \in C^{+}$.

Similarly, one can prove the analyticity  of  $\mu_{+,3}(x, z)$ in $C^+$ and $\mu_{+,1}(x, z), \mu_{-,3}(x, z)$ in  $C^-$.

\subsection{Analyticity of scattering coefficients}

\label{appendixA2}
By virtue of Eqs.~\eref{Laxpairjvzhenxingshi}, one can obtain that
\begin{equation}
	\label{A11}
	\begin{aligned}
		&[\bm{\phi}^{-1} (x,t,z)_x = -\bm{\phi}^{-1}(x,t,z) (\mathbf{X}_{\pm}(z) + \bm{\Delta}\!\mathbf{Q}_{\pm}(x,t) ), \\
		&[\bm{\phi}^{-1} (x,t,z)]_t = -\bm{\phi}^{-1}(x,t,z) (\mathbf{T}_{\pm}(z) + \bm{\Delta}\!\mathbf{T}_{\pm}(x,t) ).
	\end{aligned}
\end{equation}	
Then, inserting the relation~\eref{s16} into Eq.~\eref{A11} gives rise to
\begin{equation}
	\label{A12}
	\begin{aligned}
		&[\bm{\mu}^{-1}_{\pm}(x,t,z) \mathbf{Y}_{\pm}(z)]_x + \mi [\bm{\mu}^{-1}_{\pm}(x,t,z) \mathbf{Y}_{\pm}(z), \mathbf{J}(z)] = - \bm{\mu}^{-1}_{\pm}(x,t,z) \bm{\Delta}\!\mathbf{Q}_{\pm}(x,t) \mathbf{Y}_{\pm}(z),\\
		&[\bm{\mu}^{-1}_{\pm}(x,t,z) \mathbf{Y}_{\pm}(z)]_t + \mi [\bm{\mu}^{-1}_{\pm}(x,t,z) \mathbf{Y}_{\pm}(z), \mathbf{\Omega}(z)] = - \bm{\mu}^{-1}_{\pm}(x,t,z) \bm{\Delta}\!\mathbf{T}_{\pm}(x,t) \mathbf{Y}_{\pm}(z).
	\end{aligned}
\end{equation}	
Based on Eq.~\eref{A12},  we can obtain the total differential form
\begin{align}
	&	-d \big(  \mathbf{E}^{-1}(x,t,z) \bm{\mu}^{-1}_{\pm}(x,t,z) \mathbf{Y}_{\pm}(z) \mathbf{E}(x,t,z) \big) \notag \\
	& \hspace{2cm} =
	\big(\mathbf{E}^{-1}(x,t,z) \bm{\mu}^{-1}_{\pm}(x,t,z) \bm{\Delta}\!\mathbf{Q}_{\pm}(x,t) \mathbf{Y}_{\pm}(z) \mathbf{E}(x,t,z)  \big) dx \notag \\
	& \hspace{2.4cm} 	+\big( \mathbf{E}^{-1}(x,t,z)  \bm{\mu}^{-1}_{\pm}(x,t,z) \bm{\Delta}\!\mathbf{T}_{\pm}(x,t) \mathbf{Y}_{\pm}(z) \mathbf{E}(x,t,z)  \big) dt.
\end{align}
Thus, choosing the integration paths $(x, t) \to (\infty, t)$ and $(-\infty, t) \to (x, t)$, $\bm{\mu}^{-1}_{\pm}(x,t,z)$ can be represented as
\begin{equation}
	\begin{aligned}
	&\bm{\mu}^{-1}_{+}(x,t,z) = \mathbf{Y}^{-1}_{+}(z) + \int_{x}^{\infty}  e^{\mi \mathbf{J}(z) (x-y)} \bm{\mu}^{-1}_{+}(y,t,z) \bm{\Delta}\!\mathbf{Q}_{+}(y,t) \mathbf{Y}^{}_{+}(z)  \bm{\mu}^{-1}_{+}(y,t,z)  e^{-\mi \mathbf{J}(z) (x-y)} \mathbf{Y}^{-1}_{+}(z) dy, \notag\\
	&\bm{\mu}^{-1}_{-}(x,t,z) = \mathbf{Y}^{-1}_{-}(z) - \int_{-\infty}^{x}   e^{\mi \mathbf{J}(z) (x-y)} \bm{\mu}^{-1}_{-}(y,t,z) \bm{\Delta}\!\mathbf{Q}_{-}(y,t) \mathbf{Y}^{}_{-}(z)  \bm{\mu}^{-1}_{-}(y,t,z)  e^{-\mi \mathbf{J}(z) (x-y)} \mathbf{Y}^{-1}_{-}(z) dy. \notag
	\end{aligned}
\end{equation}

Using the same way in Appendix~\ref{appendixA1}, it can be shown that $[\bm{\mu}^{-1}_{+}]^{1}$ and $[\bm{\mu}^{-1}_{-}]^{3}$ are analytic in the region $C^{-}$, and that $[\bm{\mu}^{-1}_{-}]^{1}$ and $[\bm{\mu}^{-1}_{+}]^{3}$ are analytic in $C^{+}$, where $[\bm{\mu}^{-1}_{\pm}]^{j}$ represent the $j$th rows of  $\bm{\mu}^{-1}_{\pm}$. However,  $[\bm{\mu}^{-1}_{\pm}]^{2}$ lack the analyticity in any region of the complex $z$-plane. Based on the relation~\eref{s16}, the corresponding rows of $\bm{\phi}^{-1}_{\pm}$ have the same analyticity.
Then, by the relations $\mathbf{A}(z) = \bm{\phi}^{-1}_{+}  \bm{\phi}_{-} $ and $\mathbf{B}(z) = \bm{\phi}^{-1}_{-}  \bm{\phi}_{+}$, Theorem~\ref{th2} can be finally proved.

	
	
\subsection{Adjoint problem}
\label{appendixA4}
In this part,  we suppress the $x$-, $t$- and $z$-dependence in the involved functions.

\textbf{Proof of Proposition~\ref{pro1}}:
For  two arbitrary column vectors $u, v \in \mathbb{C}^{3}$, there are the following identities:
\begin{subequations}
	\begin{align}
		&(\bm{\sigma}_{3} u)\times (\bm{\sigma}_{3} v)   + \bm{\sigma}_{3} (u\times v )  = \mathbf{0} , \label{3a1}\\
		&(\bm{\sigma}_{3} u)\times v   + u\times (\bm{\sigma}_{3}v)  - u\times v   - (\bm{\sigma}_{3} u)\times (\bm{\sigma}_{3} v) = \mathbf{0}  , \label{3a2}\\
		&(\mathbf{Q}^* u)\times v   + u \times (\mathbf{Q}^*v)  - \bm{\sigma}_{3} \mathbf{Q} \bm{\sigma}_{3} (u\times v )  = \mathbf{0}, \label{3a3} \\
		&(\bm{\sigma}_{3} \mathbf{Q}^*_x u)\times  u   + u\times (\bm{\sigma}_{3} \mathbf{Q}^*_xv)  =- \mathbf{Q}_x \bm{\sigma}_{3} (u\times v ) ,\label{3a4}  \\
		&(\bm{\sigma}_{3} (\mathbf{Q}^*)^2 u)\times  u   + u\times (\bm{\sigma}_{3} (\mathbf{Q}^*)^2 v)  =-\mathbf{Q}^2 \bm{\sigma}_{3} (u\times v ), \label{3a5} \\
		&2 \mi q^2_0 (u\times v ) + (\mathbf{V}_0 u) \times v + u \times (\mathbf{V}_0 v ) = - \bm{\sigma}_{3} \mathbf{V}_0 \bm{\sigma}_{3} (u\times v ) \label{3a6},
	\end{align}
\end{subequations}
where '$\times$' represents the usual cross product. These identities  will be used in the following  process.
\begin{subequations}
	\label{3a7}
	\begin{align}
	&	r_x = e^{\mi \theta_{2}}  \bm{\sigma}_{3} \big\{- \mi k \big[u \times v - (\bm{\sigma}_{3} u) \times v - u \times (\bm{\sigma}_{3} v) \big]  +(\mathbf{Q}^*u) \times v + u \times (\mathbf{Q}^* v) \big\}, \label{3a71} \\
& 	r_t	=  2 k r_x + e^{\mi \theta_{2}}  \bm{\sigma}_{3} \big\{ \big[2 \mi q^2_0 (u\times v ) +(\mathbf{V}_0 u ) \times v + u \times (\mathbf{V}_0 v ) \big] - \mi \big[ ( \bm{\sigma}_{3} \mathbf{Q}^*_x u ) \times v + u \times ( \bm{\sigma}_{3} \mathbf{Q}^*_x v ) \big] \nonumber \\
		& \hspace{2.85cm} +\mi \big[(\bm{\sigma}_{3}(\mathbf{Q}^*)^2 u) \times v + u \times (\bm{\sigma}_{3}(\mathbf{Q}^*)^2 v ) \big] \big\}. \label{3a72}
	\end{align}
\end{subequations}
Applying the relations~\eref{3a1}-\eref{3a3} to Eq.~\eref{3a71} and the relations~\eref{3a4}-\eref{3a6} to Eq.~\eref{3a72}, we can obtain
\begin{subequations}
	\begin{align}
		r_x  & = e^{\mi \theta_{2}}  \bm{\sigma}_{3} \big[- \mi k   \bm{\sigma}_{3}(u\times v )  + \bm{\sigma}_{3}\mathbf{Q}\bm{\sigma}_{3}(u\times v ) \big] = (- \mi k \bm{\sigma}_{3}  + \mathbf{Q} ) r, \\
		r_t	& = 2 k \mathbf{X}  r + e^{\mi \theta_{2}}  \bm{\sigma}_{3} \big( - \bm{\sigma}_{3}\mathbf{V}_0\bm{\sigma}_{3} (u\times v ) + \mi \mathbf{Q}_x \bm{\sigma}_{3} (u\times v )  - \mi \mathbf{Q}^2 \bm{\sigma}_{3}(u\times v )   \big) \notag \\
		& = \big(2 k \mathbf{X}  + \mi \bm{\sigma}_{3}\mathbf{Q}_x  - \mi \bm{\sigma}_{3}\mathbf{Q}^2   -\mathbf{V}_0 \big)  r,
	\end{align}	\label{3a8}
\end{subequations}
which indicates that the function $r$ satisfies Lax pair~\eref{Laxpairjvzhenxingshi}.

\textbf{Proof of Lemma~\ref{th2.7}}:
To illustrate, we only verify Eq.~\eref{xcx42a} at $j=2$ (which means that $l=3, m=1$) since the other cases can be proved similarly.  Based on Eqs.~\eref{4} and~\eref{sm40}, we know that the functions $v_{\pm} = e^{\mi \theta_{2}} \bm{\sigma}_{3} \big(\tilde{\phi}_{\pm,3} \times \tilde{\phi}_{\pm,1} \big) $ satisfy the Lax pair~\eref{Laxpairjvzhenxingshi} and exhibit the asymptotic behavior
\begin{equation}
	\label{xcx17a}
		v_{\pm} \sim \gamma e^{\mi \theta_{2}}  \mathbf{Y}_{\pm,3} + o(1), \quad x \to \pm \infty.
\end{equation}
On the other hand, $v_{\pm}$ admits a linear combination of the columns of $\phi_{\pm}$, i.e.,
\begin{equation}
	\label{xcxA17}
	\begin{aligned}
		&v_{\pm} = a_{\pm}\phi_{\pm,1} + b_{\pm} \phi_{\pm,2} + c_{\pm} \phi_{\pm,3}.
	\end{aligned}
\end{equation}
Then, by requiring  Eq.~\eref{xcxA17} satisfy the asymptotic behavior in Eq.~\eref{xcx17a}, it can be determined that  $b_{\pm} = \gamma$ and $a_{\pm} = c_{\pm} = 0$.

\textbf{Proof of Lemma~\ref{co2.8}}:
Substituting Eq.~\eref{xcx42b} into Eq.~\eref{xcx38} yields
\begin{equation}
	\begin{aligned}
&		(\phi_{-,2} \times \phi_{-,3}, \,\phi_{-,3} \times\phi_{-,1}, \, \phi_{-,1} \times \phi_{-,2}) \bm{\Xi}^{-1}(z) \notag \\
& \quad\quad  = (\phi_{+,2} \times \phi_{+,3}, \,\phi_{+,3} \times\phi_{+,1}, \, \phi_{+,1} \times \phi_{+,2}) \bm{\Xi}^{-1}(z) \tilde{\mathbf{A}}(z). 		 \label{sm46}
	\end{aligned}
\end{equation}
Note that the following relation holds
\begin{equation}
	\label{ap42}
	\begin{aligned}
		(\mathbf{A}^T(z))^{-1} = \frac{1}{\det(\mathbf{A}(z))}(A_2 \times A_3, \,A_3 \times A_1, \,A_1 \times A_2),
	\end{aligned}
\end{equation}
where $A_{i}$ is the $i$th column of  $\mathbf{A}(z)$. Then, applying this relation to Eq.~\eref{sm46} gives rise to
\begin{equation}
\det (\bm{\phi}_{-}) (\bm{\phi}^T_{-})^{-1} \bm{\Xi}^{-1}(z)  = \det(\bm{\phi}_{+}) (\bm{\phi}^T_{+})^{-1} \bm{\Xi}^{-1}(z) \tilde{\mathbf{A}}(z),
\end{equation}
which, by virtue of Eq.~\eref{sm1}, can be converted into
\begin{align}
	\bm{\Xi}(z) (\mathbf{A}^{-1}(z))^T \bm{\Xi}^{-1}(z)  =   \tilde{\mathbf{A}}(z).
\end{align}


\textbf{Proof of Proposition~\ref{co2.9}}:
Based on Eq.~\eref{xcx38}, the functions $w$ and $\overline{w}$ in Eq.~\eref{xcx41} can be written as
\begin{equation}
	\begin{aligned}
		w
		&= e^{\mi \theta_{2}} \bm{\sigma}_{3} \big[  \tilde{a}_{23}(z) \tilde{\phi}_{+,2} \times \tilde{\phi}_{+,1}  + \tilde{a}_{33}(z) \tilde{\phi}_{+,3}   \times \tilde{\phi}_{+,1} \big] /\gamma(z), \\
	    \overline{w}
	    &= e^{\mi \theta_{2}} \bm{\sigma}_{3} \big[  \tilde{a}_{11}(z) \tilde{\phi}_{+,1} \times \tilde{\phi}_{+,3}  + \tilde{a}_{21}(z) \tilde{\phi}_{+,2}   \times \tilde{\phi}_{+,3} \big] /\gamma(z).
	\end{aligned}
\end{equation}
Then,  replacing the cross products in the above equations by virtue of Eq.~\eref{xcx42} yields
\begin{equation}
	\begin{aligned}
		w
		&=\big[ \tilde{a}_{23}(z) \phi_{+,3} + \tilde{a}_{33}(z) \gamma(z) \phi_{+,2} \big]/\gamma(z), \\
	    \overline{w}
	    &=\big[ - \tilde{a}_{11}(z) \gamma(z)  \phi_{+,2} + \tilde{a}_{21}(z) \phi_{+,1} \big]/\gamma(z). \\
	\end{aligned}
\end{equation}
Meanwhile, noticing from Lemma~\ref{co2.8} that $\tilde{a}_{32}(z)= - \gamma(z) b_{32}(z)$, $\tilde{a}_{33}(z)= b_{33}(z)$, $\tilde{a}_{21}(z)= \gamma(z) b_{12}(z)$, $\tilde{a}_{11}(z)= b_{11}(z)$, thus we  obtain
\begin{equation}
	\begin{aligned}
			w &=  - b_{32}(z)  \phi_{+,3} + b_{33}(z) \phi_{+,2},  \\
	        \overline{w}&=  - b_{11}(z) \phi_{+,2}  + b_{12}(z) \phi_{+,1},
		\end{aligned}
\end{equation}
which indicates that Eq.~\eref{sm48a} holds true. Similarly, one can prove the validity of Eq.~\eref{sm48b} by combining Eqs.~\eref{xcx41} and $\tilde{\bm{\phi}}_{+} = \tilde{\bm{\phi}}_{-}  \tilde{\mathbf{B}}(z)$.

\subsection{Symmetry}
\label{appendixA5}


\textbf{Proof of Lemma~\ref{le2.12}}:
In this part, the $x$- and $t$-dependence is suppressed.
Taking the derivatives of $\bm{\sigma}_{3} (\bm{\phi}^{\dagger}(z^*))^{-1}$, respectively, with respect to $x$ and $t$ yields
\begin{subequations}
	\begin{align}
		\big[\bm{\sigma}_{3} (\bm{\phi}^{\dagger}(z^*))^{-1}\big]_x
		&= -\bm{\sigma}_{3}(\bm{\phi}^{\dagger}(z^*))^{-1} \bm{\phi}^{\dagger}_{x}(z^*) (\bm{\phi}^{\dagger}(z^*))^{-1}  \nonumber \\
		&= -\bm{\sigma}_{3}(\bm{\phi}^{\dagger}(z^*))^{-1} \bm{\phi}^{\dagger}(z^*) (\mi k \bm{\sigma}_{3} + \mathbf{Q}) (\bm{\phi}^{\dagger}(z^*))^{-1}   \nonumber
\\
		&=  (- \mi k \bm{\sigma}_{3} + \mathbf{Q}) \big[\bm{\sigma}_{3} (\bm{\phi}^{\dagger}(z^*))^{-1}\big],  \\
		\big[\bm{\sigma}_{3} (\bm{\phi}^{\dagger}(z^*))^{-1}\big]_t
		&= -\bm{\sigma}_{3}(\bm{\phi}^{\dagger}(z^*))^{-1} \bm{\phi}^{\dagger}_{t}(z^*) (\bm{\phi}^{\dagger}(z^*))^{-1}  \nonumber \\
		&= - \bm{\sigma}_{3} (2 k \mathbf{X}^{\dagger}(z)  - \mi  \mathbf{Q}_x \bm{\sigma}_{3}- \mi {\mathbf{Q}}^2 \bm{\sigma}_{3} + \mathbf{V_0} ) (\bm{\phi}^{\dagger}(z^*))^{-1}  \nonumber \\
		&= (2 k \mathbf{X} + \mi  \bm{\sigma}_{3} \mathbf{Q}_x + \mi \bm{\sigma}_{3} {\mathbf{Q}}^2  - \mathbf{V_0} )\big[\bm{\sigma}_{3} (\bm{\phi}^{\dagger}(z^*))^{-1}\big],
	\end{align}
\end{subequations}
which imply that $\bm{\sigma}_{3} (\bm{\phi}^{\dagger}(z^*))^{-1}$ satisfies Lax pair~\eref{Laxpairjvzhenxingshi}. Then, based on Eq.~\eref{4},
we obtain that
\begin{equation}
	\label{ap51}
\bm{\sigma}_{3} (\bm{\phi}_{\pm}^{\dagger}(z^*))^{-1} \sim \bm{\sigma}_{3} (\mathbf{Y}^{\dagger}_{\pm}(z^*))^{-1} \mathbf{E}(z) + o(1), \quad x \to \pm \infty.
\end{equation}
From the asymptotic behaviors of $\bm{\phi}(z)$ and $\bm{\sigma}_{3} (\bm{\phi}^{\dagger}(z^*))^{-1}$ in Eqs.~\eref{4} and~\eref{ap51},  one can determine the matrix $\mathbf{C}(z)$ as given in Eq.~\eref{xcx49}.


\textbf{Proof of Lemma~\ref{le2.13}}:
Conjugating on both sides of Eq.~\eref{xcx49} and letting $z\to z^*$, one can obtain that
\begin{equation}
	\label{ap52}
	\begin{aligned}
		\bm{\phi}^*_{\pm}(z^*) = \frac{1}{\gamma(z) e^{ \mi \theta_{2}(z)} } \bm{\sigma}_{3}\big(\phi_{\pm,2}(z) \times\phi_{\pm,3}(z), \,\phi_{\pm,3}(z) \times \phi_{\pm,1}(z), \,\phi_{\pm,1}(z) \times \phi_{\pm,2}(z)\big) \mathbf{C}(z),
	\end{aligned}
\end{equation}
where the $x$- and $t$-dependence is suppressed, and  the identities~\eref{ap42} and $\mathbf{C}^*(z^*) = \mathbf{C}(z)$ are utilized.
Then,  replacing $\phi_{+,2}$ for the first column in Eq.~\eref{ap52} by virtue of Eqs.~\eref{sm48} yields
\begin{equation}
	\label{ap53}
	\begin{aligned}
		\phi^*_{+, 1}(z^*) &=  \bm{\sigma}_{3} \Big[\frac{1}{b_{33}(z)} \big( b_{32}(z) \phi_{+,3}(z)  + w(z) \big) \times \phi_{+,3}(z) \Big](z) e^{-\mi \theta_{2}(z)}, \\
		                   &=  \frac{1}{b_{33}(z)} \bm{\sigma}_{3} [w(z) \times \phi_{+,3}](z) e^{-\mi \theta_{2}(z)}. \\
	\end{aligned}
\end{equation}
Repeating the above process, we can prove Eqs.~\eref{xcx49b},~\eref{xcx49c} and~\eref{xcx49d}.

\textbf{Proof of Lemma~\ref{le2.14}}: For Eq.~\eref{xcx49} with the case `$-$',  we make the replacement $\bm{\phi}_{-}\to \bm{\phi}_{+}\mathbf{A}(z)$ by Eq.~\eref{sm1}
and derive that
\begin{equation}
	\label{xcx491a}
	\begin{aligned}
		&\bm{\phi}_{+}(x,t,z) \mathbf{A}(z) = \bm{\sigma}_{3} [\mathbf{B}(z^*) \phi^{-1}_{+}(x,t,z^*)]^{\dagger} \mathbf{C}(z). \\
	\end{aligned}
\end{equation}
Then, applying Eq.~\eref{xcx49} to $\phi_{+}(x,t,z)$ in the left-hand side of Eq.~\eref{xcx491a} gives rise to
\begin{equation}
	\begin{aligned}
		&\bm{\sigma}_{3} (\phi^{-1}_{+}(x,t,z^*))^{\dagger} \mathbf{C}(z) \mathbf{A}(z) = \bm{\sigma}_{3} (\phi^{-1}_{+}(x,t,z^*))^{\dagger}    \mathbf{B}^{\dagger}(z^*) \mathbf{C}(z),
	\end{aligned}
\end{equation}
which immediately confirms the relation~\eref{xcx50}.

\textbf{Proof of Lemma~\ref{le2.17}}:
By using the relations $k(-z) = - k(z)$, $\bm{\delta}_1 \bm{\sigma}_{3} = \bm{\sigma}_{3}\bm{\delta}_1$ and $\mathbf{Q}(-x, t)= - \bm{\delta}_1 \mathbf{Q}(x, t) \bm{\delta}_1$, we take the derivatives of $\bm{\delta}_1 \bm{\phi}(-x,t,-z)$ with respect to $x$, yielding
	\begin{align}
		\big[\bm{\delta}_1 \bm{\phi}(-x,t,-z)\big]_x
		&= - \bm{\delta}_1 [-\mi (-k) \bm{\sigma}_{3} + \mathbf{Q}(-x, t)] \bm{\phi}(-x,t,-z)  \notag \\
		&= - [-\mi (-k) \bm{\delta}_1 \bm{\sigma}_{3} - \bm{\delta}^2_1  \mathbf{Q}(x, t) \bm{\delta}_1] \bm{\phi}(-x,t,-z) \notag \\
		&=  [-\mi k \bm{\sigma}_{3} +  \mathbf{Q}(x, t) ] \bm{\delta}_1 \bm{\phi}(-x,t,-z).
	\end{align}
For the $t$-part of Lax pair,  based on the relations $\frac{d\mathbf{Q}(-x, t)}{dx} =   \bm{\delta}_{1} \mathbf{Q}_x(x, t) \bm{\delta}_{1}$ and $ \mathbf{Q}^2(-x, t) =  \bm{\delta}_{1} \mathbf{Q}^2(x, t) \bm{\delta}_{1}$, one can derive that
\begin{align}
		\big[\bm{\delta}_1 \bm{\phi}(-x,t,-z) \big]_t
		&= \bm{\delta}_1 \Big[- 2 \mi  k^2 \bm{\sigma}_{3} - 2 k \mathbf{Q}(-x, t)+  \mi \bm{\sigma}_{3} \frac{d\mathbf{Q}(-x, t)}{dx} - \mi \bm{\sigma}_{3} {\mathbf{Q}}^2(-x, t) -\mathbf{V_0} \Big] \bm{\phi}(-x,t,-z)  \notag \\
		&= [- 2 \mi  k^2 \bm{\sigma}_{3} + 2 k \mathbf{Q}(x, t) + \mi \bm{\sigma}_{3} \mathbf{Q}_x(x, t) - \mi \bm{\sigma}_{3} {\mathbf{Q}}^2(x, t) - \mathbf{V_0} ] \bm{\delta}_1 \bm{\phi}(-x,t,-z) .
	\end{align}
Hence, we acquire that $\bm{\delta}_1 \bm{\phi}(-x,t,-z)$ solves Lax pair~\eref{Laxpairjvzhenxingshi}. Besides, it can be checked that $\bm{\phi}(x,t,\hat{z})$ also satisfies Lax pair~\eref{Laxpairjvzhenxingshi} since $k(z)=k(\hat{z})$.

Furthermore, it follows from Eq.~\eref{4} that $\bm{\delta}_1  \bm{\phi}_{\pm}(-x,t,-z)$ and $\bm{\phi}_{\pm}(x,t,\hat{z})$ have the following asymptotic behaviors:
\begin{equation}
	\begin{aligned}
		&\bm{\delta}_1  \bm{\phi}_{\pm}(-x,t,-z) \sim
		\bm{\delta}_1 \mathbf{Y}(-x,t,-z) \mathbf{E}(-x,t,-z)  +  o(1), \quad x \to \pm \infty, \\
		&\bm{\phi}_{\pm}(x,t,\hat{z}) \sim
		\mathbf{Y}(x,t,\hat{z}) \mathbf{E}(x,t,\hat{z})  +  o(1), \quad x \to \pm \infty.
	\end{aligned}
\end{equation}
By comparing the asymptotic behaviors of $\bm{\phi}(x,t,z)$, $\bm{\sigma}_{3} (\bm{\phi}^{\dagger}(x,t,z^*))^{-1}$ and
$\bm{\phi}_{\pm}(x,t,\hat{z})$ as $x\to \pm \infty$,
we can obtain the matrices $\bm{\sigma}_{4}$ and $\bm{\Pi}(z)$ as given in Eq.~\eref{eq246}.

\textbf{Proof of Lemma~\ref{le2.18}}:
Based on the symmetry relations  in Eqs.~\eref{xcx56} and~\eref{xcx62}, we can write Eq.~\eref{sm1} as
\begin{subequations}
\begin{align}
& \bm{\delta}_1  \bm{\phi}_{+}(-x,t,-z) \bm{\sigma}_4 = \bm{\delta}_1  \bm{\phi}_{-}(-x,t,-z) \bm{\sigma}_4  \mathbf{A}(z), \\
& \bm{\phi}_{-}(x,t,\hat{z}) \bm{\Pi}(z) =  \bm{\phi}_{+}(x,t,\hat{z}) \bm{\Pi}(z) \mathbf{A}(z),
\end{align}
\end{subequations}
which mean that
\begin{subequations}
	\begin{align}
& \mathbf{A}(z) = \bm{\sigma}^{-1}_{4} \bm{\phi}^{-1}_{-}(-x,t,-z) \bm{\phi}_{+}(-x,t,-z) \bm{\sigma}_4 = \bm{\sigma}_{4} B(-z) \bm{\sigma}_4, \\
& \mathbf{A}(z) = \bm{\Pi}(z)^{-1} \bm{\phi}^{-1}_{+}(x,t,\hat{z}) \bm{\phi}_{-}(x,t,\hat{z}) \bm{\Pi}(z) = \bm{\Pi}(z)^{-1} \mathbf{A}(\hat{z}) \bm{\Pi}(z).
\end{align}
\end{subequations}

\textbf{Proof of Lemma~\ref{le2.19}}:
By virtue of Eq.~\eref{xcx56}, we replace $\phi^*_{-,3}(x,t,z^*)$ and $\phi^*_{+,1}(x,t,z^*)$ in Eq.~\eref{xcx54} and obtain that
\begin{align}
		w(x, t, z) &= e^{\mi \theta_{2}(-x,t,-z)} \bm{\sigma}_{3} \big[ (- \bm{\delta}_{1}  \phi^*_{+,3}(-x,t,-z^*) )  \times ( \bm{\delta}_{1}  \phi^*_{-,1}(-x,t,-z^*)  ) \big] /\gamma(-z)  \notag \\
		         &= - \bm{\delta}_{1}  e^{\mi \theta_{2}(-x,t,-z)} \bm{\sigma}_{3} \big[  \phi^*_{+,3}(-x,t,-z^*)   \times    \phi^*_{-,1}(-x,t,-z^*)   \big] /\gamma(-z)  \notag \\
		         &=  \bm{\delta}_{1}  \overline{w}(-x,t,-z),
\end{align}
where we have used the relations
\begin{align}
& \theta_{2}(x,t,z) = \theta_{2}(-x,t,-z), \quad \gamma(z) = \gamma(-z), \notag \\
& (\bm{\delta}_{1} \phi^*_{+,3}(-x,t,-z^*) )  \times (\bm{\delta}_{1} \phi^*_{-,1}(-x,t,-z^*) ) = \bm{\delta}_{1} ( \phi^*_{+,3}(-x,t,-z^*) \times  \phi^*_{-,1}(-x,t,-z^*) ). \notag
\end{align}
Similarly, based on the relations $\theta_{2}(x,t,\hat{z}) = \theta_{2}(x,t,z)$ and $\big(\frac{\mi \sqrt{2} q_{0} }{z} \big)^2 \gamma(\hat{z}) = \gamma(z) $, replacing $\phi^*_{-,3}(x,t,z^*)$ and $\phi^*_{+,1}(x,t,z^*)$ in Eq.~\eref{xcx54} by  Eq.~\eref{xcx62} yields
\begin{align}
	w(x, t, z)
			& = e^{\mi \theta_{2}(x,t,\hat{z})} \bm{\sigma}_{3} \Big[\Big(- \mi \frac{\sqrt{2} q_{0} }{z}  \phi^*_{-,1}(x,t,\hat{z}^* ) \Big)  \times \Big(\mi \frac{\sqrt{2} q_{0} }{z}   \phi^*_{+,3}(x,t,\hat{z}^*) \Big)  \Big] /\gamma(z) \notag \\
			& = - \Big( \frac{\mi \sqrt{2} q_{0} }{z} \frac{\mi \sqrt{2} q_{0} }{z} \gamma(\hat{z}) /\gamma(z) \Big) e^{\mi \theta_{2}(x,t,\hat{z})} \bm{\sigma}_{3} \big[\phi^*_{-,1}(x,t,\hat{z}^*)  \times \phi^*_{+,3}(x,t,\hat{z}^*)   \big] /\gamma(\hat{z}) \notag \\
			& = - \overline{w}(x,t,\hat{z}).
\end{align}

%


\section{}
\label{appendixB}
\renewcommand{\theequation}{B.\arabic{equation}}
\setcounter{equation}{0}

\subsection{Discrete spectrum}
\label{appendixA8}

\textbf{Proof of Eq.~\eref{xcx73}}:  From Eq.~\eref{sm48a}, the auxiliary eigenfunctions $w $ and $\overline{w}$ can be represented as
\begin{equation}
	\begin{aligned}
		&w= b_{33}(z)  \phi_{+,2} - b_{32}(z) \phi_{+,3}, \quad
		\overline{w} = b_{12}(z) \phi_{+,1}  - b_{11}(z)\phi_{+,2}.
	\end{aligned}
\end{equation}
Substituting the above equation into $\det (\mathbf{\Phi}^\pm) $ gives
\begin{equation}
	\begin{aligned}
		\det ( \mathbf{\Phi}^+) & =\det \big(\phi_{-,1}, b_{33}(z)  \phi_{+,2} - b_{32}(z) \phi_{+,3},\phi_{+,3}\big) \\
		&= \det\big(\phi_{-,1}, b_{33}(z) \phi_{+,2},\phi_{+,3}\big),\\
		\det ( \mathbf{\Phi}^-) & =\det \big(\phi_{+,1}, - b_{12}(z) \phi_{+,1}  + b_{11}(z) \phi_{+,2},\phi_{-,3}\big)\\
		&= \det \big(\phi_{+,1},  b_{11}(z)\phi_{+,2},\phi_{-,3}\big).
	\end{aligned}
\end{equation}
Then, by virtue of Eq.~\eref{sm1}, the above two determinants  can be simplified as
\begin{equation}
	\begin{aligned}
		\det ( \mathbf{\Phi}^+)	                   & = \det \big( a_{11}(z) \phi_{+,1}, b_{33}(z)\phi_{+,2},\phi_{+,3}\big)
= a_{11}(z) b_{33}(z) \gamma(z)  e^{\mi\theta_{2}(z)}, \\
		\det ( \mathbf{\Phi}^-)	                   & =  \det \big(\phi_{+,1},  b_{11}(z)\phi_{+,2},a_{33}(z) \phi_{+,3}\big) =  a_{33}(z) b_{11}(z)  \gamma(z) e^{\mi\theta_{2}(z)}.
	\end{aligned}
\end{equation}

\textbf{Proof of Lemma~\ref{le3.3}}: (i) Suppose that $\xi_i$ is a simple zero of $a_{11}(z)$ and lies on the circle $\Gamma$. Thus, it follows from
Eq.~\eref{xcx73} that $\det (\mathbf{\Phi}^{+}(x,t,z)) = 0$, which means the linear dependence among $\phi_{-,1}(x,t,\xi_i)$, $w(x,t,\xi_i)$ and $\phi_{+,3}(x,t,\xi_i)$.
Note that the boundary conditions~\eref{4} imply that $\phi_{+,3}(x,t,\xi_i) , \phi_{-,1}(x,t,\xi_i) \neq \mathbf{0}$, so there exist two constants $\alpha_1$ and $\alpha_2$ such that
\begin{equation}
\begin{aligned}
	w(x,t,\xi_i) & = \alpha_1 \phi_{-,1}(x,t,\xi_i) + \alpha_2 \phi_{+,3}(x,t,\xi_i).  \\
\end{aligned}
\end{equation}
When $w(x,t,\xi_i) \neq \mathbf{0}$, Eqs.~\eref{xcx54} imply the linear independence between $\phi_{-,1}(x,t,\xi_i)$ and $\phi_{+,3}(x,t,\xi_i)$.
Meanwhile, Lemma~\ref{le2.18} shows that $\xi_i$ is also a simple zero of $b_{33}$. Then, Eq.~\eref{xcx49a} shows that $w(x,t,\xi_i)$ and
$\phi_{+,3}(x,t,\xi_i)$  are linearly dependent. Therefore, we conclude that $\alpha_1 = 0$. Again, repeating the same process by Eq.~\eref{xcx49d},
one can derive that $\alpha_2 = 0$.  This results in a contradiction, thereby implying that $w(x,t,\xi_i) =\mathbf{0}$.
Based on the symmetries in Eqs.~\eref{xcx49},~\eref{xcx56} and~\eref{xcx62}, we can prove the remaining results in Eq.~\eref{xcx76}.

(ii) At the discrete eigenvalues $\xi_i$ ($\Re(\xi_i)\neq 0$, $1 \leq i \leq n_1$),
Eqs.~\eref{xcx54} and~\eref{xcx76} suggest that $\phi_{\pm, 1}$ and $\phi_{\pm, 3}$ obey the following relations:
\begin{equation}
	\label{ap46}
	\begin{aligned}
\phi_{-, 1}(x,t,\xi_i)    =  c_{i}   \phi_{+, 3}(x,t,\xi_i), \quad   \phi_{+, 1}(x,t,\xi_i^*) = \overline{c}_{i} \phi_{-, 3}(x,t,\xi_i^*),
	\end{aligned}
\end{equation}
where $c_{i}$ and $\overline{c}_i$ are constants.  Then, applying the symmetry~\eref{xcx56} to Eq.~\eref{ap46} yields
\begin{equation}
\label{ap47}
\begin{aligned}
& \phi_{+, 1}(x,t,-\xi_i)   = -c_{i} \phi_{-, 3}(x,t,-\xi_i), \quad  \phi_{-, 1}(x,t,-\xi_i^*) = -\overline{c}_{i} \phi_{+, 3}(x,t,-\xi_i^*).
\end{aligned}
\end{equation}
Further noticing that $\hat{\xi}_i = \xi_i^*$ ($1 \leq i \leq n_1$), we apply the symmetry~\eref{xcx62} to the first relation in Eq.~\eref{ap46}:
\begin{equation}
	\label{xcxB7}
	\begin{aligned}
		&\phi_{-, 3}(x,t,\xi_i^*)  =  - c_{i}   \phi_{+, 1}(x,t,\xi_i^*).
	\end{aligned}
\end{equation}
Substituting it into the second relation of Eq.~\eref{ap46} immediately yields $ c_{i} \overline{c}_{i} = -1$ ($1 \leq i \leq n_1$).

In order to prove $\overline{c}_{i} c_{i}^* \frac{ a'_{33}(\xi_i^*)}{ b'_{11}(\xi_i^*)} =  1$ ($1 \leq i \leq n_1$), we use the relations~\eref{xcx49b} and~\eref{xcx49c}, i.e.,
\begin{subequations}
\label{ap50}
\begin{align}
& - \bm{\sigma}_{3} [w \times \phi_{-,3}](x,t,z) e^{-i \theta_{2}(x,t,z)} = f(x,t,z) a_{33}(z), \label{ap50a} \\
& - \bm{\sigma}_{3} [w \times \phi_{+,1}](x,t,z) e^{-i \theta_{2}(x,t,z)} = g(x,t,z) b_{11}(z), \label{ap50b}
\end{align}
\end{subequations}
where $f(x,t,z):=\phi^*_{-, 1}(x,t,z^*)$ and $g(x,t,z):=\phi^*_{+, 3}(x,t,z^*)$.
Then, by using the second relation in Eq.~\eref{ap46} and $w(x,t,\xi_i^*) = 0$,
calculation of the derivative for the left-hand side of Eq.~\eref{ap50a} with respect to $z$ at $\xi_i$ gives
\begin{align}
	\label{apa51}
		{[\rm{LHS\,of}~\eref{ap50a}]'}_{z = \xi_i^*}
		=& - \bm{\sigma}_{3} \Big\{ [w'\times \phi_{-,3}](x,t,\xi_i^*)  e^{-\mi \theta_{2}(x,t,\xi_i^*)} + [w\times \phi'_{-,3}](x,t,\xi_i^*)  e^{-\mi \theta_{2}(x,t,\xi_i^*)}\notag\\
	 & \,\,\, \,\,\,\,\,\,\quad -\mi [w\times \phi_{-,3}](x,t,\xi_i^*) \theta'_{2}(x,t,\xi^*_i)e^{-\mi \theta_{2}(x,t,\xi_i^*)} \Big\}  \notag \\
=& - \frac{1}{\overline{c}_{i}} \bm{\sigma}_{3} {\big\{ [w \times \phi_{+,1}](x,t,z) e^{-\mi \theta_{2}(x,t,z)} \big\}}'_{z = \xi_i^*}  = \frac{1}{\overline{c}_{i}} [g(x,t,z) b_{11}(z)]'_{z = \xi_i^*}.
\end{align}
On the other hand, by using the first relation of Eq.~\eref{ap46} and  $a_{33}(\xi_i^*)= b_{11}(\xi_i^*)= 0$, it can be deduced that
\begin{align}
\label{apa52}
	{[\rm{RHS\,of}~\eref{ap50a}]}'_{z = \xi_i^*} & = f(x,t,\xi_i^*) a'_{33}(\xi_i^*)
	= c_i^* \phi^*_{+, 3}(x,t,\xi_i) a'_{33}(\xi_i^*)  \notag \\
	& = c_i^* \frac{ a'_{33}(\xi_i^*)}{ b'_{11}(\xi_i^*)} [g(x,t,z) b_{11}(z)]'_{z = \xi_i^*}.
\end{align}
Combining Eqs.~\eref{apa51} and~\eref{apa52} immediately produces $\overline{c}_{i} c_{i}^* \frac{ a'_{33}(\xi_i^*)}{ b'_{11}(\xi_i^*)} =  1$.

(iii) In particular, for the pure imaginary eigenvalues $\xi_{0,j}$ ($\Re(\xi_{0,j}) = 0, 1 \leq j \leq m_1$), one can also derive Eqs.~\eref{ap46} and~\eref{ap47} and thus have $ c_{0,j} \overline{c}_{0,j} = -1$. Meanwhile,  $\xi_{0,j} = -\xi_{0,j}^*$ implies that  $\overline{c}_{0,j}=-c_{0,j} $. Therefore, we arrive at $c^2_{0,j} = 1$ ($1 \leq j \leq m_1$).
%
%

\textbf{Proof of Lemma~\ref{le3.5}}:
(i) For the discrete eigenvalues $z_i$ ($\Re(z_i)\neq 0$, $1 \leq i \leq n_2$), in view of $a_{11}(z_i)= a_{11}(-z^*_i)=b_{33}(-\hat{z}_i)= b_{33}(\hat{z}^*_i)= 0$, Eq.~\eref{xcx49a} and~\eref{xcx49d} imply that Eq.~\eref{xcx81} hold, where $d_i$, $\breve{d}_{i}$, $\hat{d}_i$ and $\overline{d}_{i}$ are independent of $x$ and $t$. Then, based on the symmetry~\eref{xcx56}, we implement the transformations $\phi_{-, 1}(x,t,z)\to \delta_{1}  \phi_{+,1}(-x,t,-z)$, $\phi_{+, 3}(x,t,z)\to -\delta_{1}  \phi_{-,3}(-x,t,-z)$, $w(x, t, z)\to \delta_{1}  \overline{w}(-x,t,-z)$, and then make the parity transformation $x \to -x$. As a result, the four relations in Eq.~\eref{xcx81a} can be derived.

Next, in order to prove the constraint relations in Eq.~\eref{xcx81c}, we  apply Eq.~\eref{xcx62} and the second relation in Eq.~\eref{2.49} to the first and second relations of Eq.~\eref{xcx81a}, yielding
\begin{equation}
	\begin{aligned}
		- w(x,t,-\hat{z}_i) = -d_{i}    \frac{\mi\sqrt{2} q_{0}  }{z_i}   \phi_{+,3}(x,t,-\hat{z}_i), \quad
		- w(x,t,\hat{z}^*_i) = \breve{d}_{i}  \frac{ \mi\sqrt{2} q_{0}  }{z^*_i}   \phi_{+,3}(x,t,\hat{z}^*_i). \\
	\end{aligned}
\end{equation}
Comparing them with the last two relations of Eq.~\eref{xcx81} gives rise to
$\hat{d}_i= \frac{\mi \sqrt{2} q_0}{z_i} d_i$ and $\overline{d}_i= -\frac{\mi \sqrt{2} q_0}{z^*_i} \breve{d}_{i}$.
Then, substituting the first relation of~\eref{xcx81} into Eq.~\eref{xcx49a} with $z=z_i$ and conjugating on both sides yields
\begin{equation}
\label{B12}
	\begin{aligned}
		&\phi_{+, 1}(x,t,z^*_i)= \frac{d^*_{i} }{b^*_{33}(z_i)} e^{\mi \theta^*_{2}(x,t,z_i)} \bm{\sigma}_{3} [\phi^*_{-,1} \times \phi^*_{+,3}](x,t,z_i).
	\end{aligned}
\end{equation}
Using the second relation of~\eref{xcx81a}, Eq.~\eref{B12} becomes
\begin{equation}
	\label{hlj13}
	\begin{aligned}
		&\overline{w}(x,t,z^*_i) = \frac{d^*_{i} \breve{d}_i }{b^*_{33}(z_i)} e^{\mi \theta^*_{2}(x,t,z_i)} \bm{\sigma}_{3} [\phi^*_{-,1} \times \phi^*_{+,3}](x,t,z_i).
	\end{aligned}
\end{equation}
By comparing Eq.~\eref{hlj13} with~\eref{xcx54b}, one can derive $\frac{d^*_{i} \breve{d}_{i} }{b^*_{33}(z_i)} = \frac{1}{\gamma(z_i^*)}$. Meanwhile, we have $\overline{d}_i= -\frac{\mi \sqrt{2} q_0}{z^*_i} \breve{d}_{i} = - \frac{\mi \sqrt{2} q_0}{z^*_i}  \frac{b^*_{33}(z_i)}{\gamma(z_i^*) d_i^*}$.

(ii) For the discrete eigenvalues $z_{0,j}$ ($\Re(z_{0,j})= 0$,  $1 \leq j \leq m_2$), Eqs.~\eref{xcx81} and~\eref{xcx81a} still hold true. In view of $z_{0,j} = - z^*_{0,j}$,  there must be $d_{0,j} = \breve{d}_{0,j}$ and $\hat{d}_{0,j} = \overline{d}_{0,j}$, and Eqs.~\eref{xcx81} and~\eref{xcx81a} is reducible to Eq.~\eref{xcx81d}. In this case, the first relation of Eq.~\eref{xcx81c} is equivalent to $|d_{0,j}|^2 = \frac{\gamma(z_{0,j}^*) }{b^*_{33}(z_{0,j})}$ and thus Eq.~\eref{xcx81e} can be proved.

\subsection{Asymptotic behaviors}
\label{appendixA9}
\textbf{Proof of Lemma~\ref{co3.6}}:

As before, the dependence on $x$, $t$ and $z$ is omitted for the involved functions. By referring to Eqs.~\eref{Laxpairjvzhenxingshib}, \eref{2.7}, \eref{xcx17} and~\eref{sm99},  we expand $\mathbf{Y}^{\rm{adj}}_{\pm}$ (the adjugate matrix of $\mathbf{Y}_{\pm}$), $\mi \mathbf{J} $, $\mi \bm{\Omega}$ and $\bm{\Delta}\!\mathbf{T}_{\pm}$ as
\begin{equation}
\begin{aligned}
		& \mathbf{Y}^{\rm{adj}}_{\pm} = \mathbf{Y}^{\rm{adj}}_{\pm,0} + \frac{1}{z} \mathbf{Y}^{\rm{adj}}_{\pm,-1} + \frac{1}{z^2} \mathbf{Y}^{\rm{adj}}_{\pm,-2}, \\
		&\mi \mathbf{J}= z \mathbf{J}_{1} + \frac{1}{z} \mathbf{J}_{-1}, \quad
		\mi \bm{\Omega}= z^{2} \bm{\Omega}_{2} +  \bm{\Omega}_{0} + \frac{1}{z^{2}} \bm{\Omega}_{-2}, \\
		&\bm{\Delta}\!\mathbf{T}_{\pm} = z \bm{\Delta}\!\mathbf{T}_{1}^{\pm}  +  \bm{\Delta}\!\mathbf{T}_{0}^{\pm}  + \frac{1}{z} \bm{\Delta}\!\mathbf{T}_{-1}^{\pm},
	\end{aligned}
\end{equation}
where
\begin{equation}
	\begin{aligned}
		& \mathbf{Y}^{\rm{adj}}_{\pm,0}= \left(
		\begin{array}{ccc}
			\frac{q^{*}_{\pm}}{q_0} & \frac{q^*_{\mp}}{q_0} & 0 \\
			-\frac{\sqrt{2} q_{\mp}}{q_0} & \frac{\sqrt{2} q_{\pm}}{q_0} & 0 \\
			0 & 0 & \sqrt{2} \\
		\end{array}
		\right),  \quad
		\mathbf{Y}^{\rm{adj}}_{\pm,-1}  = \left(
		\begin{array}{ccc}
			0 & 0 & 2 \mi q_0 \\
			0 & 0 & 0 \\
			-\mi \sqrt{2} q^*_{\pm} & -\mi \sqrt{2} q^*_{\mp} & 0 \\
		\end{array}
		\right), \\
		&  \mathbf{Y}^{\rm{adj}}_{\pm,-2}  = \left(
		\begin{array}{ccc}
			0 & 0 & 0 \\
			2 \sqrt{2} q_{\mp} q_0 & -2 \sqrt{2} q_{\pm} q_0 & 0 \\
			0 & 0 & 0 \\
		\end{array}
		\right),\\
		&\mathbf{J}_{1} = \bm{\Omega}_{2} = \diag\Big(-\frac{\mi}{2},-\frac{\mi}{2},\frac{\mi}{2}\Big), \quad
		\bm{\Omega}_{0} = \diag(0,-\mi q_0^2,0), \\
		&\mathbf{J}_{-1} = \frac{1}{2 q_0^2} \bm{\Omega}_{-2} =  \diag(\mi q_0^2,-\mi q_0^2,-\mi q_0^2), \quad \bm{\Delta}\!\mathbf{T}_{1}^{\pm}  =    \bm{\Delta}\!\mathbf{Q}_{\pm}, \\
		& \bm{\Delta}\!\mathbf{T}_{0}^{\pm}  =  \mi \bm{\Delta}\!\mathbf{Q}_{\pm,x} \bm{\sigma}_{3} + +\mi \bm{\sigma}_{3}{\mathbf{Q}}^2 - \mi \bm{\sigma}_{3}\mathbf{Q}^2_{\pm}, \quad
		\bm{\Delta}\!\mathbf{T}_{-1}^{\pm}  = 2 q^2_{0}  \bm{\Delta}\!\mathbf{Q}_{\pm}.  \\
	\end{aligned}
\end{equation}

Then, substituting the above formulas into Eqs.~\eref{xcx83} yields
\begin{align}
		& \Big[\Big( \mathbf{Y}^{\rm{adj}}_{\pm,0} + \frac{1}{z} \mathbf{Y}^{\rm{adj}}_{\pm,-1} +...\Big) \Big(\bm{\mu}^{\pm}_{0} +  \frac{1}{z} \bm{\mu}^{\pm}_{-1} +...\Big) \Big]_x  \notag \\
		& \,\,\,\,\,\, + \Big[ \Big(\mathbf{Y}^{\rm{adj}}_{\pm,0} + \frac{1}{z} \mathbf{Y}^{\rm{adj}}_{\pm,-1} +...\Big) \Big(\bm{\mu}^{\pm}_{0} +  \frac{1}{z} \bm{\mu}^{\pm}_{-1} \Big), \Big(z \mathbf{J}_{1} + \frac{1}{z} \mathbf{J}_{-1} +...\Big)  \Big] \notag \\
		&  = \Big(\mathbf{Y}^{\rm{adj}}_{\pm,0} + \frac{1}{z} \mathbf{Y}^{\rm{adj}}_{\pm,-1} +...\Big) \bm{\Delta}\!\mathbf{Q}_{\pm} \Big(\bm{\mu}^{\pm}_{0} +  \frac{1}{z} \bm{\mu}^{\pm}_{-1} +...\Big),\label{ap68}
	\end{align}
and
\begin{equation}
	\label{ap69}
	\begin{aligned}
		& \Big[ \Big(\mathbf{Y}^{\rm{adj}}_{\pm,0} + \frac{1}{z} \mathbf{Y}^{\rm{adj}}_{\pm,-1} +...\Big) \Big(\bm{\mu}^{\pm}_{0} +  \frac{1}{z} \bm{\mu}^{\pm}_{-1} +...\Big)  \Big]_t  \\
		& \,\,\,\,\,\,+ \Big[ \Big(\mathbf{Y}^{\rm{adj}}_{\pm,0} + \frac{1}{z} \mathbf{Y}^{\rm{adj}}_{\pm,-1} +...\Big) \Big(\bm{\mu}^{\pm}_{0} +   \frac{1}{z}\bm{\mu}^{\pm}_{-1} +...\Big), \Big(z^{2} \bm{\Omega}_{2} +  \bm{\Omega}_{0} + \frac{1}{z^{2}} \bm{\Omega}_{-2} \Big)  \Big] \\
		& = \Big(\mathbf{Y}^{\rm{adj}}_{\pm,0} + \frac{1}{z} \mathbf{Y}^{\rm{adj}}_{\pm,-1} +...\Big) \Big(z \bm{\Delta}\!\mathbf{T}_{1}^{\pm}  +  \bm{\Delta}\!\mathbf{T}_{0}^{\pm}  + \frac{1}{z} \bm{\Delta}\!\mathbf{T}_{-1}^{\pm} +...\Big)  \Big(\bm{\mu}^{\pm}_{0} +  \frac{1}{z} \bm{\mu}^{\pm}_{-1} +...\Big).
	\end{aligned}
\end{equation}
Comparing the first few highest power terms of $z$  on both sides of Eqs.~\eref{ap68} and~\eref{ap69} yields
\begin{subequations}
	\begin{align}
		&  \big[\mathbf{Y}^{\rm{adj}}_{\pm,0} \bm{\mu}^{\pm}_{0},  \mathbf{J}_{1} \big] = 0, \label{ap70a} \\
		& \big( \mathbf{Y}^{\rm{adj}}_{\pm,0} \bm{\mu}^{\pm}_{0} \big)_x + \big[\mathbf{Y}^{\rm{adj}}_{\pm,-1} \bm{\mu}^{\pm}_{0} + \mathbf{Y}^{\rm{adj}}_{\pm,0} \bm{\mu}^{\pm}_{-1},  \mathbf{J}_{1} \big] = \mathbf{Y}^{\rm{adj}}_{\pm,0} \bm{\Delta}\!\mathbf{Q}_{\pm} \bm{\mu}_{\pm,0}, \label{ap70b} \\	
	&  \big[\mathbf{Y}^{\rm{adj}}_{\pm,-1} \bm{\mu}^{\pm}_{0} + \mathbf{Y}^{\rm{adj}}_{\pm,0} \bm{\mu}^{\pm}_{-1},  \mathbf{J}_{1} \big]  = \mathbf{Y}^{\rm{adj}}_{\pm,0} \bm{\Delta}\!\mathbf{Q}_{\pm} \bm{\mu}^{\pm}_{0}. \label{ap71b}
	\end{align}
\end{subequations}

From Eqs.~\eref{ap70b} and~\eref{ap71b}, we deduce that $\big( \mathbf{Y}^{\rm{adj}}_{\pm,0} \bm{\mu}^{\pm}_{0} \big)_x = 0$, which means $\bm{\mu}^{\pm}_{0}$ is independent of $x$. Meanwhile, considering the asymptotic behaviors of $\bm{\mu}_{\pm}$ in Eq.~\eref{4}, we can obtain
\begin{equation}
	\label{hljB24a}
	\begin{aligned}
		\bm{\mu}^{\pm}_{0} = \left(
		\begin{array}{ccc}
			\frac{q_{\pm}}{q_0} & -\frac{q^{*}_{\mp}}{\sqrt{2} q_0}  & 0 \\
			\frac{q_{\mp}}{q_0} &  \frac{q^{*}_{\pm}}{\sqrt{2}q_0} & 0  \\
			0 & 0 & \sqrt{2} \\
		\end{array}
		\right).
	\end{aligned}
\end{equation}
Bringing~\eref{hljB24a} back into~\eref{ap70b} gives rise to Eqs.~\eref{xcx84} and~\eref{xcx84b} together with~\eref{xcx86}.

Based on the above results, we can use the symmetry~\eref{xcx62} to directly obtain the asymptotic behaviors of $\mu_{\pm,1}$ and $\mu_{\pm,3}$ as $z\to 0$, i.e., Eqs.~\eref{xcx85} and~\eref{xcx85b}.

\subsection{Trace formulas}
\label{appendixC3}

\textbf{Proof of Lemma~\ref{le4.4}}:
Note that $a_{33}(z)$ and $b_{33}(z)$ are, respectively, analytic in the regions $C^-$ and $C^+$,  and recall that $a_{33}(z)$ admits simple zeros  $\{-\xi_i,  \xi_i^*\}^{n_1}_{i=1}$, $\{-\xi_{0,j}\}^{m_1}_{i=1}$, $\{\hat{z}_i,  -\hat{z}_i^*\}^{n_2}_{i=1}$ and $\{\hat{z}_{0,j}\}^{m_2}_{i=1}$, while $b_{33}(z)$ has simple zeros $\{\xi_i,  -\xi_i^*\}^{n_1}_{i=1}$, $\{\xi_{0,j}\}^{m_1}_{i=1}$, $\{-\hat{z}_i,  \hat{z}_i^*\}^{n_2}_{i=1}$ and $\{-\hat{z}_{0,j}\}^{m_2}_{i=1}$.  Then, we introduce the following functions:
\begin{equation}
	\label{sl39}
	\begin{aligned}
		&\beta_{1}(z) = a_{33}(z) \displaystyle\prod_{i = 1}^{n_1} \Big(\frac{ z - \xi_i  }{z- \xi^*_i }   \frac{ z + \xi^*_i }{z + \xi_i }\Big)
		\displaystyle\prod_{i = 1}^{m_1}  \frac{ z - \xi_{0,j} }{z + \xi_{0,j} }
		\displaystyle\prod_{i = 1}^{n_2} \Big(\frac{ z - \hat{z}^*_i  }{z + \hat{z}^*_i}   \frac{ z +\hat{z}_i }{z -\hat{z}_i }\Big)
		\displaystyle\prod_{i = 1}^{m_2} \frac{ z +\hat{z}_{0,j} }{z -\hat{z}_{0,j} },  \\
		&\beta_{2}(z) = b_{33}(z) \displaystyle\prod_{i = 1}^{n_1} \Big(\frac{ z - \xi^*_i  }{z- \xi_i }   \frac{ z + \xi_i }{z + \xi^*_i }\Big)
		\displaystyle\prod_{i = 1}^{m_1} \frac{ z + \xi_{0,j} }{z - \xi_{0,j} }
		\displaystyle\prod_{i = 1}^{n_2} \Big(\frac{  z + \hat{z}^*_i }{z -\hat{z}^*_i }   \frac{ z -\hat{z}_i  }{z+ \hat{z}_i }\Big)
		\displaystyle\prod_{i = 1}^{m_2}  \frac{ z -\hat{z}_{0,j} }{z +\hat{z}_{0,j} },
	\end{aligned}
\end{equation}
which admits the analyticity in the regions $C^-$  and $C^+$, respectively,  and have no zeros in the corresponding regions.

From $  \mathbf{A}(z) \mathbf{B}(z)  = \mathbf{I}$ for all $z \in \Sigma_o$, we deduce that
\begin{equation}
	\label{B.21}
	\begin{aligned}
		&\frac{1}{a_{33}(z) b_{33}(z)} = 1 + \frac{a_{13}(z) b_{31}(z)}{a_{33}(z) b_{33}(z)} + \frac{a_{23}(z) b_{32}(z)}{a_{33}(z) b_{33}(z)}.
	\end{aligned}
\end{equation}
According to
\begin{align}
& \rho_{1}(z) = - \frac{b_{31}( \hat{z} )}{b_{33}( \hat{z} ) } = - \frac{a^*_{31}(z^*)}{a^*_{11}(z^*)}= \frac{a^*_{13}(\hat{z}^*)}{a^*_{33}(\hat{z}^*)} , \notag \\
& \rho_{2}(z)  =   \frac{\mi \sqrt{2} q_0 }{ z } \frac{a_{23}( \hat{z})}{a_{33}(\hat{z}) }= - \frac{\mi \sqrt{2} q_0 }{ z \gamma(\hat{z}) }  \frac{b^*_{32}(\hat{z}^*)}{b^*_{33}(\hat{z}^*)}, \notag
\end{align}
Eq.~\eref{B.21} can be written as
\begin{equation}
	\begin{aligned}
		&\frac{1}{a_{33}(z) b_{33}(z)} = 1 -  \rho_{1}( \hat{z}) \rho^*_{1}( \hat{z}^*)   - \frac{\hat{z}^2 \gamma^*( \hat{z}^*) }{2 q^2_{0}}\rho_{2}( \hat{z})  \rho^*_{2}( \hat{z}^*).
	\end{aligned}
\end{equation}

Then, the relation $\beta_{1} \beta_{2} = a_{33} b_{33}$ implies that
\begin{equation}
	\begin{aligned}
		&\log \beta_{1}  + \log \beta_{2} =  - \log \Big( 1 -  \rho_{1}( \hat{z}) \rho^*_{1}( \hat{z}^*)   - \frac{\hat{z}^2 \gamma^*( \hat{z}^*) }{2 q^2_{0}}\rho_{2}( \hat{z})  \rho^*_{2}( \hat{z}^*) \Big),\quad z \in \Sigma_o. \label{EqB23}
	\end{aligned}
\end{equation}
From the limiting conditions $\lim\limits_{z \to \pm \infty} \beta_{1} =  \lim\limits_{z \to \pm \infty} \beta_{2} = 1$, applying Plemelj's formula to Eq.~\eref{EqB23} yields
\begin{equation}
	\label{sm41b}
	\begin{aligned}
		&\log \beta_{1}=  - \frac{1}{2 \pi \mi}  \int_{\Sigma}^{ }  \frac{\log \Big( 1 -  \rho_{1}( \hat{z}) \rho^*_{1}( \hat{z}^*)   - \frac{\hat{z}^2 \gamma^*( \hat{z}^*) }{2 q^2_{0}}\rho_{2}( \hat{z})  \rho^*_{2}( \hat{z}^*) \Big) }{s-z}ds,\quad z \in C^-, \\
		&\log \beta_{2} = \frac{1}{2 \pi \mi}  \int_{\Sigma}^{ }  \frac{\log \Big( 1 -  \rho_{1}( \hat{z}) \rho^*_{1}( \hat{z}^*)   - \frac{\hat{z}^2 \gamma^*( \hat{z}^*) }{2 q^2_{0}}\rho_{2}( \hat{z})  \rho^*_{2}( \hat{z}^*) \Big) }{s-z}ds,\quad z \in C^+, \\
	\end{aligned}
\end{equation}
Finally, substituting Eq.~\eref{sl39} into Eqs.~\eref{sm41b},  we arrive at Eqs.~\eref{xcx108a} and~\eref{xcx108b}.

\section{}
\label{appendixC}
\renewcommand{\theequation}{C.\arabic{equation}}
\setcounter{equation}{0}

\subsection{The construction of RHP}
\label{appendixC1}

\textbf{Proof of Lemma~\ref{le4.1}}:
For brevity, the $x$-, $t$- and $z$-dependence in $\phi_{\pm,j}$ and $w$ is omitted here. It follows from  Eq.~\eref{sm1} that
\begin{subequations}
	\label{cp1}
	\begin{align}
		&\phi_{+, 1}  = \frac{\phi_{-, 1}}{a_{11}(z)}  - \frac{ a_{21}(z)}{a_{11}(z)} \phi_{+, 2} -  \frac{ a_{31}(z) }{a_{11}(z)} \phi_{+, 3}, \label{cp1a} \\
		&\frac{\phi_{-, 3} }{a_{33}(z)} =  \frac{ a_{13}(z) }{a_{33}(z)} \phi_{+, 1}  +  \frac{ a_{23}(z) }{a_{33}(z)} \phi_{+, 2} + \phi_{+, 3}. \label{cp1b}
	\end{align}
\end{subequations}
Based on Eq.~\eref{sm48a}, we can replace $\phi_{+, 2}$ in Eq.~\eref{cp1a} with $ \frac{1}{b_{33}(z)} ( b_{32}(z) \phi_{+,3} + w )$:
\begin{equation}
	\label{cp2}
	\begin{aligned}
\phi_{+, 1}	= \frac{\phi_{-, 1} }{a_{11}(z)}  - \frac{ a_{21}(z) }{a_{11}(z)} \frac{ w }{b_{33}(z)} - \Big(\frac{ a_{31}(z) }{a_{11}(z)} + \frac{ b_{32}(z) }{b_{33}(z)} \frac{ a_{21}(z) }{a_{11}(z)} \Big)\phi_{+, 3}.
	\end{aligned}
\end{equation}
Then, using the first relation of~\eref{sm48a} and Eq.~\eref{cp2} to remove $\phi_{+, 1}$ and $\phi_{+, 2}$ in Eq.~\eref{cp1b} yields
\begin{equation}
	\label{cp3}
	\begin{aligned}
\frac{\phi_{-, 3} }{a_{33}(z)} =&  \frac{ a_{13}(z) }{a_{33}(z)} \frac{\phi_{-, 1} }{a_{11}(z)}  + \Big(\frac{ a_{23}(z) }{a_{33}(z)}  - \frac{ a_{13}(z) }{a_{33}(z)} \frac{ a_{21}(z) }{a_{11}(z)} \Big) \frac{ w }{b_{33}(z)} \\
		&- \big(\frac{ a_{13}(z) }{a_{33}(z)} \frac{ a_{31}(z) }{a_{11}(z)}   +   \frac{ a_{13}(z) }{a_{33}(z)} \frac{ b_{32}(z) }{b_{33}(z)} \frac{ a_{21}(z) }{a_{11}(z)}  - \frac{ a_{23}(z) }{a_{33}(z)}  \frac{b_{32}(z)}{b_{33}(z)}-  1 \big) \phi_{+, 3}.
	\end{aligned}
\end{equation}
On the other hand, combining the second relation of~\eref{sm48a} and Eq.~\eref{cp2} gives
\begin{equation}
     \label{cp4}
	\begin{aligned}
		\frac{\overline{w} }{b_{11}(z)} 	= &  \frac{b_{12}(z)}{b_{11}(z)} \frac{\phi_{-, 1} }{a_{11}(z)}  -   \Big(\frac{b_{12}(z)}{b_{11}(z)}  \frac{ a_{21}(z) }{a_{11}(z)} + 1 \Big) \frac{ w }{b_{33}(z)}  \\
 & - \Big(\frac{ a_{21}(z) }{a_{11}(z)} \frac{ b_{12}(z) }{b_{11}(z)} \frac{ b_{32}(z) }{b_{33}(z)} +  \frac{ a_{31}(z) }{a_{11}(z)} \frac{ b_{12}(z) }{b_{11}(z)}  + \frac{b_{32}(z)}{b_{33}(z)} \Big)\phi_{+, 3}.  \\
	\end{aligned}
\end{equation}

Equivalently, Eqs.~\eref{cp2},~\eref{cp3} and~\eref{cp4} can be written in the matrix form
\begin{equation}
	\label{C.5}
	\begin{aligned}
		&\Big(\phi_{+,1}, - \frac{\overline{w} }{b_{11}(z)}, \frac{\phi_{-,3}}{a_{33}(z)} \Big)  = \Big(\frac{\phi_{-,1}}{a_{11}(z)}, \frac{w }{b_{33}(z)}, \phi_{+,3} \Big) \mathbf{G}(z),
	\end{aligned}
\end{equation}
with
\begin{equation}
	\begin{aligned}
		&G_{11} = 1, \quad G_{12} = -\frac{b_{12}(z)}{b_{11}(z)}, \quad G_{13} = \frac{ a_{13}(z) }{a_{33}(z)} \\
		&G_{21} = - \frac{ a_{21}(z) }{a_{11}(z)}, \quad G_{22} = \frac{b_{12}(z)}{b_{11}(z)}  \frac{ a_{21}(z) }{a_{11}(z)} + 1, \quad
		G_{23} = \frac{ a_{23}(z) }{a_{33}(z)}  - \frac{ a_{13}(z) }{a_{33}(z)} \frac{ a_{21}(z) }{a_{11}(z)} \\
		&G_{31} = - \Big(\frac{ a_{31}(z) }{a_{11}(z)} + \frac{ b_{32}(z) }{b_{33}(z)} \frac{ a_{21}(z) }{a_{11}(z)} \Big), \quad
		G_{32} = \frac{ a_{21}(z) }{a_{11}(z)} \frac{ b_{12}(z) }{b_{11}(z)} \frac{ b_{32}(z) }{b_{33}(z)} +  \frac{ a_{31}(z) }{a_{11}(z)} \frac{ b_{12}(z) }{b_{11}(z)}  + \frac{b_{32}(z)}{b_{33}(z)}, \\
		&G_{33} = - \Big(\frac{ a_{13}(z) }{a_{33}(z)} \frac{ a_{31}(z) }{a_{11}(z)}   +   \frac{ a_{13}(z) }{a_{33}(z)} \frac{ b_{32}(z) }{b_{33}(z)} \frac{ a_{21}(z) }{a_{11}(z)}
		- \frac{ a_{23}(z) }{a_{33}(z)}  \frac{b_{32}(z)}{b_{33}(z)}-  1 \Big).
	\end{aligned}
\end{equation}
Recalling  the relations in Eqs.~\eref{Eq250}--\eref{xcx71}, we have
\begin{equation}
	\begin{aligned}
		&\frac{b_{12}(z)}{b_{11}(z)} = \frac{\rho^*_{2}(z^*)}{\gamma(z)},\quad \frac{ a_{31}(z) }{a_{11}(z) } = -\rho^*_{1}(z^*),\quad \frac{ a_{21}(z) }{a_{11}(z) } = \rho_{2}(z),\\
		&\frac{ a_{23}(z) }{a_{33}(z) }  = \frac{\rho_{2}(\hat{z})\hat{z}}{\mi \sqrt{2} q_{0}},
		\quad \frac{ a_{13}(z) }{a_{33}(z) } = \rho^*_{1}(\hat{z}^*),\quad
		\frac{ b_{32}(z) }{b_{33}(z) }  =  \frac{\mi \sqrt{2} q_{0}}{\hat{z}} \frac{\rho_{2}^*(\hat{z}^*)}{\gamma(\hat{z})}.
	\end{aligned}
\end{equation}
Thus, Eq.~\eref{C.5} can be finally written as Eq.~\eref{xcx91}.

\subsection{Proof of Theorem~\ref{Thm5.1}}
\label{appendixC4}

Based on Lemmas~\ref{le4.2} and~\ref{le4.2b}, the first and second columns of $\mathbf{M}^{-}$ in the reflectionless case can be represented as
\begin{align}
		M^-_{1}(x,t,z)  = &
		\left(
		\begin{array}{ccc}
			\frac{q_{+}}{q_0}  \\
			\frac{q_{-}}{q_0}  \\
			\frac{ 2 \mi  q_0 }{z}  \\
		\end{array}
		\right) + \displaystyle\sum_{i = 1}^{n_1} \bigg( \frac{C_{i}}{z-\xi_i} M^{+}_{3} (x,t,\xi_i) +   \frac{  \overline{C}_{i} }{z+\xi^*_i} M^{+}_{3} (x,t,-\xi^*_i)    \bigg) \notag\\
		&+ \displaystyle\sum_{i = 1}^{n_2} \bigg(\frac{ E_{i}}{z - z_i} M^{+}_{2} (x,t,z_i) + \frac{  \breve{E}_{i} }{z + z^*_i}  M^{+}_{2} (x,t,-z^*_i)    \bigg) \notag \\
		&+ \displaystyle\sum_{i = 1}^{m_1} \frac{C_{0,j}}{z-\xi_{0,j}} M^{+}_{3} (x,t,\xi_{0,j})
		+ \displaystyle\sum_{i = 1}^{m_2}\frac{ E_{0,j}}{z - z_{0,j}} M^{+}_{2} (x,t,z_{0,j}), \label{xcx99} \\
		M^-_{2}(x,t,z)  = &
		\left(
		\begin{array}{ccc}
			- \frac{q^{*}_{-}}{\sqrt{2} q_0}  \\
			\frac{q^{*}_{+}}{\sqrt{2} q_0}   \\
			0  \\
		\end{array}
		\right) + \displaystyle\sum_{i = 1}^{n_2} \bigg( \frac{\hat{E}_{i}}{z+\hat{z}_i} M^{+}_{3} (x,t,-\hat{z}_i) +   \frac{  \overline{E}_{i} }{z-\hat{z}^*_i} M^{+}_{3} (x,t,\hat{z}^*_i)    \bigg)\notag\\
		&+ \displaystyle\sum_{i = 1}^{n_2} \bigg(\frac{ F_{i}}{z + z_i} M^{-}_{1} (x,t,-z_i) + \frac{ \breve{F}_{i} }{z - z^*_i} M^{-}_{1} (x,t,z^*_i)    \bigg) \notag \\
		&+ \displaystyle\sum_{i = 1}^{m_2}  \frac{\hat{E}_{0,j}}{z+\hat{z}_{0,j}} M^{+}_{3} (x,t,-\hat{z}_{0,j})
		+ \displaystyle\sum_{i = 1}^{m_2} \frac{ F_{0,j}}{z + z_{0,j}} M^{-}_{1} (x,t,-z_{0,j}).
\end{align}
By combing Eq.~\eref{xcx64} and Lemma~\ref{le2.19}, we can get
\begin{equation}
M^{+}_{3}(x,t,z) = -\frac{\mi \sqrt{2} q_{0} }{z} M^{-}_{1}(x,t,\hat{z}), \quad  M^{+}_{2}(x,t,z) =  M^{-}_{2}(x,t,\hat{z}). \label{EqC9}
\end{equation}
Then, by virtue of~\eref{EqC9},  the $(1,1)$- and $(1,2)$-elements of $\mathbf{M}^-$ can be derived as
\begin{align}
		M^-_{11}(x,t,z)  =&
		\frac{q_{+}}{q_0}
		- \displaystyle\sum_{i = 1}^{n_1} \bigg( \frac{C_{i}}{z-\xi_i} \frac{\mi \sqrt{2} q_{0} }{\xi_i}M^{-}_{11}(x,t,\xi_i^*)
		+   \frac{  \overline{C}_{i} }{z+\xi^*_i}   \frac{\mi \sqrt{2} q_{0} }{-\xi^*_i} M^{-}_{11}(x,t,-\xi_i)    \bigg) \notag \\
		&+ \displaystyle\sum_{i = 1}^{n_2} \bigg(\frac{ E_{i}}{z - z_i}  M^{-}_{12} (x,t,\hat{z}_i)
		+ \frac{  \breve{E}_{i} }{z + z^*_i}  M^{-}_{12} (x,t,-\hat{z}^*_i)  \bigg) \notag \\
		&- \displaystyle\sum_{j =1}^{m_1} \frac{C_{0,j}}{z-\xi_{0,j}} \frac{\mi \sqrt{2} q_{0} }{\xi_{0,j}} M^{-}_{11}(x,t,-\xi_{0,j})
		+ \displaystyle\sum_{j =1}^{m_2} \frac{ E_{0,j}}{z - z_{0,j}}  M^{-}_{12} (x,t,\hat{z}_{0,j}), \label{C11}\\
		M^-_{12} (x,t,z) =&
		- \frac{q^{*}_{-}}{\sqrt{2} q_0}
		- \displaystyle\sum_{i = 1}^{n_2} \Big[
		\Big(\frac{  \overline{E}_{i} }{z-\hat{z}^*_i}  \frac{\mi \sqrt{2} q_{0} }{\hat{z}^*_i }
		- \frac{ \breve{F}_{i} }{z - z^*_i} \Big) M^{-}_{11} (x,t,z^*_i) \notag \\
		& \quad +
		\Big(\frac{\hat{E}_{i}}{z+\hat{z}_i} \frac{\mi \sqrt{2} q_{0} }{-\hat{z}_i}
		-   \frac{ F_{i}}{z + z_i}  \Big) M^{-}_{11}(x,t,-z_i)
		\Big] \notag \\
		&- \displaystyle\sum_{j = 1}^{m_2} \Big(\frac{  \hat{E}_{0,j} }{z+\hat{z}_{0,j}}  \frac{\mi \sqrt{2} q_{0} }{-\hat{z}_{0,j} }
		- \frac{ F_{0,j} }{z + z_{0,j}} \Big) M^{-}_{11} (x,t,-z_{0,j}). \label{C12}
\end{align}

If we sequentially select $z$ from $\{-\xi_i,\xi_i^*\}^{n_1}_{i=1}\cup\{-\xi_{0,j} \}^{m_1}_{j=1}\cup\{-z_i,z_i^*\}^{n_2}_{i=1}\cup\{-z_{0,j} \}^{m_2}_{j=1}$ for Eq.~\eref{C11} and from  $\{\hat{z}_i, -\hat{z}_i^*\}^{n_2}_{i=1}$ and $\{\hat{z}_{0,j} \}^{m_2}_{j=1}$ for Eq.~\eref{C12},  then one can get a system of equations involving $N_1+2N_2$ ($N_1 = 2n_1+m_1$, $N_2 = 2n_2+m_2$) unknowns. Thus, solving the resultant equations by Crammer's rule yields
\begin{equation}
	\begin{aligned}
		&M^{-}_{11}(x,t,\xi_i^*)  = \frac{\det(\mathbf{S}^{\rm{ext}}_i)}{\det(\mathbf{S})}, \quad
		M^{-}_{11}(x,t,-\xi_i)   = \frac{\det(\mathbf{S}^{\rm{ext}}_{i+n_1})}{\det(\mathbf{S})}, \quad 1 \leq  i \leq n_1, \\
		&M^{-}_{11}(x,t,-\xi_{0,j})   = \frac{\det(\mathbf{S}^{\rm{ext}}_{j+2n_1})}{\det(\mathbf{S})}, \quad 1 \leq  j \leq m_1, \\
		&M^{-}_{12} (x,t,\hat{z}_i)   = \frac{\det(\mathbf{S}^{\rm{ext}}_{i+N_1})}{\det(\mathbf{S})}, \quad
		M^{-}_{12} (x,t,-\hat{z}^*_i)   = \frac{\det(\mathbf{S}^{\rm{ext}}_{i+N_1+n_2})}{\det(\mathbf{S})}, \quad 1 \leq  i \leq n_2, \\
		&M^{-}_{12} (x,t,\hat{z}_{0,j})   = \frac{\det(\mathbf{S}^{\rm{ext}}_{j+N_1+2n_2})}{\det(\mathbf{S})}, \quad 1 \leq  j \leq m_2, \\
	\end{aligned}
\end{equation}
where $\mathbf{S}^{\rm{ext}}_k = (S_1,..., S_{k-1}, -\mathbf{K}, S_{k+1},..., S_{N_1+2N_2})$ with $S_j$ being the $j$th column of $\mathbf{S}$. Then, Eq.~\eref{xcx107} can be written as
\begin{align}
	\label{xcx79}
	q(x,t)  =& q_{+} +
	\mi \frac{\sqrt{2}}{2}
	\bigg[ \displaystyle\sum_{i = 1}^{n_1}
	\Big(
	D_{i}  \frac{\det(\mathbf{S}^{\rm{ext}}_i)}{\det(\mathbf{S})}+
	\overline{D}_{i}  \frac{\det(\mathbf{S}^{\rm{ext}}_{i+n_1})}{\det(\mathbf{S})}
	\Big)
	+
	\displaystyle\sum_{i = 1}^{n_2}
	\Big(
	\hat{F}_{i}  \frac{\det(\mathbf{S}^{\rm{ext}}_{i+N_1})}{\det(\mathbf{S})}+
	\overline{F}_{i} \frac{\det(\mathbf{S}^{\rm{ext}}_{i+N_1+n_2})}{\det(\mathbf{S})}
	\Big)  	 \notag \\
	&+ \displaystyle\sum_{j =1}^{m_1} D_{0,j} \frac{\det(\mathbf{S}^{\rm{ext}}_{j+2n_1})}{\det(\mathbf{S})}
	+ \displaystyle\sum_{j = 1}^{m_2} \hat{F}_{0,j}  \frac{\det(\mathbf{S}^{\rm{ext}}_{j+N_1+2n_2})}{\det(\mathbf{S})}
	\bigg] \notag \\
	= & q_{+} \bigg[ 1 +  \frac{1}{\det(\mathbf{S})} \displaystyle\sum_{k= 1}^{N_1 +2 N_2} (-1)^{k+2} H_{k}
	\det(\mathbf{K}, S_1,..., S_{k-1}, S_{k+1},..., S_{N_1+2N_2})  \bigg],
\end{align}
which is exactly the Laplace expansion of $\det(\mathbf{S}^{\rm{aug}})$, where $H_k$ is defined in Eq.~\eref{Eq57}.  \hfill\endproof

\end{appendix}

\end{CJK}

\end{document}